\documentclass[letterpaper,11pt,fleqn]{article}

\usepackage{jheppub}
\notoc

\usepackage{graphicx}
\usepackage[figuresright]{rotating}
\usepackage{bm,amsmath,amssymb}
\usepackage[mathscr]{eucal}
\usepackage{color}
\usepackage{caption}
\usepackage[normalem]{ulem}
\usepackage{setspace}

\long\def\comment#1{ }
\newcommand{\eqn}[1]{Eq.~\eqref{#1}}
\newcommand{\beq}{\begin{eqnarray}}
  \newcommand{\eeq}{\end{eqnarray}}

\newcommand{\nn}{\nonumber\\}
\newcommand{\dif}{{\rm d}}
\newcommand{\rmd}{{\rm d}}
\newcommand{\rme}{{\rm e}}
\newcommand{\rmi}{{\rm i}}

\newcommand{\rmI}{{\rm I}}
\newcommand{\rmJ}{{\rm J}}
\newcommand{\rmc}{{\rm c}}
\newcommand{\del}{\partial}

\newcommand{\order}[1]{\mcal{O}{(#1)}}
\newcommand{\mcal}{\mathcal}

\newcommand{\bx}{\bm{x}}
\newcommand{\by}{\bm{y}}
\newcommand{\bu}{\bm{u}}

\newcommand{\bz}{\bm{z}}

\newcommand{\br}{\bm{r}}

\newcommand{\abar}{\bar{\alpha}_s}

\newcommand{\xbj}{x_{\rm\scriptscriptstyle Bj}}

\newcommand{\sdla}{{\rm \scriptscriptstyle DLA}}

\newcommand{\gammap}{\gamma_{\scriptscriptstyle \mathbb{P}}}
\newcommand{\omegap}{\bar{\omega}_{\scriptscriptstyle \mathbb{P}}}
\newcommand{\betap}{\beta_{\scriptscriptstyle \mathbb{P}}}
\newcommand{\Dp}{D_{\scriptscriptstyle \mathbb{P}}}
\newcommand{\thetap}{\theta_{\scriptscriptstyle \mathbb{P}}}

\newcommand{\calA}{\mathcal{A}}

\newcommand{\minus}{\!-\!}

\newcommand{\kt}{k_\perp} 

\title{\Large Non-linear evolution in QCD at high-energy beyond leading order}

\author[a]{B.~Duclou\'e,}

\author[a]{E.~Iancu,}

\author[b]{A.H.~Mueller,}

\author[a]{G.~Soyez}

\author[c]{and D.N.~Triantafyllopoulos\,}

\affiliation[a]{Institut de physique th\'{e}orique, Universit\'{e} Paris Saclay, CNRS, CEA, F-91191 Gif-sur-Yvette, France}

\affiliation[b]{Department of Physics, Columbia University, New York, NY 10027, USA}

\affiliation[c]{European Centre for Theoretical Studies in Nuclear Physics and Related Areas (ECT*)\\and Fondazione Bruno Kessler, Strada delle Tabarelle 286, I-38123 Villazzano (TN), Italy}

\emailAdd{bertrand.ducloue@ipht.fr}
\emailAdd{edmond.iancu@ipht.fr}
\emailAdd{ahm4@columbia.edu}
\emailAdd{gregory.soyez@ipht.fr}
\emailAdd{trianta@ectstar.eu}

\abstract{ The next-to-leading order (NLO) Balitsky-Kovchegov (BK) equation describing the high-energy evolution of the scattering between a dilute projectile and a dense target suffers from instabilities unless it is supplemented by a proper resummation of the radiative corrections enhanced by (anti-)collinear  logarithms. Earlier studies have shown that if one expresses the evolution in terms 
of the rapidity of the dilute projectile, the dominant anti-collinear contributions can be resummed to all orders. However, in applications to physics, the results must be re-expressed  in terms of the rapidity of the dense target. We show that although they lead to stable evolution equations, resummations expressed in the rapidity of the dilute projectile show a strong, unwanted, scheme dependence  when their results are translated in terms of the  target rapidity. Instead, in this paper, we work directly in the rapidity of the dense target where anti-collinear contributions are absent but where new,  collinear, instabilities arise. These are milder since disfavoured by  the typical BK evolution. We propose several prescriptions for resumming these new double logarithms and find only little scheme dependence. The 
resummed equations are non-local in rapidity and can be extended to full NLO accuracy.}

\begin{document}
\maketitle

\newpage

\begin{spacing}{1.14}
\tableofcontents
\end{spacing}

\newpage

\section{Introduction}
\label{sec:intro}

The non-linear evolution equations in QCD at high energy --- the Balitsky-JIMWLK\footnote{This acronym stays for Balitsky, Jalilian-Marian, Iancu, McLerran, Weigert, Leonidov and Kovner.} hierarchy  \cite{Balitsky:1995ub,JalilianMarian:1997jx,JalilianMarian:1997gr,Kovner:2000pt,Iancu:2000hn,Iancu:2001ad,Ferreiro:2001qy} and its mean field approximation known as the Balitsky-Kovchegov (BK) equation  \cite{Balitsky:1995ub,Kovchegov:1999yj} --- represent an essential ingredient of our current theoretical description of high-energy hadronic scattering from first principles. To provide a realistic phenomenology --- say, in relation with hadron-hadron collisions at RHIC and the LHC, or electron-nucleus deep inelastic scattering (DIS) at the Electron-Ion Collider that is currently under study ---, these equations must be known to next-to-leading order (NLO) accuracy at least. Their versions at leading order (LO), with the inclusion of unitarity corrections, have been established about two decades ago, but their extension beyond LO appeared to be extremely subtle. Not only the calculation of the NLO corrections to the BK equation \cite{Balitsky:2008zza} and, subsequently, to the full B-JILWLK hierarchy \cite{Balitsky:2013fea,Kovner:2013ona,Kovner:2014lca,Lublinsky:2016meo}, turned out to be a {\it tour de force}, but when trying to use such NLO results  in practice, the situation appeared to be very deceiving. 

The NLO BK equation turned out to be unstable \cite{Lappi:2015fma}, due to the presence of large and negative NLO corrections enhanced by double collinear logarithms, i.e.~corrections of relative order $\abar\ln^2(Q^2/Q_0^2)$, where  $\abar \equiv \alpha_s N_c/\pi$ ($\alpha_s$ is the QCD coupling and $N_c$ the number of colors) and $Q^2$ and $Q_0^2$ are the characteristic transverse momentum scales in the dilute projectile ($Q^2$) and the dense target ($Q_0^2$). Such logarithms are indeed large, since $Q^2\gg Q_0^2$ for the ``dilute-dense''  collisions to which the BK equation is meant to apply. For instance, for the case of DIS at high energy, or small Bjorken $x$, $Q^2$ is the virtuality of the photon exchanged in the $t$-channel and $Q_0^2$ is an intrinsic scale in the hadronic target at low energy: $Q_0\sim\Lambda_{_{\rm QCD}}$ if the target is a proton, or $Q_0\sim$ the saturation momentum in the McLerran-Venugopalan model \cite{McLerran:1993ni,McLerran:1993ka} if the target is a large nucleus. Within the NLO BK evolution, these double logarithms are generated by integrating over {\it anti-collinear} configurations, where the transverse momentum of the emitted gluon is much smaller than that of its parent. (In the dipole picture of the  evolution, this corresponds to the case where both daughter dipoles have transverse sizes much larger than that of their parent.) Such ``hard-to-soft'' emissions are indeed the typical ones, since the global evolution proceeds from the hard scale $Q^2$ of the projectile down to the soft scale  $Q_0^2$ of the target.

As a matter of fact, this instability of the strict NLO approximation is hardly a surprise and it has indeed been anticipated \cite{Triantafyllopoulos:2002nz,Avsar:2011ds} on the basis of previous experience with the NLO version \cite{Fadin:1995xg,Fadin:1997zv,Camici:1996st,Camici:1997ij,Fadin:1998py,Ciafaloni:1998gs} of the  Balitsky-Fadin-Kuraev-Lipatov (BFKL) equation \cite{Lipatov:1976zz,Kuraev:1977fs,Balitsky:1978ic} (the linear version of the BK equation, valid so long as the scattering is weak), where similar problems were identified and eventually cured \cite{Kwiecinski:1997ee,Salam:1998tj,Ciafaloni:1998iv,Ciafaloni:1999yw,Ciafaloni:2003rd,Vera:2005jt}. In the terminology of Ref.~\cite{Salam:1998tj}, the instability of the NLO BK equation is a consequence of the ``wrong choice for the energy scale''. In a language which is better adapted to our current analysis, this refers to the choice of the rapidity variable which plays the role of the ``evolution time'' in the high-energy evolution. Roughly speaking, this variable must scale like the logarithm of the center-of-mass energy squared, but its precise definition starts to matter at NLO. 

The rapidity generally used in relation with the BK equation is that of the projectile, that we shall denote as $Y$; this looks indeed natural, given that this equation has been constructed by following the evolution of the wavefunction of the dilute projectile with the emission of softer and softer gluons (or, equivalently, with increasing $Y$). This is nevertheless a ``bad choice'' from the viewpoint of the NLO analysis in Ref.~\cite{Salam:1998tj}, in the sense that the typical (``hard-to-soft'') evolution with $Y$ includes gluon emissions which violate the correct time-ordering of the fluctuations, i.e.~the condition that a daughter gluon must have a shorter lifetime than its parent. On the other hand, the proper time-ordering is automatically respected if, instead of $Y$, one orders the successive emissions according to the {\it target} rapidity, that we shall denote as $\eta$. The precise definitions for $Y$ and $\eta$ will be given in the next sections, where we shall see that $\eta$ is indeed a right measure of the rapidity phase-space available in DIS (since related to Bjorken $x$, via $\eta= \ln(1/x_{\rm\scriptscriptstyle Bj})$), whereas $Y$ is always larger than $\eta$, namely $Y=\eta+\ln(Q^2/Q_0^2)$, corresponding to the fact that the projectile rapidity is overcounting the energy phase-space. The large and negative NLO corrections enhanced by the double transverse logarithm are intended to compensate for this overcounting at NLO. Similar corrections, i.e.~terms of order $[\abar\ln^2(Q^2/Q_0^2)]^n$ with $n\ge 1$ and with alternating signs, occur in the higher orders and jeopardise the convergence and also the stability of the perturbative expansion for the evolution equation in $Y$.

In the context of the BFKL dynamics and for an asymmetric collision with $Q^2\gg Q_0^2$, it is natural to associate the whole evolution with the target  --- that is, to evolve in $\eta$ ---  and thus avoid the complications with the violation of time ordering; this is the ``correct  choice for the energy scale'' advocated in \cite{Salam:1998tj}. But in the framework of the non-linear evolution, where the NLO corrections to both BK and B-JIMWLK equations were explicitly computed by studying the evolution of the projectile, it looks more natural to evolve in $Y$ and try and cure the instability problem via all-order resummations of the radiative corrections enhanced by the double transverse logarithms. Two methods have been proposed in that sense \cite{Beuf:2014uia,Iancu:2015vea}, which use different recipes for enforcing time-ordering in the evolution with $Y$. Ref.~\cite{Beuf:2014uia} has introduced kinematical constraints  leading to an evolution equation which is similar to the LO BK equation (in the sense of having the same splitting kernel), but is {\it non-local} in $Y$. In Ref.~\cite{Iancu:2015vea} on the other hand,  the double-logarithmic corrections have been resummed in the kernel and the ensuing equation, dubbed ``collinearly-improved BK'', is still {\it local} in $Y$.  These two methods are equivalent in so far as the resummation of the leading double transverse logs is concerned, but differ from each other at the level of subleading terms (i.e.~corrections with a larger power for $\abar$ than for  the double-log $\ln^2(Q^2/Q_0^2)$). Both procedures solve the instability problem: the associated numerical solutions are indeed  stable, as explicitly demonstrated in  \cite{Iancu:2015joa,Albacete:2015xza,Lappi:2016fmu}.

Besides the change in the structure of the differential equation, a proper enforcement of the time-ordering condition should also modify the formulation of the initial value problem for the evolution in $Y$.  The corresponding modifications have not been properly implemented in the original literature. For instance,  \cite{Beuf:2014uia} failed to recognise that the non-local version of the BK equation should be solved as a {\it boundary-value} problem, rather than as an initial-value one. Concerning the local resummation in \cite{Iancu:2015vea}, this can be still formulated as  an initial-value problem, but the initial value (at $Y=0$) itself must receive double-logarithmic corrections to all orders, similarly to the kernel. The need for such an additional resummation was recognised in  \cite{Iancu:2015vea}. However the recipe for the initial condition that was proposed in  \cite{Iancu:2015vea} is not accurate enough: it is correct in a leading-logarithmic approximation for the transverse logs (see \cite{Iancu:2015vea} for details), but not also to full BFKL accuracy. Correcting these inconsistencies in the formulation of the initial value problem for the resummed evolution in $Y$ represented our original motivation for the present study. However, during our study, we have discovered even more severe conceptual problems, which made us understand that the evolution in $Y$ is intrinsically ill-behaved and should be replaced with an evolution in $\eta$ --- similarly to what was done for the NLO BFKL equation \cite{Salam:1998tj,Ciafaloni:1999yw,Ciafaloni:2003rd}.

To understand the additional difficulties, we first observe that the aforementioned inconsistencies with the formulation of the initial condition should only affect the evolution at relatively low values of $Y$, but not also the asymptotic behavior at large $Y$. For instance, different resummation methods should give similar predictions for the saturation exponent $\lambda_s$, which controls the growth of the saturation momentum with the rapidity. By ``similar'' we mean that different predictions must differ by a quantity of $\order{\abar^2}$ : $\delta\lambda_s=c\abar^2$ with $c\sim\order{1}$. Yet, as we shall shortly explain, this expectation is not met in practice. Note that, since the ``correct'' evolution ``time'' is $\eta$ (and not $Y$), physical quantities like the saturation exponent, the shape of the saturation front (and the associated property of geometric scaling  \cite{Stasto:2000er,Iancu:2002tr, Mueller:2002zm,Munier:2003vc,Munier:2003sj}), or the DIS structure functions at small Bjorken $x$, should also be studied in $\eta$. Hence, even if one starts by solving the BK equation in $Y$, one must re-express the results in terms of $\eta$ before inferring any physical conclusion.

\comment{The difference between using $Y$ and $\eta$ formally starts to matter only at NLO; in practice though, the difference $Y\minus\eta=\ln(Q^2/Q_0^2)$ is quite large, so this change of variables has important consequences already at LO.  As a matter of fact, the effect of the resummation in $Y$ should be largely equivalent (up to $\order{\abar^2}$--corrections) to solving the LO BK equation in $\eta$: indeed, the main effect of the resummation in $Y$ is to enforce the proper time-ordering of the fluctuations, which is the same as their ordering in $\eta$.}

In this context, Fig.~\ref{SresumY} (right) shows $\bar\lambda_s$ --- the  saturation exponent for the evolution  in $\eta$  --- as obtained from the resummed evolution in $Y$ using three different methods: the ``local'' resummation proposed in \cite{Iancu:2015vea} and two prescriptions for the ``non-local'' resummation in  \cite{Beuf:2014uia} (see Sect.~\ref{sec:BKnlY} for details). In principle, these various methods are equivalent to the accuracy of interest, so their results for $\bar\lambda_s$ should agree with each other up to corrections of $\order{\abar^2}$. Yet, the curves shown in Fig.~\ref{SresumY} (right) appear to strongly deviate from each other (and also from the corresponding result $\bar\lambda_0\simeq 4.88\abar$ of the LO BK evolution in $\eta$) and this deviation increases with $\abar$: one can write  $\delta\bar\lambda_s=c(\abar)\abar^2$, where $c(\abar)$ rises with $\abar$ and is significantly larger than 1 already for $\abar=0.25$. This strong scheme-dependence is likely related to the existence of important subleading corrections, beyond the leading double-logarithms resummed by all these methods.


We thus conclude that, even after performing resummations which are tantamount to enforcing time-ordering, the evolution in $Y$ is still lacking predictive power. This observation motivates us to reformulate the (NLO and beyond) BK evolution as an evolution with the target rapidity $\eta$. Instead of going through a tedious NLO computation of gluon emissions in the background of the dense gluon distribution of the target, we shall deduce the NLO corrections to the BK equation in $\eta$ from the corresponding corrections to the evolution in $Y$ via a mere change of variables. This is a straightforward procedure in (strict) perturbation theory,  to be described in Sect.~\ref{sec:NLOeta} at NLO level.  As expected, the main consequence of this change of variables is to eliminate the double anti-collinear logarithms responsible for the failure of the NLO BK equation in $Y$. 


The resulting NLO version of the BK equation in $\eta$, presented in Sect.~\ref{sec:NLOeta} (see  \eqn{nlobketa}),  is the true starting point of our analysis. (The first 2 sections of this paper will mainly serve to illustrate the problems with the evolution in $Y$.) Since time-ordering is now built-in, one may expect this equation to predict a smooth evolution, which is well behaved and free of instabilities. However, this turns out not to be the case: as we will demonstrate in Sect.~\ref{sec:nloinst}, via both analytic and numerical arguments, the NLO BK evolution in $\eta$ does still exhibit an instability, albeit somewhat milder (and more slowly developed) than the one for the respective evolution in $Y$. This instability is again related to NLO corrections enhanced by double transverse logarithms, but of a different kind: these are {\it collinear} logarithms associated with ``soft-to-hard'' emissions in which the transverse momentum of the emitted gluon is much larger than that of its parent. (In the dipole picture, one of the daughter dipoles  is much smaller than the other one and than their common parent.) Such emissions are atypical in the problem at hand (which explains why the associated instabilities are relatively mild), yet they are allowed by the non-locality of the BFKL (dipole) kernel in the transverse plane, which leads to ``BFKL diffusion'', i.e.~to excursions via dipole configurations of any size. 

These instabilities could have been anticipated from the experience with the NLO BFKL equation: in that context too, and after selecting the ``correct energy scale'' (= evolution variable), the strict NLO approximation is still unstable (e.g.~it yields a complex saddle point leading to oscillating solutions) and calls for collinear resummations (see \cite{Salam:1999cn} for a pedagogical discussion). 

In this paper, we shall proceed to resummations of the (leading) double collinear logs to all orders. The guiding principle for such resummations is the condition that successive emissions which are ordered in $\eta$ must be also ordered in their longitudinal momenta\footnote{This is the counterpart of the condition used in the evolution with $Y$, namely the fact that the emissions ordered in $Y$ must be also ordered in their lifetimes, i.e.~in $\eta$.} (i.e.~in $Y$). As for the evolution with $Y$  \cite{Beuf:2014uia,Iancu:2015vea}, we shall propose two strategies for the collinear resummations --- one leading to equations which are non-local in $\eta$ but with the standard BFKL kernel (see Sect.~\ref{sec:shift}), the other one leading to a local equation, but with a kernel which receives all-order corrections (cf. Sect.~\ref{sec:localeta}). As expected, the (local and non-local) resummations in $\eta$ show only little scheme dependence: the respective predictions for the saturation exponent agree with each other within the expected $\order{\abar^2}$--accuracy (see Fig.~\ref{fig:nloeta}). This confirms the fact that, by trading $Y$ for $\eta$ as the evolution ``time'', we have restored the predictive power of the (resummed) perturbation theory.

We shall nevertheless discard the local version of the resummation since, as we shall see, it does not properly encode the approach towards saturation. (The ``soft-to-hard'' evolution and the associated resummations in $\eta$ also impact the non-linear dynamics, unlike the resummations in $Y$ which matter only at weak scattering; see the discussion in Sect.~\ref{sec:lt}.) The non-local equations  in $\eta$ can be extended to full NLO accuracy by adding the missing NLO corrections (not enhanced by double collinear logarithms); this will be explained in Sect.~\ref{sec:match}.

The fact of evolving in $\eta$ also alleviates the problem of the initial condition that was present for the evolution in $Y$: our resummed equations in $\eta$ can be unambiguously formulated as initial-value problems, although this requires some care due to their non-locality in the evolution ``time'' $\eta$; this will be explained in Sect.~\ref{sec:IC}.

Among the resummed equations which are non-local in $\eta$, we will find it natural to select one of them, whose expansion to $\order{\abar^2}$ shows the closest resemblance to the strict NLO equation displayed in \eqn{nlobketa}. This will be shown in  \eqn{bketato}, that we repeat here for convenience (see also Eqs.~\eqref{bketao} and \eqref{bketafin} for other versions of this equation whose respective virtues will be explained in due time). This is an equation for the dipole $S$-matrix $\bar{S}_{\bx\by}(\eta)$, whose structure is quite similar to that of the LO BK equation, except for the non-locality in the rapidity arguments of the $S$-matrices describing the scattering of the daughter dipoles:
\begin{align}
        \label{bkfin}
        \frac{\del \bar{S}_{\bx\by}(\eta)}{\del \eta} = 
        \frac{\abar}{2\pi}
        \int \frac{\dif^2 \bz \,(\bx\minus\by)^2}{(\bx \minus\bz)^2 (\bz \minus \by)^2}\,
        \Theta\big(\eta \minus \delta_{\bx\by\bz}\big)
        \big[\bar{S}_{\bx\bz}(\eta \minus \delta_{\bx\bz;r})\bar{S}_{\bz\by}(\eta \minus \delta_{\bz\by;r}) 
        \minus \bar{S}_{\bx\by}(\eta) \big].
\end{align}
In this equation, $\delta_{\bx\by\bz}={\rm max}\{\delta_{\bx\bz;r},\delta_{\bz\by;r}\}$, $r=|\bx\!-\!\by|$, and the rapidity shifts are defined as
\beq
\label{delta10}
\delta_{\bx\bz;r}\equiv \ln\frac{{\rm max}\{(\bx\!-\!\bz)^2, r^2\}}{(\bx\!-\!\bz)^2}\,,\quad
\delta_{\bz\by;r}\equiv \ln\frac{{\rm max}\{(\bz\!-\!\by)^2, r^2\}}{(\bz\!-\!\by)^2}\,.
\eeq
They are non-vanishing (meaning that the collinear resummation plays a role) only in the case where the transverse size of one of the daughter dipoles (either $|\bx\!-\!\bz|$, or $|\bz\!-\!\by|$) is much smaller than the size $r$ of the parent dipole.

 For this  ``canonical'' equation we shall present a rather complete analysis in Sects.~\ref{sec:BFKLeta} and \ref{sec:etafronts}. Notably, in Sect.~\ref{sec:chi} we shall discuss the relation between our collinear resummation of the BK equation in transverse coordinate space and the corresponding procedure (the ``$\omega$--shift'') used in the context of the NLO BFKL equation in Mellin space \cite{Salam:1998tj,Ciafaloni:1998iv,Ciafaloni:1999yw,Ciafaloni:2003rd}. Also, in Sect.~\ref{sec:sat} we shall present rather detailed, semi-analytic and numerical, studies of the solutions to \eqn{bketato}, including the pre-asymptotic corrections to the saturation exponent, the saturation anomalous dimension, the quality of geometric scaling, and the effects of including a running coupling. We shall find that, as a consequence of the collinear resummation and of the running of the coupling, the evolution is considerably slowed down:  the effective ($\eta$-dependent) saturation exponent takes typical values $\bar\lambda_s=0.2\div 0.3$, which are consistent with the phenomenology (see  Fig.~\ref{fig:rc}. (right)).

\section{Balitsky-Kovchegov evolution through NLO: a brief summary}
\label{sec:bkgen}

This first section does not contain any new result, but only a collection of informations about the (leading-order and next-to-leading order) BK equation that will be useful for the subsequent discussion. This summary will also give us the opportunity to introduce our notations and  explain the kinematics.

\subsection{The Balitsky-Kovchegov equation at leading order}
\label{sec:lobk}

We consider the high-energy scattering between a dilute projectile --- a quark-antiquark color dipole propagating towards the positive direction of the longitudinal axis with a large momentum $q^+$ --- and a dense target --- a nucleon or a nucleus moving in the opposite direction with a longitudinal momentum $q_0^-$ (per nucleon). The scattering will be treated in the eikonal approximation, so that the transverse coordinates  $\bx$ and $\by$ of the quark and the antiquark are not modified by the collision (and hence the same is true for the dipole transverse size  $r=|\br|$, with $\br=\bx-\by$). 

The target is characterized by a transverse momentum scale $Q_0$, which plays the role of a saturation momentum (the typical scale for strong scattering) at {\it low} energy: a dipole with size $r\sim 1/Q_0$ would strongly scatter off the target already for a low energy $q^+\sim  q_0^+$, with $q_0^+\equiv Q_0^2/2 q_0^-$. In reality,  we are interested in much higher energies $q^+\gg q_0^+$, where the typical scale for the onset of strong scattering is the $Y$-dependent saturation momentum $Q_s(Y)$ and is much harder: $Q_s^2(Y)\gg Q_0^2$. Here, $Y$ is the (boost-invariant) rapidity difference between the projectile and the target,
 \begin{equation}
         Y \equiv \ln \frac{q^+}{q_0^+} = \ln \frac{2 q^+ q_0^-}{Q_0^2} = \ln \frac{s}{Q_0^2}\,,
 \end{equation} 
with $s=2 q^+ q_0^-$ the center of mass (COM) energy squared, assumed to be very large: $s\gg Q_0^2$.  

The scale $Q_s^2(Y)$ is rapidly increasing with $Y$ (roughly, like an exponential; see below), due to {\it quantum evolution}, i.e.~due to the successive emissions of softer and softer gluons, which carry only a small fraction of the longitudinal momentum of their parent: each such an emission occurs with a probability of  $\order{\abar Y}$, with $\abar = \alpha_s N_c/\pi$, where $\alpha_s$ is the QCD coupling and $N_c$ the number of colors. Depending upon the choice of a Lorentz frame, such emissions can be viewed as additional Fock space components in the wavefunction of the color dipole, or of the dense nucleus, or both. Physical observables like the $S$-matrix $S_{\bx\by}(Y)$ for elastic scattering, are boost-invariant (they depend only upon the rapidity difference $Y$), but the physical picture of the evolution and the associated mathematical description depend upon the frame, due to the dilute-dense asymmetry. This picture becomes much simpler if the evolution is fully associated with the dilute projectile, since in that case one can neglect non-linear effects like gluon saturation in the evolution of the dipole wavefunction, but only include them (as unitarity corrections) in the evolution of the scattering amplitude. That description applies in a frame where the dipole carries most of the total energy, i.e.~$q^+\gg q_0^-$. This is the description that we shall use throughout this paper, first to leading-order (LO), then to next-to-leading order (NLO) and ultimately when performing specific resummations to all orders.

At LO and in a suitable mean-field description of the gluon distribution in the target (which in particular requires the multi-color limit $N_c\gg 1$), the evolution of the elastic $S$-matrix with increasing $Y$ is described by the LO BK equation \cite{Balitsky:1995ub,Kovchegov:1999yj}, which reads
\begin{align}
\label{lobk}
\frac{\partial S_{\bx\by}(Y)}{\partial Y} = \,&
 \frac{\abar}{2 \pi}
 \int
 \frac{\dif^2 \bz \,(\bx-\by)^2}{(\bx -\bz)^2 (\bz - \by)^2}
 \big[S_{\bx\bz}(Y) S_{\bz\by}(Y) - S_{\bx\by}(Y) \big].
\end{align}
This equation depicts the dipole evolution at large $N_c$ as the splitting of the original dipole $(\bx, \by)$ into two new dipoles, $(\bx, \bz)$ and $(\bz, \by)$, where the variable $\bz$ is truly the transverse coordinate of the emitted gluon at the time where this interacts with the target
The kernel of this equation describes the probability density for dipole splitting. The first term within the square brackets, which is quadratic in $S$, describes a situation in which the emitted gluon (equivalently, the system of two daughter dipoles) exists at the time of scattering, so both dipoles interact with the target; for brevity, this term will be referred as the ``real'' term (in the sense of really measuring the scattering of the soft gluon). The term linear in $S$, which is negative and will be referred to as ``virtual'', measures the decrease in the probability to have the original dipole at the time of scattering. \eqn{lobk} should be solved as an initial value problem: given (generally, a model for) the $S$-matrix $S_{\bx\by}(Y_0)$ at some relatively low rapidity $Y_0$, this equation uniquely determines the $S$-matrix at any $Y\ge Y_0$. In this paper, we shall choose $Y_0=0$, for simplicity.  When the initial condition will be explicitly needed, we shall mostly use the McLerran-Venugopalan model \cite{McLerran:1993ni,McLerran:1993ka},  which applies (in the sense of a mean field approximation and for relatively low energy) to a large nucleus with atomic number $A\gg 1$. This reads
\begin{equation}
	\label{SMV}
S_{\bx\by}^{(0)} =
\exp \left( - \frac{r^2 Q_{\rm A}^2}{4}\, \ln \frac{4}{r^2 \Lambda_{_{\rm QCD}}^2} \right), 	
\end{equation}
valid for $r^2 \Lambda_{_{\rm QCD}}^2/4 \ll 1$. The scale $Q_{\rm A}^2$ represents the average color charge density of the valence quarks per unit transverse area and grows with $A$ like $Q_{\rm A}^2\propto A^{1/3}$. The saturation momentum $Q_s^2$ in this model is defined by the condition that the exponent be of $\mcal{O}(1)$ when $r = 2/ Q_s$; this implies
\begin{equation}
	\label{QsMV}
	Q_s^2=
	Q_{\rm A}^2 \ln \frac{Q_s^2}{\Lambda_{_{\rm QCD}}^2} \simeq 
	Q_{\rm A}^2 \ln \frac{Q_{\rm A}^2}{\Lambda_{_{\rm QCD}}^2}.
\end{equation}
showing that $Q_s^2$ is strictly larger than the scale $Q_{\rm A}^2$ appearing in the exponent of Eq.~\eqref{SMV}.

 The elastic $S$-matrix would be equal to one in the absence of any scattering, so for discussing the effects of the scattering it is preferable to work with the scattering {\it amplitude} $T_{\bx\by}(Y) \equiv 1- S_{\bx\by}(Y)$: this is small when the projectile is small enough to resolve the dilute tail of the target wavefunction, while it approaches the unitarity limit $T_{\bx\by}(Y) = 1$ when the projectile is becoming sufficiently large to probe the saturated components of the target\footnote{Strictly speaking, this suggestive physical picture, in which the unitarization is related to gluon saturation in the target, holds in a frame where most of the total energy and hence the high-energy evolution are carried by the target. In the dipole frame in which we shall develop our formalism, $Q_s(Y)$ is simply the characteristic scale for the onset of unitarity corrections in the dipole-target scattering. This scale ``knows'' about both colliding systems: about the target, via the initial condition $Q_0^2=Q_s^2(Y=0)$, and about the projectile, via the dependence upon $Y$.}. These two regimes are separated by the saturation momentum $Q_s(Y)$, which is an increasing function of $Y$, as already mentioned, and whose leading behavior will be given below. As manifest in \eqn{lobk}, $T=1$ is a fixed point of the BK equation, meaning that unitarity is indeed preserved.

In the remaining part of this section, we shall assume a homogeneous target so that the amplitude depends upon the dipole size $r$ of the dipole (but not also upon the impact parameter $(\bx+\by)/2$); we shall then write $T_{\bx\by}(Y) \equiv T(Y,r)$. Notice that if this property is satisfied in the initial condition at $Y=0$, then it will be preserved for all $Y$ by the evolution equation given in \eqref{lobk}. Although, even in this simplified case, Eq.~\eqref{lobk} has not been analytically solved, one can construct a piecewise asymptotic solution for $\abar Y \gg 1$ in the two interesting regimes at $r^2 Q_s^2(Y) \ll 1$ and $r^2 Q_s^2(Y) \gg 1$. 

When $r^2 Q_s^2(Y) \ll 1$, the amplitude is weak, $T(Y,r)\ll 1$, and  \eqn{lobk} can be linearized in $T$, thus yielding the (leading-order) BFKL equation \cite{Lipatov:1976zz,Kuraev:1977fs,Balitsky:1978ic}
\begin{align}
\label{lobfkl}
\frac{\partial T_{\bx\by}(Y)}{\partial Y} = \,&
 \frac{\abar}{2 \pi}
 \int
 \frac{\dif^2 \bz \,(\bx-\by)^2}{(\bx -\bz)^2 (\bz - \by)^2}
 \big[T_{\bx\bz}(Y) + T_{\bz\by}(Y) - T_{\bx\by}(Y) \big].
\end{align}
Remarkably, one can use this equation also to study the approach towards saturation and in particular to determine the asymptotic behavior of the saturation momentum at large $Y$ \cite{Gribov:1984tu,
Iancu:2002tr,Mueller:2002zm}: to that aim, it suffices to supplement the BFKL equation with the saturation condition that $T(Y,r)\sim\order{1}$ when $r\sim 1/Q_s(Y)$ \cite{Gribov:1984tu,Iancu:2002tr}, or, more precisely, with a saturation boundary in the $(r,Y)$ phase-space \cite{Mueller:2002zm} (this last construction also allows for a study of the subasymptotic corrections). The deep reason why such a relatively simple analysis works is the fact that the growth of the saturation momentum with $Y$ is driven by the BFKL increase in the dilute  tail of the amplitude at $r\gg 1/Q_s(Y)$ --- a property often referred to as ``the pulled front'' (or ``traveling waves'') and related to a correspondence between high-energy evolution in QCD and reaction-diffusion problems in statistical physics \cite{Munier:2003vc,Iancu:2004es}.

The BFKL equation \eqref{lobfkl} is scale invariant (actually, even conformal invariant), so one can define a ``characteristic'' or ``eigenvalue'' function $\omega_0(\gamma)$ by the action of its r.h.s. on an amplitude which is a pure power, $T(r)\sim r^{2\gamma}$ with $0<\gamma <1$; this yields \cite{Kuraev:1977fs,Balitsky:1978ic}
\begin{align}
\label{omega0}
        \omega_0(\gamma) & = 
        \frac{1}{r^{2\gamma}}\,
        \frac{\abar}{2\pi}
        \int
 \frac{ \dif^2 \bz \, r^2}{z^2 |\br \minus \bz|^2}\,
 \left(z^{2\gamma} + |\br-\bz|^{2\gamma} -r^{2\gamma}\right)
 \nonumber\\*[0.2cm]
  & = \abar [2 \psi(1) - \psi(\gamma) -\psi(1-\gamma)]
  \equiv \abar \chi_0(\gamma), 
\end{align}
where $\psi(\gamma) = \dif \ln \Gamma(\gamma)/\dif \gamma$. Then the solution to the BFKL equation can be expressed as the line integral in the complex-plane
\begin{equation}
        \label{lobfklsol}
        T(Y,r) = \int_{\frac{1}{2}- \rmi \infty}^{\frac{1}{2}+\rmi \infty} 
        \frac{\dif \gamma}{2 \pi \rmi}\, T(Y=0,\gamma)
        \exp\big[\abar \chi_0(\gamma) Y - \gamma \rho\big],
\end{equation}
where we have also introduce a logarithmic variable for the dipole transverse size: $\rho \equiv \ln (1/r^2 Q_0^2)$. Here, $T(Y=0,\gamma)$ is the Mellin transform of the initial condition $T(Y=0,r)$ and in general we have
\begin{equation}
        \label{tmellin}
        T(Y,\gamma) = \int_{-\infty}^{\infty} \dif \rho\, T(Y,\rho) \exp(\gamma \rho). 
\end{equation}

For sufficiently large values of $Y$, one can estimate the inverse Mellin transform Eq.~\eqref{lobfklsol} via the saddle point method. Here, we are interested in the special saddle point (to be denoted as $\gamma_0$) which controls the approach towards saturation; this is determined by requiring that both the exponent in Eq.~\eqref{lobfklsol} and its first derivative w.r.t.~$\gamma$ must vanish when $\gamma=\gamma_0$ (see e.g.~Ref.~\cite{Mueller:2002zm} for details). One thus finds that $\gamma_0$ is a number independent of $\abar$ (sometimes referred to as the ``saturation anomalous dimension'') that reads \cite{Gribov:1984tu}  
\begin{equation}
        \label{logamma}
        \chi_0'(\gamma_0) = \frac{\chi_0(\gamma_0)}{\gamma_0}
        \,\Rightarrow\, \gamma_0 \simeq 0.628.
\end{equation}
The system of the two aforementioned conditions also determines the leading asymptotic behavior of the saturation momentum. A more elaborate analysis, which at the same time takes properly into account the presence of the saturation boundary \cite{Mueller:2002zm} (or equivalently by studying the analogy to the traveling waves \cite{Munier:2003sj}), also fixes the first preasymptotic term for large $\abar Y$ and one finds\footnote{Knowing the functional form of the preasymptotic term is particularly useful when one solves numerically, as it helps in fitting reliably the numerical data.}
\begin{equation}
\label{dlogqs}
\frac{\dif \ln Q_s^2}{\dif Y} = \lambda_0        - \frac{3}{2 \gamma_0}\frac{1}{Y}\,,\qquad
 \lambda_0 \equiv \abar\frac{\chi_0(\gamma_0)}{\gamma_0}\,.
\end{equation}
This number $ \lambda_0$ is generally referred to as the ``asymptotic saturation exponent'' (here, evaluated to leading order). Via the same methods, one can also obtain an analytic approximation to the amplitude in the vicinity of the saturation line; this reads \cite{Mueller:2002zm} 
\begin{equation}
        \label{tabove}
        T(Y,r) = \big(r^2 Q_s^2\big)^{\gamma_0}
        \bigg(\ln \frac{1}{r^2 Q_s^2} + c\bigg)
        \exp \bigg[ - \frac{\ln^2 \big(r^2 Q_s^2\big)}{D_0 Y} \bigg],
\end{equation}
where $c$ is a positive constant of order $\mcal{O}(1)$ and $D_0 = 2 \abar \chi''_0(\gamma_0)$ is the ``diffusion'' coefficient.  This approximation is valid in the regime $Q_s^2 \ll 1/r^2 \ll Q_s^2 \exp (D_0 Y)$. In particular, when $\ln (r^2 Q_s^2) \ll \sqrt{D_0 Y}$  the diffusive factor in Eq.~\eqref{tabove} can be set equal to unity. Then the amplitude shows geometrical scaling \cite{Stasto:2000er,Iancu:2002tr, Mueller:2002zm,Munier:2003vc,Munier:2003sj}, i.e.~it becomes a function of just one variable, the dimensionless quantity $r^2 Q_s^2$.

We shall later by interested in the limiting form of the $S$-matrix deeply at saturation, i.e.~for very large dipoles sizes, such that $r^2 Q_s^2 \gg 1$. In that regime the $S$-matrix approaches the black-disk limit $S(Y,r) \to 0$, hence we can neglect the term quadratic in $S$ in Eq.~\eqref{lobk}. This is indeed the case so long as the two daughter dipoles are themselves large compared to $1/Q_s$, that is, for values of $\bz$ which obey $|\bx-\bz|^2,|\bz-\by|^2\gtrsim 1/Q_s^2$. Moreover, the integration over $\bz$ becomes logarithmic if one of the daughter dipoles is much smaller than the parent dipole, that is, one has either $|\bx-\bz|^2\ll r^2$, or $|\bz-\by|^2 \ll r^2$. Adding both possibilities, the BK equation reduces to
\begin{equation}
        \label{bkblack}
        \frac{\del S(Y,r)}{\del Y} = -\abar S(Y,r) \int_{1/Q_s^2}^{r^2}
        \frac{\dif z^2}{z^2} = -\abar \ln [r^2 Q_s^2(Y)]  S(Y,r).
\end{equation}
To the accuracy of interest, it is enough to use the dominant $Y$-dependence of the saturation scale, that is $Q_s^2(Y)\propto \rme^{\lambda_0 Y}$, which in turn implies $\ln [r^2 Q_s^2(Y)]\simeq \lambda_0(Y\minus Y_s)$, with $Y_s$ the rapidity scale at which $Q_s^2(Y_s)=1/r^2$.  \eqn{bkblack} holds only for $Y>Y_s$, hence it can be integrated to yield
 \beq\label{sbelow}
 S(Y,r)\, 
 \simeq\, S(Y_s,r) 
 \exp\left\{-\frac{\abar\lambda_0}{2}(Y-Y_s)^2\right\}\,
 \simeq\,
  \exp\left\{-\frac{\abar}{2\lambda_0}\ln^2\big[r^2 Q_s^2(Y)\big]\right\}, 
 \eeq
where $S(Y_s,r)\sim\order{1}$. From the above derivation, it should be clear that the exponent in \eqn{sbelow} is known only to double logarithmic accuracy: subleading terms, e.g.~of $\order{Y\minus Y_s}$, are not under control. The functional form in Eq.~\eqref{sbelow} is generally known as the Levin-Tuchin formula \cite{Levin:1999mw}. Its precise form with $\lambda_0$ as in Eq.~\eqref{dlogqs} corresponds to the prediction of the BK equation (in the large $N_c$ limit) and in fact it has been numerically confirmed to high accuracy \cite{Alvioli:2012ba}. The coefficient in the exponent is known to receive finite-$N_c$ corrections \cite{Mueller:2002pi,Iancu:2011nj} and, more importantly, $\order{1}$ corrections from dipole number fluctuations \cite{Mueller:1996te,Iancu:2003zr}. Given Eqs.~\eqref{tabove} and \eqref{sbelow}, one sees that the amplitude exhibits geometric scaling everywhere in the region $\Lambda^2_{_{\rm QCD}} \ll 1/r^2 \ll Q_s^2 \exp\big(\sqrt{D_0 Y}\big)$, a feature which is indeed confirmed by numerical solutions.

\subsection{NLO BK evolution in $Y$}
\label{sec:nlobky}

At NLO we must also resum terms of size $\abar (\abar Y)^n$ in the presence of the strong target field. This leads to the  NLO BK equation \cite{Balitsky:2008zza} which for our purposes reads
\begin{align}
 \label{nlobk}
 \hspace*{-0.8cm}
 \frac{\partial S_{\bx\by}(Y)}{\partial Y} = \,&
 \frac{\abar}{2 \pi}
 \int
 \frac{\dif^2 \bz \,(\bx\minus\by)^2}{(\bx \minus\bz)^2 (\bz \minus \by)^2}\,
 \bigg\{ 1 + \abar
 \bigg[\bar{b}\, \ln (\bx \minus \by)^2 \mu^2 
 -\bar{b}\,\frac{(\bx \minus\bz)^2 - (\by \minus\bz)^2}{(\bx \minus \by)^2}
 \ln \frac{(\bx \minus\bz)^2}{(\by \minus\bz)^2}
 \nn*[0.2cm]
 & \hspace*{3.4cm}
 +\frac{67}{36} - \frac{\pi^2}{12} - 
 \frac{1}{2}\ln \frac{(\bx \minus\bz)^2}{(\bx \minus\by)^2} \ln \frac{(\by \minus\bz)^2}{(\bx \minus\by)^2}\bigg] 
 \bigg\}
 \left[S_{\bx\bz}(Y) S_{\bz\by}(Y) - S_{\bx\by}(Y) \right]
 \nn*[0.2cm]
  & \hspace*{-1.8cm} +
\frac{\abar^2}{8\pi^2}
 \int \frac{\dif^2 \bu \,\dif^2 \bz}{(\bu \minus \bz)^4}
 \bigg\{\minus 2
 + \frac{(\bx \minus\bu)^2 (\by \minus\bz)^2 + 
 (\bx \minus \bz)^2 (\by \minus \bu)^2
 - 4 (\bx \minus \by)^2 (\bu \minus \bz)^2}{(\bx \minus \bu)^2 (\by  \minus \bz)^2 - (\bx \minus \bz)^2 (\by \minus \bu)^2}
 \ln \frac{(\bx \minus \bu)^2 (\by  \minus \bz)^2}{(\bx \minus \bz)^2 (\by \minus \bu)^2}
 \nn*[0.2cm]
 & \hspace*{1.3cm} +
 \frac{(\bx \minus \by)^2 (\bu \minus \bz)^2}{(\bx \minus \bu)^2 (\by  \minus \bz)^2}
 \left[1 + \frac{(\bx \minus \by)^2 (\bu \minus \bz)^2}{(\bx \minus \bu)^2 (\by  \minus \bz)^2 - (\bx \minus \bz)^2 (\by \minus \bu)^2} \right]
 \ln \frac{(\bx \minus \bu)^2 (\by  \minus \bz)^2}{(\bx \minus \bz)^2 (\by \minus \bu)^2}\bigg\}
 \nn*[0.2cm]
 & \hspace*{1.4cm} \left[S_{\bx\bu}(Y) S_{\bu\bz}(Y) S_{\bz\by}(Y) - S_{\bx \bu}(Y) S_{\bu \by}(Y)\right],
 \end{align}
where $\bar{b} = (11 N_c -2 N_{\rm f})/12 N_c$, with $N_f$ the number of flavors, and where $\mu$ is a renormalization scale at which the coupling should be evaluated.

In writing \eqn{nlobk} we have neglected two types of terms. First, we have not written terms which involve more complicated (than the dipole) color structures and are $1/N_c^2$ suppressed and this allows us to deal with a closed equation. Second we have dropped the terms proportional to $N_{\rm f}/N_c$ \cite{Balitsky:2006wa,Kovchegov:2006vj} (apart those included in the definition of $\bar{b}$). The latter don't bring any new difficulty and could be easily included in \eqn{nlobk}, however they vanish in the regime of weak scattering. In any case, both types of terms do not play any role on the aspects to be discussed in this paper.  

To derive the NLO contributions, i.e.~those proportional to $\abar^2$ in the r.h.s.~of \eqn{nlobk}, one has considered two consecutive gluon emissions. These are both soft with respect to the projectile dipole $(\bx,\by)$, but they are not strongly ordered with respect to each other, that is, they have similar longitudinal momenta. Therefore, although the first emission is taken as eikonal, the kinematics in the vertex for the second emission must be treated exactly. (Still, one must notice that the scattering of the ensuing partonic system with the nuclear target is eikonal.) After the longitudinal integration is performed, the NLO terms can be collected in two pieces. One piece involves a single (2-dimensional) integration over the  transverse coordinate $\bz$, and does not change the structure of the  LO BK equation. It is only the respective kernel which receives corrections of order $\mcal{O}(\abar^2)$, and in particular those corrections proportional to $\bar{b}$ which are associated with the running of the QCD coupling. In the other piece all the partonic fluctuations scatter with the target and hence one remains with two transverse convolutions, over $\bz$ and $\bu$. There, the structure is $S_{\bx\bu} S_{\bu\bz} S_{\bz\bu}$, since we have assumed the large-$N_c$ limit in which the parent dipole and the two daughter gluons are equivalent to the three dipoles $(\bx,\bu)$, $(\bu,\bz)$ and $(\bz,\by)$. The virtual structure $-S_{\bx\bu} S_{\bu\by}$ stands for the case that the gluon at $\bz$ be both emitted and reabsorbed either before or after the scattering and its presence is necessary to render innocuous the potential UV singularity due to the $1/(\bu-\bz)^4$ factor of the kernel.

\subsection{Large anti-collinear logarithms at NLO in $Y$-evolution}
\label{sec:colly}

In principle, one would like to solve \eqn{nlobk} in order to calculate $\mcal{O}(\abar)$ corrections on top of the LO solution. Nonetheless, this equation as it stands is plagued with various shortcomings. There are various NLO terms which are enhanced by large logarithms in certain corners of the transverse space and which eventually render invalid the strict expansion in $\abar$. The terms proportional to $\bar{b}$, although they multiply logarithms which can get large, are very familiar and in fact they do not pose any serious difficulty. Choosing the running coupling scale $\mu$ as the hardest scale of the splitting process, i.e.~taking for example $\mu^2 = r_{\rm min}^{-2}$ with  $r_{\rm min}={\rm min}\{|\bx-\by|,|\bx-\bz|,|\bz-\by|\}$, the terms under consideration when added together never become large.

The remaining transverse logarithms do not (and should not) cancel by an appropriate choice of $\mu$, since they are of different origin. These ``anti-collinear'' logarithms arise when transverse sizes among successive emissions are very disparate and the respective NLO corrections get large in the regime where the scattering is still weak, i.e.~when $T\ll 1$. More precisely we consider the strongly ordered regime
\begin{equation}
\label{collso}
	1/Q_s \gg |\bz-\bx| \simeq |\bz-\by| \simeq |\bz-\bu|
	\gg |\bu-\bx| \simeq |\bu-\by| \gg |\bx-\by|=r,
\end{equation} 
which means the parent dipole is the smallest one, a gluon is emitted very far away from it at $\bu$ and a second one even further at $\bz$, but all the formed dipoles have sizes smaller than the inverse saturation scale and thus scatter weakly with the target nucleus. In this hard-to-soft evolution the dominant NLO contribution in the single integration piece in \eqn{nlobk} comes from the double logarithm which becomes
\begin{equation}
	- 
 \frac{1}{2}\ln \frac{(\bx -\bz)^2}{(\bx -\by)^2} 
 \ln \frac{(\by-\bz)^2}{(\bx -\by)^2}
 \simeq 
 - 
 \frac{1}{2}\ln^2 \frac{(\bx \minus\bz)^2}{r^2}. 
 \end{equation}  
At the same time, since we are in the linear regime, and since larger dipoles interact much stronger than smaller ones, we can approximate
\begin{equation}
	S_{\bx\bz} S_{\bz\by} - S_{\bx\by}
	\simeq - T_{\bx\bz} - T_{\bz\by} + T_{\bx\by} \simeq -2 T_{\bx\bz},
\end{equation}
i.e.~only the real terms matter. Moreover, the first line in the square bracket in the double integration in \eqn{nlobk} leads to a single collinear logarithm when the integration over $\bu$ is done in the regime \eqref{collso} \cite{Iancu:2015joa}. Putting everything together and letting for convenience $|\bx-\bz|\to z$, we arrive at
\begin{equation}
\label{colleq}
	\frac{\partial T(Y,r)}{\partial Y} = 
	\abar \int_{r^2}^{1/Q_s^2} \dif z^2\,
	\frac{r^2}{z^4}
	\left[1 - \abar \left(\frac{1}{2} \ln^2 \frac{z^2}{r^2} 
	+ \frac{11}{12} \ln \frac{z^2}{r^2}   \right)
	 \right] T(Y,z)
\end{equation}
valid in the collinear regime in \eqn{collso}. Now it becomes apparent that when the daughter dipoles are sufficiently large, the NLO corrections get comparable to (or larger than) the LO contribution. Thus, the perturbative expansion in $\abar$ has no predictive accuracy and this is one of the major shortcomings of \eqn{nlobk}. For example, let us assume a GBW type initial condition with the dilute tail $T(Y=0,r) = r^2 Q_s^2$, and perform just a single iteration in \eqn{colleq}. The integration  becomes logarithmic and gives 
\begin{equation}
\label{colleqiter}
	\Delta T(Y,r) = \abar Y r^2 Q_s^2\ln \frac{1}{r^2 Q_s^2}
	\left(1 - \frac{\abar}{6} \ln^2\frac{1}{r^2 Q_s^2} 
	- \frac{11}{12} \frac{\abar}{2} \ln\frac{1}{r^2 Q_s^2}
	\right).
\end{equation}   
Thus, when $r^2 Q_s^2$ gets small, not only $\Delta T$ becomes large, but it is also negative and thus the solution will develop an instability, as indeed  seen in numerical studies  \cite{Lappi:2015fma}. In this work we shall deal only with the double logarithms, which are obviously the dominant ones. Still, eventually one needs to take care of the single logarithms as well. The latter are related to DGLAP physics, as can be inferred from the value 11/12 of the coefficient; a procedure for their resummation has been proposed in \cite{Iancu:2015joa}.    

\section{Time ordering and collinear resummation in the dipole evolution with $Y$}
\label{sec:to} 

In this section, we shall analyse the physical origin of the time-ordering of successive emissions and its consequences for the high-energy evolution of the right-moving projectile (the dipole) with increasing $Y$. We shall first discuss the double-logarithmic approximation (DLA) where the implementation of the time-ordering (TO) condition is unambiguous and naturally leads to an evolution equation formulated as a boundary value problem non-local in $Y$. Alternatively, this equation can be equivalently rewritten (modulo an analytic continuation) as an initial-value problem local in $Y$, where both the kernel and the initial condition at $Y=0$ resum to all orders radiative corrections enhanced by double anti-collinear logarithms. Last but not least, the DLA evolution in $Y$ with TO will be shown to be equivalent with the standard (unconstrained) DLA evolution with decreasing Bjorken $x$, or increasing the rapidity $\eta\equiv \ln(1/x_{\rm\scriptscriptstyle Bj})$ of the  left-moving target: the two evolutions are simply related to each other via a change of variables from $Y$ to $\eta$.

Then we will study the possibility to extend the evolution in $Y$ with TO to the full BK dynamics, including the LO BFKL kernel and the non-linear effects responsible for gluon saturation in the target and the unitarization of the scattering amplitude. We will present and amend previous proposals in that sense, which build upon either the non-local  \cite{Beuf:2014uia}, or the local \cite{Iancu:2015vea}, version of the DLA equation. Such extensions are unavoidably ambiguous, but one may hope that the scheme dependence remains small --- say, an effect of $\order{\alpha_s^2}$ on the value of the saturation exponent.  As explained in the Introduction, the physical information can only be read from the  evolution with $\eta$, so it will be appropriate to compare the respective saturation exponents after changing the rapidity variable from $Y$ to $\eta$. This comparison however turns out to be deceptive: the various resummation schemes that we shall consider are found to lead to widely different predictions for  the saturation exponent in the evolution with $\eta$.

\subsection{Time ordering in the double logarithmic approximation}

Before we discuss the physical origin of time ordering, let us briefly explain the emergence of the  double logarithmic approximation (DLA) in the context of the LO BK equation for dipole-hadron scattering. DLA is formally the leading order pQCD approximation to the BFKL equation (the linearized version of the BK equation \eqref{lobk}) in the regime where both the phase-space for the high-energy evolution, as measured by the  rapidity difference $Y$, and the phase-space for transverse momentum (or virtuality) evolution, as measured by the ``collinear'' logarithm $\rho\equiv\ln(Q^2/Q^2_0)$ are large, in the sense that $Y\gg 1$, $\rho\gg 1$ and $\abar Y\rho\gg 1$. Its ``naive'' formulation which neglects time-ordering resums to all orders the radiative corrections of order $(\abar Y\rho)^n$. These corrections are associated with ``soft'' and ``anti-collinear'' gluon emissions, i.e.~emissions such that both the longitudinal momentum and the transverse momentum of the emitted gluon are strongly decreasing from one emission to the next one:
\beq\label{DLAphase}
q^+\gg k_1^+\gg k_2^+\gg\dots\gg q_0^+\,,\qquad Q^2\gg k_{1\perp}^2\gg k_{2\perp}^2
\gg\dots\gg Q_0^2\,,
\eeq
with obvious notation. In the transverse coordinate representation in which the BK equation is most naturally written, this corresponds to daughter dipoles which are much larger than the parent one, at each successive dipole splitting. Normally, such anticollinear splittings are disfavored by the rapid decay of the dipole kernel for large daughter dipoles (recall that
$r=|\bm{x}-\bm{y}|$)
 \begin{align}\label{Mlargez}
 \frac{(\bm{x}-\bm{y})^2}
 {(\bm{x}-\bm{z})^2(\bm{z}-\bm{y})^2}\simeq \frac{r^2}{(\bz-\bx)^4}\quad\mbox{
when $|\bz-\bx|\simeq |\bz-\by|\gg r$}\,,\end{align} 
but in the context of the BFKL evolution this decrease is compensated by the fact that the dipole scattering amplitude $T_{\bx\by}(Y) \equiv 1-S_{\bx\by}(Y)$ is rapidly increasing with the dipole size, due to ``color transparency'' for small dipoles: $T_{\bx\by}\propto r^{2\gamma}$, where $\gamma=1$ at tree-level (e.g.~in the MV model) and it remains equal to one when the evolution is computed at DLA (see below). Starting with the LO BFKL equation \eqref{lobfkl}, the ``naive'' (in the sense of no time-ordering) version of DLA is obtained by, first, factorizing out the dominant $r^2$ behavior of the dipole amplitude, via the rewriting 
 \begin{equation}
	T_{\bx\by}(Y) \equiv r^2 Q_0^2 \mcal{A}(Y,r^2)\,,
\end{equation}    
and then performing approximations which exploit the fact that the daughter dipoles are much larger than the parent one. That is, the dipole kernel is simplified as in \eqn{Mlargez} and for the dipole amplitudes one can keep just the two ``real'' terms, which describe the scattering of the daughter dipoles and which give equal contributions to DLA: $T_{\bx\bz}(Y)\simeq T_{\bz\by}(Y)=\bar z^2Q_0^2 \mcal{A}(Y,\bar z^2) $, where $\bar z\equiv |\bz-\by|\simeq |\bz-\bx|\gg r$. Notice that, for the time being,  we ignore the dependence of the reduced amplitude $ \mcal{A}(Y,r^2)$ upon the dipole impact parameter $\bm{b}\equiv(\bx+\by)/2$, to simplify notations. (This dependence will be restored when going beyond DLA, later on.) We thus find a simple equation,
 \beq\label{intDLA}
 { \calA(Y,r^2) } =  \calA^{(0)}(r^2) +
 {\abar}\int_0^Y\rmd Y_1 \int_{r^2}^{1/Q_0^2}\frac{\rmd \bar z^2}{\bar z^2}\, \calA(Y_1,z^2)\,,\eeq
 that we have  directly written in integral form. The inhomogeneous term in the r.h.s. is the tree-level amplitude and plays the role of an initial condition for the evolution with increasing $Y$:  $\calA^{(0)}(r^2)=\calA(Y=0,r^2)$. This integral equation can be solved (at least, formally) via iterations: $ \calA=
 \sum_{n=0}^{\infty}\calA^{(n)}$, with $\calA^{(n)}$ of order $\abar^n$.
For instance, for the simple
initial condition  $ \calA(0,r^2)=1$, one finds 
 \beq
 \label{A1}
 \calA(Y,\rho)=\sum_{n\ge 0}\frac{(\abar Y\rho)^n}{(n!)^2}\
  =\rmI_0(2\sqrt{\abar Y\rho})\,, \eeq
where $\rho\equiv\ln 1/{(r^2Q_0^2)}$ and $\rmI_0$ is a modified Bessel function.

Implicit in the above argument is the fact that all the gluons produced up to a given step in the evolution can act as sources for new emissions in the subsequent steps. This in turn requires that successive emissions be strictly ordered in time, i.e.~any fluctuation should have a lifetime smaller than its predecessors. In general, the lifetime of a right moving fluctuation is given by $\tau_{k} \sim 1/k^- = 2k^+/k_{\perp}^2$.  Accordingly, the time-ordering (TO) condition amounts to
\beq\label{TOcond}
\frac{2q^+}{Q^2}\gg \frac{2k_1^+}{ k_{1\perp}^2}\gg  \frac{2k_2^+}{ k_{2\perp}^2}\gg\dots\gg \frac{2q_0^+}{Q_0^2}\,,\eeq
where the leftmost inequality is the condition that the lifetime of the first gluon fluctuation be much smaller than
the coherence time  $\tau_q=2q^+/Q^2$ of the incoming dipole. Similarly, the rightmost inequality shows that, in order to 
significantly scatter, a fluctuation must live (much) longer than the width $\tau_0\simeq 1/q^-= {2q_0^+}/{Q_0^2}$ of the left-moving target. When computing the Feynman graphs for soft gluon emissions, this time-ordering is effectively enforced by the energy denominators (see e.g.~the discussion in \cite{Beuf:2014uia,Iancu:2015vea}). But clearly, this condition is violated by the solution to the DLA equation  \eqref{intDLA}, which involves unrestricted integrations over the phase-space \eqref{DLAphase}. To enforce TO to the accuracy of interest, it suffices to modify the integration limits in  \eqn{intDLA} according to \eqn{TOcond}, that is,
\begin{equation}
\label{adla}
	\mcal{A}(q^+,Q^2) = \mcal{A}^{(0)}(Q^2)
	+\abar \int^{Q^2}_{Q_0^2} \frac{\dif \kt^2}{\kt^2}
	\int_{q_0^+ (\kt^2 /Q_0^2)}^{q^+ (\kt^2/Q^2)} \frac{\dif k^+}{k^+} \mcal{A}(k^+,\kt^2),
\end{equation}
where we temporarily use the momentum variables $k^+$ and $\kt^2\equiv 1/\bar z^2$ (instead of $Y_1$ and $\bar z^2$), together with obvious notations like $Q^2=1/r^2$ and $\mcal{A}(q^+,Q^2) \equiv\calA(Y,r^2)$, to better emphasize the relation to the TO conditions \eqref{TOcond}.

Yet, at DLA, it is more economical to use logarithmic variables for both the longitudinal and the transverse phase-space: recalling the notations $\rho=\ln(Q^2/Q^2_0)$ and $Y=\ln (q^+/q_0^+)$ and similarly defining $\rho_1 \equiv \ln (\kt^2 /Q_0^2)$ and $Y_1 \equiv \ln (k^+/q_0^+)$, we can rewrite \eqn{adla} as\footnote{On this occasion, we would like to correct a mistake in one of earlier works \cite{Iancu:2015vea}: in that paper, the lifetime of a fluctuation was ordered w.r.t. to its parent dipole, but not also w.r.t. the target size; that is, the lower limit on $k^+$ in the analog of \eqn{adla} --- which is Eq.~(16) from Ref.~\cite{Iancu:2015vea}  --- was incorrectly written as $q_0^+$; similarly, the lower limit in the integral over $Y_1$ in \eqn{adlarho} was taken to be zero (see Eq.~(17) in Ref.~\cite{Iancu:2015vea}) instead of the correct value $\rho_1$. \label{foot:erratum}.}:
\begin{equation}
\label{adlarho}
	\mcal{A}(Y,\rho) = \mcal{A}^{(0)}(\rho)
	+\abar \int_{0}^{\rho} \dif \rho_1
	\int_{\rho_1}^{Y-\rho+\rho_1} \dif Y_1\, \mcal{A}(Y_1,\rho_1),
\end{equation}
where a step function $\Theta(Y-\rho)$, standing for the TO condition between the two end points in \eqn{TOcond}, is implicitly assumed. It is very important to notice that, in the context of this equation, the tree-level (reduced) amplitude $\mcal{A}^{(0)}(\rho) $ plays the role of a boundary condition at $Y=\rho$,
\begin{equation}
\label{bv}
	\mcal{A}^{(0)}(\rho) = \mcal{A}(Y=\rho,\rho),	
\end{equation}
which means that we are actually dealing with a {\it boundary value problem}. (Notice that, within the integrand, the function $ \mcal{A}(Y_1,\rho_1)$ is also needed only for $Y_1 > \rho_1$, so this boundary value problem is indeed well defined.) Moreover, \eqn{adlarho} is {\it non-local} in the projectile rapidity $Y$, because of the transverse dependence in the limits of the $Y_1$ integration. This becomes perhaps clearer after taking a derivative w.r.t.~$Y$ to deduce a differential version of this equation:
\begin{equation}
\label{adladiff}
	 \frac{\del  \mcal{A}(Y,\rho) }{\del Y}=
	 \abar \Theta(Y-\rho)\int_{0}^{\rho} \dif \rho_1\,\mcal{A}(Y-\rho+\rho_1,\rho_1).
\end{equation} 

Despite being formulated as a boundary value problem, the DLA evolution with TO is still simple enough to be solved  (at least for sufficiently simple expressions for the function $\mcal{A}^{(0)}(\rho)$) by iterating the integral equation \eqref{adla} (or  \eqref{adlarho}). But this is actually not needed: by inspection of the above equations, it is easy to see that the boundary value problem with TO can be equivalently rewritten as an initial value problem without TO --- i.e.~as the ``naive'' DLA equation  --- via the following change of the rapidity variable and the corresponding redefinition of the amplitude:
\beq\label{abarred}
Y\,\to\,\eta\equiv Y-\rho\,,\quad\quad 
\bar{\mcal{A}}(\eta, \rho) \equiv \mcal{A}(Y=\eta+\rho,\rho).
\eeq
The new function $\bar{\mcal{A}}(\eta, \rho) $ obeys the simple equation (for $\eta\ge 0$ of course)
\begin{equation}
	\label{abardla}
	\bar{\mcal{A}}(\eta,\rho) =\mcal{A}^{(0)}(\rho) 
	+\abar \int_{0}^{\rho} \dif \rho_1
	\int_{0}^{\eta} \dif \eta_1 \bar{\mcal{A}}(\eta_1,\rho_1),
\end{equation}
which  is similar to the ``naive'' DLA  equation \eqref{intDLA}, except for the replacement of $Y$ by $\eta$.
In particular, it describes an initial-value problem, with the initial condition $ \bar{\mcal{A}}(\eta=0,\rho)=\mcal{A}^{(0)}(\rho)$.

The fact that the DLA evolution becomes local when reformulated in terms of $\eta$ is easy to understand: ordering in $\eta$ is tantamount to ordering in the lifetime of the fluctuations; e.g.~the integration variable $\eta_1$ in \eqn{abardla} is recognized as
\begin{equation}\label{eta1def}
 	\eta_1= Y_1-\rho_1\ = \ln \frac{k_1^+}{q_0^+}-\ln\frac{k_{1\perp}^2}{Q_0^2} = \ln\frac{\tau_k}{\tau_0}\,,
 \end{equation} 
 and similarly $\eta=\ln(\tau_q/\tau_0)$. Hence
by integrating $\eta_1$ over the interval $0<\eta_1<\eta$, one ensures the proper ordering $\tau_0\ll\tau_k\ll\tau_q$ for the respective time scales. In other terms, by ordering the quantum fluctuations of the right-moving projectile in $\eta$ and $\kt^2$ (rather than $Y$ and $\kt^2$), the respective phase-space is properly counted, including all the kinematical constraints that matter to DLA.  

Alternatively, since $\tau_k=1/k^-$, the ordering in $\eta$ is also equivalent to an ordering in the variable $k^-$, which is increasing from the projectile towards the target. This variable is the light-cone energy of the fluctuations of the right-moving projectile, but can also be viewed as the longitudinal momentum for the fluctuations of the left-moving target. Thus, one can also interpret \eqn{abardla} as the standard DLA evolution of the {\em target}, as formulated in the variables $k^-$ and $\kt^2$: $k^-$ is strongly decreasing, while $\kt^2$ is strongly increasing, from one emission to the next one. This  {\em collinear}  evolution needs no special constraint since time-ordering in the corresponding LC time, i.e.~$x^-$, is automatically satisfied: the lifetime $\sim 2k^-/\kt^2$ of the fluctuations is strongly decreasing along the evolution.

As well known, the target rapidity is also the right variable to study DIS, since directly related to the kinematical variable $x_{\rm\scriptscriptstyle Bj}\equiv Q^2/s$ (Bjorken $x$) used in the experiments. One has indeed
 	 \begin{equation}\label{etadef}
 	\eta= Y-\rho\ = \ln \frac{q^+}{q_0^+}-\ln\frac{Q^2}{Q_0^2} = \ln \frac{2 q^+ q_0^-}{Q^2} = \ln \frac{s}{Q^2}
	= \ln \frac{1}{x_{\rm\scriptscriptstyle Bj}},
 \end{equation} 
 In particular, the condition $\eta\ge 0$ (i.e.~$Y\ge\rho$) corresponds to the kinematical boundary $x_{\rm\scriptscriptstyle Bj}\le 1$.  So, by solving an evolution equation in $\eta$, one can directly use the results to make predictions for observables like the DIS structure functions. On the contrary, when working in the $Y$ variable, one needs to re-express the final results in terms of $\eta\equiv Y-\rho$ in order to make contact with the phenomenology and, more generally, to have a meaningful physical interpretation.
 
This discussion makes clear that, at the level of DLA, there is no real advantage in working in the $Y$-representation: the evolution equation looks simpler in $\eta$ and this is also the variable in terms of which we need the final results. But here we are interested in a  dynamics which is much more complicated than DLA, namely the BK evolution at next-to-leading order (NLO) accuracy and even beyond. At leading-order (LO), one can still use the LO BK equation and merely interpret the associated rapidity variable as the target rapidity $\eta$, despite the fact that this equation has been constructed by studying the evolution of the dipole projectile\footnote{As an additional argument in this sense, one may recall the fact that the LO BK equation also follows from the JIMWLK evolution of the gluon distribution of the target. But this argument is not fully compelling in cases where the difference $\rho=Y-\eta$ is large, as in the problem at hand. Indeed, when constructing the JIMWLK equation for a left-moving target, the quantum fluctuations have not been ordered in the light-cone momentum $k^-$ --- as one would naturally do in the context of the linear, BFKL, equation --- but rather in the light-cone energy $k^+=2k^-/\kt^2$. (This was more convenient for the treatment of multiple scattering off the strong background field representing saturated gluons.) So, in that sense, also the JMWLK equation has been obtained by working in $Y$, and not in $\eta$.}. 
But the NLO corrections are only known for the dipole evolution with $Y$ and they include the  problematic double-(anti)collinear logarithm which leads to instabilities, as we have seen. As already recognized in the literature \cite{Beuf:2014uia,Iancu:2015vea}, this double collinear logarithm, together with similar corrections which occur in higher orders --- namely, corrections to the BK kernel that are of relative order $(\abar\rho^2)^n$ with $n\ge 1$ --- are related to the time-ordering of the successive gluon emissions and they can be resummed to all orders by simply enforcing TO in the LO BK equation. In what follows we shall present  a couple of strategies in that sense, which also allow to match with the remaining NLO corrections.

But before that, it is useful to use DLA in order to convince ourselves that the double collinear logs which appear at NLO are indeed related to time-ordering. As we shall shortly see, this relation is quite subtle, due to a fundamental difference in the way that these corrections are encoded in the NLO BK equation and in our above treatment of the DLA evolution, respectively: in the first case, they appear as corrections to the {\it kernel}, cf. Sect.~\ref{sec:nlobky}; in the second, the DLA kernel remains unchanged, but the TO condition modifies the {\it phase-space} for the evolution.

Specifically, we shall compare the perturbative estimates for the dipole amplitude to $\order{\abar^2}$ as produced, on one hand, by the DLA evolution with TO and, on the other hand, by the NLO BK evolution in \eqn{colleq} (in which we shall keep only the double collinear logarithm, for consistency with DLA). We use the simplest expression for the tree-level amplitude, namely $\mcal{A}^{(0)}(\rho)=1$. For the DLA evolution, the NLO result can be obtained either via two iterations of the integral equation \eqref{adlarho}, or by first solving the corresponding problem in $\eta$, which is simpler and has the  advantage of also giving the all-order result, and then replacing $\eta=Y-\rho$. Using the second method together with \eqn{A1}, one finds (for $Y>\rho$)
\begin{align}
	\label{DLANLO}
	\calA(Y,\rho)=\rmI_0(2\sqrt{\abar (Y\minus\rho)\rho})=1 + \abar (Y\minus\rho)\rho +
	 \frac{(\abar (Y\minus\rho)\rho)^2}{4}\,
	+\order{\abar^3}\,. \end{align}
It is convenient to first look at the terms linear in $Y$, that should naively correspond to one step in the NLO BK evolution (evaluated at DLA of course):
\beq
 \label{DLAlinY}
	\Delta\calA(Y,\rho)=\abar Y \rho\left(1-\frac{\abar\rho^2}{2}\right).
	\eeq
After multiplication by $r^2Q_0^2$, this should be compared to \eqn{colleqiter}, in which one can replace $Q_s\to Q_0$ for that purpose. Clearly, there is a mismatch between the coefficients in front of the double collinear logarithm in these two expressions: this is equal to $-1/2$ in \eqref{DLAlinY} but to $-1/6$ in \eqref{colleqiter}.  This mismatch might suggest that our present DLA calculation, which looks indeed very simple, is unable to correctly capture the double-collinear log at NLO. But this is actually not true: the correct result for $\calA(Y,\rho)$ to $\order{\abar^2}$ is the one appearing in \eqn{DLANLO}. This does not imply the existence of an error in the NLO calculation of the BK kernel: the latter is correctly given by  \eqn{colleq} to the accuracy of interest. What went wrong though, is the fact that, in obtaining  \eqn{colleqiter}, the NLO BK equation in  \eqn{colleq} has been solved as an initial value problem with the initial condition formulated at $Y=0$, i.e.~$\mcal{A}(Y=0,\rho)=1$. However, from our present discussion in this section, we know that,  as a consequence of time-ordering, the evolution in $Y$  starts being effective only for $Y>\rho$ and hence it must be formulated as a boundary value problem at $Y=\rho$. That is, a step function $\Theta(Y-\rho)$ with $\rho=\ln(1/r^2Q_s^2)$ must be implicitly understood in the r.h.s. of \eqn{colleq}. If one solves this boundary value problem with the boundary condition $\mcal{A}(Y=\rho,\rho)=1$, then \eqn{colleqiter} gets replaced by (we factor out an overall factor $r^2Q_s^2$ to comply with the present conventions)
\beq
 \label{DLAlinY1}
	\Delta\calA(Y,\rho)\Big |_{\mbox{\small 1-step}}=\abar\int_\rho^Y \dif Y_1 \int_0^\rho\dif \rho_1\left(1-\frac{\abar\rho_1^2}{2}\right)=
	\abar (Y-\rho) \rho\left(1-\frac{\abar\rho^2}{6}\right).
	\eeq
This is still not the same as  \eqn{DLAlinY}, but it does not have to:  \eqref{DLAlinY} is just a piece of the complete result to $\order{\abar^2}$ as appearing in \eqn{DLANLO}. To obtain the corresponding result for the ``NLO'' BK equation, i.e.~\eqn{colleq} interpreted as a boundary value problem, one must also perform the {\it second} iteration, which contributes to $\order{\abar^2}$ as well. To the accuracy of interest, this can be computed with the LO kernel and must involve only the  $\order{\abar}$ piece of the result given by the first iteration, that is  $\mcal{A}^{(1)}(Y,\rho)=\abar (Y-\rho) \rho$. (Incidentally, this piece has been properly reproduced by the first iteration in \eqn{DLAlinY1}, as expected.) One thus finds:
\beq
 \label{DLAlinY2}
	\Delta\calA(Y,\rho)\Big |_{\mbox{\small 2-step}}=\abar^2\int_\rho^Y \dif Y_1 \int_0^\rho\dif \rho_1\left(Y_1-\rho_1\right)
	\rho_1=
	\abar^2 (Y-\rho) \rho\left(\frac{\rho(Y+\rho)}{4}-\frac{\rho^2}{3}\right).
	\eeq
It is now easy to check that the sum of the results \eqref{DLAlinY1}  and \eqref{DLAlinY2} produced after 2 iterations coincides, as it should, with the NLO prediction of the DLA evolution with TO, cf. \eqn{DLANLO}. 

This example illustrates the fact that only a part of the radiative corrections associated with time-ordering --- namely, that part corresponding to the relative TO of the successive gluon emissions --- can be encoded into a renormalization of the kernel of the evolution equation, which is computable in perturbation theory\footnote{When computing the second iteration of the integral equation \eqref{adlarho}, it is easy to distinguish the effects of the global time constraints from those of the relative time ordering between the 2 gluon emissions. One can then check that the NLO effect of the latter is indeed equal to the double-collinear logarithm occurring in the NLO BK kernel, as exhibited in  \eqn{colleq}; see e.g.~Eq.~(13) in \cite{Iancu:2015vea} and the related discussion.}. But the corrections associated with the {\it global} time constraints --- the absolute upper limit  $\tau_q=2q^+/Q^2$  introduced by the coherence time of the incoming dipole and the absolute lower limit   $\tau_0=2q_0^+/Q^2_0$ representing the width of the target --- can only be taken into account by reformulating the evolution as a boundary-value problem, instead of an initial-value one. In particular, \eqn{DLANLO} also contains terms which are independent of $Y$ and start already at LO --- notice the $\order{\abar}$-correction $(-\abar\rho^2)$ --- and which could not be generated by an initial value problem formulated at $Y=0$. Such terms are manifestly introduced by the global time constraints alluded to above.

This being said, it is intuitively clear that for sufficiently high energies, such that $\abar(Y-\rho)\gg 1$, the effects of the global time constraints should be comparatively less important and that the asymptotic behavior at large $Y$ (and $\rho$ still large enough, $\abar\rho^2\gtrsim 1$, for the collinear resummation to be important) is rather controlled by the properly-resummed kernel alone --- or, equivalently (at least at DLA) by the rapidity shift $Y\to Y-\rho+\rho_1$ in the argument of the dipole amplitude in the r.h.s. of \eqn{adladiff}.

 \subsection{BK equation with time-ordering}
 \label{sec:BKnlY}
 
 In this section, we shall construct a generalization of the non-local equation \eqref{adladiff} which correctly accounts for the leading-order BK dynamics and at the same time resums all-orders radiative corrections enhanced by double collinear logarithms. As at DLA, this generalized (``collinearly-improved'') BK equation will be obtained by enforcing time-ordering for the successive gluon emissions. A similar construction has been originally presented in \cite{Beuf:2014uia}, which however missed the importance of formulating the ensuing non-local equation in $Y$ as a boundary-value problem.
 
The main difference w.r.t.~the previous discussion of the DLA evolution is the fact that the successive soft gluon emissions are not ordered in transverse momenta (or sizes) anymore: the daughter dipoles can be either larger, or smaller, than the parent one --- although the {\it typical} evolution for the ``dilute-dense'' physical problem at hand is still a ``hard-to-soft''  (or ``anticollinear'') evolution with increasing dipole sizes. But of course the emissions are still strongly ordered in longitudinal momenta and they will be required to be strongly ordered in lifetimes as well. As usual, we shall use $\bx$, $\by$ and $\bz$ to denote the transverse coordinates of the parent quark, antiquark, and emitted gluon respectively. For a collinear splitting in which one of the two daughter dipoles is much smaller than the other one, it is the size of this smallest dipole which should be related to the transverse momentum of the emitted gluon, via the uncertainty principle:
\beq\label{ktr<}
\kt^2\,\simeq\,\frac{1}{r_<^2}\qquad\mbox{with}\quad
r_<=\mbox{min}(|\bz-\bx|,\,|\bz-\by|)\,.\eeq
Indeed, if e.g.~$|\bz-\bx|\ll |\bz-\by|\simeq |\bx-\by|$, then the gluon has most likely been emitted by the quark at $\bx$. With this identification, the strong ordering conditions for the first gluon emission read
\beq\label{TO1}
q^+\gg k^+\gg q_0^+\,,\qquad
\frac{2q^+}{Q^2}\gg \frac{2k^+}{ k_\perp^2}\gg  \frac{2q_0^+}{Q_0^2}\,.\eeq

As in the case of DLA, these constraints are most naturally implemented at the level of the integral version of the BK equation (recall Eqs.~\eqref{adla} and \eqref{adlarho}). Consider first the lower limits in the two inequalities in \eqn{TO1}, i.e.~$ k^+\gg q_0^+$ and $ k^+\gg q_0^+ (\kt^2/Q_0^2)$; using our usual logarithmic variables, that is,
 $\rho_1 \equiv \ln (\kt^2 /Q_0^2)$ and $Y_1 \equiv \ln (k^+/q_0^+)$, we can rewrite these conditions as
 \beq\label{TOlow}
 Y_1 > 0
 \quad \& \quad 
 Y_1 > \rho_1 
 \quad \Longleftrightarrow\quad 
 Y_1 > \Theta( \rho_1) \rho_1\,.
 \eeq
Consider similarly the upper limits, which involve the momenta $q^+$ and $Q^2$ of the parent dipole; they amount to
\beq\label{TOup}
 Y >  Y_1
 \quad\&\quad 
 Y-\rho+\rho_1 >  Y_1
 \quad \Longleftrightarrow\quad 
 Y-\Theta(\rho- \rho_1)(\rho- \rho_1) > Y_1\,.
 \eeq
 These considerations immediately suggest the following integral form for the BK equation with TO:
 \begin{align}
	\label{BKTO}
	 S_{\bx\by}(Y)= S^{(0)}_{\bx\by}+
	\frac{\abar}{2\pi}
	\int \frac{\dif^2 \bz \,(\bx\minus\by)^2}{(\bx \minus\bz)^2 (\bz \minus \by)^2}\int\limits_
	{\Theta( \rho_1) \rho_1}^{Y-\Theta(\rho- \rho_1)(\rho- \rho_1)}\dif Y_1
	\big[S_{\bx\bz}(Y_1)
	S_{\bz\by}(Y_1) \minus S_{\bx\by}(Y_1) \big],
\end{align}
where $S^{(0)}_{\bx\by}$ denotes the respective estimate at tree-level (say, as given by the MV model) and we recall that $\rho_1$ stands for $\rho_1=\ln(1/r_<^2Q_0^2)$.


For the integral term in the r.h.s. of the above equation to be non-zero, the upper limit of the rapidity integral must be larger than the lower limit, $Y-\Theta(\rho- \rho_1)(\rho- \rho_1) > \Theta( \rho_1) \rho_1$, which in turn implies three different conditions depending upon the value of $\rho_1$:

\texttt{(i)} if $\rho_1 > \rho$, meaning $r_< \ll r$ (one small daughter dipole) $\Longrightarrow$ $Y>\rho_1$;

\texttt{(ii)} if $\rho > \rho_1 > 0$, meaning $r \ll  r_<  \ll 1/Q_0$ (large daughter dipoles) $\Longrightarrow$ $Y>\rho$;

\texttt{(iii)} if $\rho_1 < 0$, meaning $r_< \gg 1/Q_0$ (very large daughter dipole) $\Longrightarrow$ $Y-\rho>|\rho_1|$. 

In all these three cases, $Y$ must be larger than $\rho$, similarly to our previous finding at DLA. As in that case,
\eqn{BKTO} represents a boundary-value problem, with the boundary condition formulated at $Y=\rho$: $S_{\bx\by}(Y=\rho)=S^{(0)}_{\bx\by}$. For given values $Y$ and $\rho$ satisfying $Y>\rho$, the additional conditions above introduce limitations on the minimal value (condition  \texttt{(i)}) and respectively the maximal value (condition  \texttt{(iii)}) of the size $r_<$ of the smallest daughter dipole. 

Taking a derivative in \eqn{BKTO} w.r.t.~$Y$ and taking into account the constraints aforementioned, we arrive at a differential equation non-local in rapidity:
\begin{align}
	\label{BKTOdiff}
	\hspace*{-.5cm}
	\frac{\del S_{\bx\by}(Y)}{\del Y} = 
	\frac{\abar}{2\pi}\Theta(Y-\rho) &
	\int \frac{\dif^2 \bz \,(\bx\minus\by)^2}{(\bx \minus\bz)^2 (\bz \minus \by)^2}
	\Theta\left(Y-\Theta(\rho_1- \rho)\rho_1\right)\!
	\Theta\left(Y-\Theta(- \rho_1)(\rho + |\rho_1|)\right)\nonumber\\*[0.2cm]
	&\times \big[S_{\bx\bz}(Y \minus \Delta_{\bx\by\bz})
	S_{\bz\by}(Y \minus \Delta_{\bx\by\bz}) \minus S_{\bx\by}(Y\minus \Delta_{\bx\by\bz}) \big]\!,
\end{align}
where the rapidity shift $\Delta_{\bx\by\bz}$ is defined as
\begin{align}
\label{Delta}
\Delta_{\bx\by\bz} \equiv \Theta(\rho- \rho_1)(\rho- \rho_1) =\Theta(r_< -r)
\ln\frac{r_<^2}{r^2}
= {\rm max}\left\{ 0,\ln \frac{{\rm min}
\{(\bx\!-\!\bz)^2,(\bz\!-\!\by)^2\}}{(\bx\!-\!\by)^2} \right\}.	
\end{align}
The various rewritings in the r.h.s. of \eqn{Delta} are intended to emphasize that this shift is non-zero if and only if the daughter dipoles are (much) larger than the parent one.

\eqn{BKTOdiff}  can be further simplified  to the accuracy of interest by neglecting the rapidity shift in the ``virtual'' term, that is, by replacing $S_{\bx\by}(Y\minus \Delta_{\bx\by\bz}) \to S_{\bx\by}(Y)$. Indeed, we know that this virtual term does not contribute to DLA, hence the Taylor-series expansion of the shift $\Delta_{\bx\by\bz}$ in its rapidity argument cannot generate radiative corrections enhanced by double-collinear logs. (This can be checked via techniques that we will later develop in the $\eta$-representation; see also the discussion in  \cite{Beuf:2014uia}.) This discussion points towards an ambiguity inherent in our present construction of the collinearly-improved BK equation: the resummation of higher order corrections that is performed by this equation is ambiguous beyond the double-logarithmic accuracy. This ambiguity can in principle be fixed, order by order in perturbation theory, by comparing the strict perturbative expansion of \eqn{BKTOdiff}  --- as obtained via a Taylor expansion 
of the rapidity shift --- to the perturbative calculation of the BK kernel to the order of interest  \cite{Beuf:2014uia}. Later on, we shall explicitly perform such a matching  to NLO, i.e.~to $\order{\abar^2}$, but only in the $\eta$-representation, which is more useful in practice.

A similar ambiguity applies to the value $\Delta_{\bx\by\bz}$ of the rapidity shift: in the previous arguments, this was merely constrained via the uncertainty principle, so its value is not unique: any function which, for large daughter dipoles, is approximately equal to $\ln(r_<^2/r^2)$ and which rapidly vanishes for $r_< \ll r$, would be acceptable in that sense.  Changing one such a function for another should result in a correction  of $\order{\abar^2}$ (without double-logarithmic enhancement), or higher.
We shall shortly consider a different choice for the shift, with the purpose of numerically studying the scheme dependence of this non-local BK equation.

Returning to \eqn{BKTOdiff}, it is useful to notice that the product of the first two step functions can be more compactly written as
\beq
\Theta(Y-\rho) \Theta\left(Y-\Theta(\rho_1- \rho)\rho_1\right)=\Theta(Y-\rho_{\rm min}),
\eeq
where $\rho_{\rm min}$ is the {\it largest} among $\rho$ and $\rho_1$, meaning that it is built with the {\it smallest} among the three dipoles involved in the splitting:
\begin{align}	\label{rmin}
\rho_{\rm min}
\equiv
\ln\frac{1}{r_{\rm min}^2Q_0^2}\quad\mbox{with}\quad
	r_{\rm min} = 
	{\rm min} \{|\bx\!-\!\by|,|\bx\!-\!\bz|,|\by\!-\!\bz| \}.
\end{align}
Also, the third step function, which is effective only when $\rho_1< 0$, can be safely ignored in the problem at hand: negative values for $\rho_1$ correspond to very large daughter dipoles, with size $r_< \gg 1/Q_0$. Such dipoles are at saturation already at tree-level, so they will be deeply at saturation after allowing for the evolution with $Y$. Accordingly, their contribution to the evolution is strongly suppressed and can be neglected. We are finally led to the following, non-local, version of the BK equation:
\begin{align}
	\label{bkyto}
	\hspace*{-.5cm}
	\frac{\del S_{\bx\by}(Y)}{\del Y} = 
	\frac{\abar}{2\pi}&
	\int \frac{\dif^2 \bz \,(\bx\minus\by)^2}{(\bx \minus\bz)^2 (\bz \minus \by)^2}
	\Theta(Y-\rho_{\rm min}) 
	\big[S_{\bx\bz}(Y \minus \Delta_{\bx\by\bz})
	S_{\bz\by}(Y \minus \Delta_{\bx\by\bz}) \minus S_{\bx\by}(Y) \big]\!,
\end{align}
 \eqn{bkyto} looks similar to the one derived in \cite{Beuf:2014uia}, but it differs from the latter in the argument of
the step-function, which in \cite{Beuf:2014uia} was written as $\Theta(Y-\Delta_{\bx\by\bz})$.  This amounts to treating  the analog of  \eqn{bkyto} as an initial-value problem, with the initial condition formulated at $Y=0$, rather than as a boundary-value problem.  This difference should be important for the evolution at the early stages, but not for its asymptotic properties at $\abar(Y-\rho)\gg 1$.

\begin{figure}[t] 
\centerline{\hspace*{-.3cm}
  \includegraphics[width=0.46\textwidth]{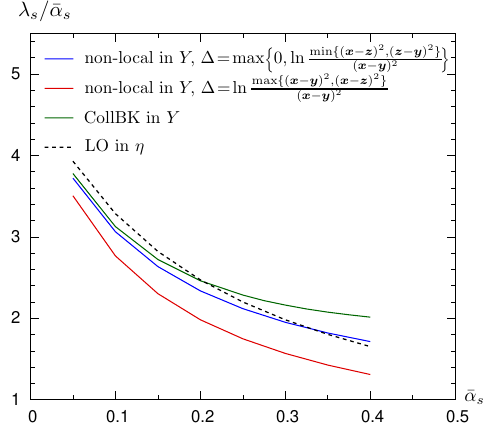}\qquad
\includegraphics[width=0.46\textwidth]{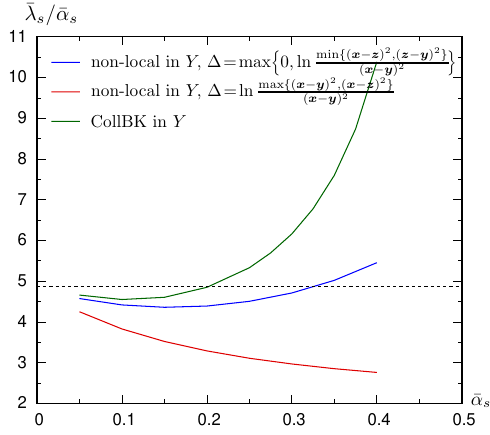}}
 \caption{\small Left: The asymptotic speed of the front (divided by $\abar$) in $Y$-evolution as a function of $\abar$ and for three different evolution schemes. The black-dashed line corresponds to the LO result $\bar{\lambda}_0/\abar \simeq 4.88$ transformed to $Y$-evolution according to the inverse of Eq.~\eqref{speedslope}. Right: The asymptotic speed of the same fronts when they are expressed as a function $\eta$ and $\rho$. Such a speed is consistent with Eq.~\eqref{speedslope}. In both figures, only the scheme corresponding to the red curve gives a physically acceptable solution.}
 \label{SresumY}
\end{figure}

A boundary-value problem like that exhibited in  \eqn{bkyto}, where the definition of the boundary  is {\it dynamical}, i.e.~it depends upon other variables (here, dipole sizes) that are modified by the evolution, represents a formidable mathematical problem which is very difficult to solve in practice. However, so long as we are interested only in asymptotic properties of the solution at large $Y$, such that $\abar(Y-\rho)\gg 1$, one can replace  \eqn{bkyto} by the initial-value formulation proposed in \cite{Beuf:2014uia}.
In what follows we shall perform such a numerical study --- namely, we shall compute the asymptotic value of the saturation exponent --- with two prescriptions for the rapidity shift: the one shown in \eqn{Delta}  and the one obtained by replacing the ``real'' term in  \eqn{bkyto} as follows
\beq\label{delta2}
S_{\bx\bz}(Y \minus \Delta_{\bx\by\bz})
	S_{\bz\by}(Y \minus \Delta_{\bx\by\bz}) \,\longrightarrow\,
	S_{\bx\bz}(Y \minus \Delta_{\bx\bz;r})
	S_{\bz\by}(Y \minus \Delta_{\bz\by;r}),\eeq 
where
\beq\label{deltaopt}
\Delta_{\bx\bz;r}\equiv {\rm max}\left\{ 0,\ln \frac{(\bx\!-\!\bz)^2}{r^2} \right\}.	
\eeq
For each of these 2 prescriptions, we have numerically solved \eqn{bkyto} as an initial-value problem with the initial condition given by the GBW model: $S_{\bx\by}(Y=0)=\rme^{- r^2Q_0^2/4}$. The solutions are stable, as expected, and the asymptotic speed $\lambda_s$ of the saturation fronts is considerably reduced as compared to the LO BK solution in $Y$, due to the reduction of the evolution phase-space introduced by the rapidity shift. However, as already mentioned, the physical interpretation and also the applications to the phenomenology involve the saturation fronts in $\eta$, that is, the function 
\beq
 {\bar S}_{\bx\by}(\eta)\equiv
 S_{\bx\by}(Y=\eta+\rho).
	\eeq
Hence, after solving the non-local BK equation in $Y$, we have replotted the results in terms of $\eta=Y-\rho$ and extracted the corresponding saturation exponent, to be denoted as $\bar\lambda_s$.	 In Fig.~\ref{SresumY} we display the asymptotic results for the saturation exponents divided by $\abar$ for the saturation fronts in $Y$ (left figure) and respectively in $\eta$ (right figure) as functions of $\abar$. In practice, these values have been 
extracted by fitting the numerical results for $\ln Q_s(Y)$ with the function shown in \eqn{dlogqs} within the range $5 < \abar Y < 25$ (and similarly for the evolution in $\eta$). We also show for comparison the corresponding prediction of the LO BK equation in $\eta$: in the right figure, this corresponds to the flat dotted line $\bar{\lambda}_0/\abar\simeq 4.88$, 
whereas in the left figure it shows a rather strong dependence upon $\abar$, as introduced by the change of variables from $\eta$ to $Y$. The additional curve in these figures, denoted as ``collBK'', will be discussed in the next section.

At this point, we open a parenthesis to present an analytic argument relating the (asymptotic) values of $\bar\lambda_s$ and $\lambda_s$ of the saturation exponent in the two representations. To that aim, we recall from Sect.~\ref{sec:nlobky} that the saturation fronts exhibit geometric scaling within a wide range of $\rho$ around the saturation scale $\rho_s\simeq\lambda_s Y$; for $\rho\gtrsim \rho_s$,
 the dipole amplitude can be approximated as
\begin{equation}
\label{yfront}
	T(Y,\rho) \approx \exp\big[-\gamma_s( \rho -\lambda_s Y)\big].
\end{equation}
Via the variables change $Y = \eta +\rho$, we find that an analogous scaling form holds in terms of $\eta$ and $\rho$:
\begin{equation}
\label{etafront}
	\bar T(\eta,\rho) \equiv T(Y= \eta +\rho, \rho) \approx \exp\big[-\gamma_s( \rho -\lambda_s (Y+\eta))\big] = \exp\big[-\bar{\gamma}_s( \rho -\bar{\lambda} \eta)\big],
\end{equation}
with the following values for the asymptotic speed and slope of the front in $\eta$:
\begin{equation}
	\label{speedslope}
	\bar{\lambda}_s = \frac{\lambda_s}{1-\lambda_s}
	\quad \textrm{and} \quad
	\bar{\gamma}_s = \gamma_s (1-\lambda_s).
\end{equation}
Since $\lambda_s$ is proportional to $\abar$, we see that for extremely small $\abar$ there is only a tiny difference between the two representations, consistent with the fact that a change in the rapidity variable (or equivalently in the energy scale) is a NLO effect. However, for  the typical values of $\abar$ relevant for phenomenology, the relations in \eqn{speedslope} lead to substantial differences between the two sets of values: they predict that the front in $\eta$ is {\it faster} ($\bar{\lambda}_s > {\lambda}_s$) and {\it less steep}  ($\bar{\gamma}_s <{\gamma}_s$) than the front in $Y$. 

We have checked that the relations \eqref{speedslope}  are indeed well satisfied by our findings in  Fig.~\ref{SresumY} (we shall later present numerical estimates for the slope $\bar{\gamma}_s$). 

We now close the parenthesis and return to a comparison between the numerical results obtained with the two prescriptions for $\Delta$, as displayed in Fig.~\ref{SresumY}. Looking first at the left figure, which refers to fronts in $Y$, it looks like the respective curves are relatively close to each other and also to the LO result in $\eta$ (replotted in terms of $Y$, of course); in particular, all these three curves lie well below the LO result in $Y$, that is, $\lambda_0/\abar\simeq 4.88$, and the deviation from the latter is monotonically increasing with $\abar$. However, after changing variable from $Y$ to $\eta$, the differences between the various curves are amplified by the division with ${1-\lambda_s}$, cf. \eqn{speedslope} ---  a relatively small number which decreases with $\abar$. (Notice that $\lambda_s$ is increasing with $\abar$ for all the curves  in Fig.~\ref{SresumY} (left), even though this increase is slower than linear.) As a consequence, the results for $\bar\lambda_s$ predicted by the two prescriptions for $\Delta$ look very different from each other. The one in \eqn{deltaopt} predicts an evolution which is considerably slower than at LO, with the difference   $(\bar\lambda_0-\bar\lambda_s)/\abar$ increasing monotonically with $\abar$. On the contrary, the original prescription in \eqn{Delta} yields a value for $\bar\lambda_s$ which stays closer to the LO result and, especially, is not monotonous with $\abar$: it first slightly decreases and then increases and shoots over $\bar\lambda_0$ for $\abar\gtrsim 0.35$. This is unphysical and we shall explain in Sect.~\ref{sec:shift} why it happens. Moreover, the difference between the respective predictions is considerably larger than the expected scheme dependence $\sim \order{\abar^2}$: e.g., for $\abar=0.3$, Fig.~\ref{SresumY} (right) shows a difference $\delta\bar\lambda_s\simeq 1.7\abar\simeq 5.6\abar^2$, where the coefficient 5.6 is unnaturally large.

As we shall discover in the next section, the situation becomes even less satisfactory after also considering the local version of the collinear resummation in $Y$.


\comment{}

\subsection{BK equation with collinearly-improved kernel}
\label{sec:bcy}

In this section, we shall describe an alternative formulation of the collinear resummation, originally proposed  in \cite{Iancu:2015vea}, which is closer in spirit to the usual philosophy of the perturbation theory and also to the corresponding treatment of the NLO BFKL equation. This method leads to an equation which is local in $Y$ and formulated as an initial-value problem, but where both the kernel and the initial condition at $Y=0$ include all-order resummations of the double-collinear logarithms. However, this method meets with a serious difficulty concerning the formulation of the initial condition beyond the double-logarithmic approximation, that was overlooked in the original analysis  in \cite{Iancu:2015vea} and which hinders its applications in practice.

To explain the general idea, let us first observe that the explicit solution to the DLA equation with TO that we have obtained in \eqn{DLANLO} (for the special boundary condition $\calA(Y=\rho,\rho)=1$)
admits an analytic continuation in the non-physical regime at $0\le Y<\rho$, as given by its series expansion:
\beq
 \label{Acont}
 \calA(Y,\rho)\equiv \sum_{n\ge 0}\frac{[\abar (Y-\rho)\rho]^n}{(n!)^2}\
  =\Theta(\rho-Y)\rmJ_0(2\sqrt{\abar (\rho-Y)\rho})\,+\,\Theta(Y-\rho)
  \rmI_0(2\sqrt{\abar (Y-\rho)\rho})\,, \nonumber\\
  \eeq
where $\rmJ_0(x)$ is the ordinary Bessel function of the first kind and is the analytic continuation of the modified Bessel function to purely imaginary values of its argument: $\rmJ_0(x)= \rmI_0(\rmi x)$, with real $x$. The r.h.s. of \eqn{Acont} represents the physical amplitude only for $Y\ge \rho$, but we use the same notation $ \calA(Y,\rho)$ also for its analytic continuation to $Y<\rho$, to avoid a proliferation of symbols.

Recall that the physical amplitude obeys the non-local evolution equation \eqref{adladiff} that for the present purposes will be rewritten in integral form:
	\beq   \mcal{A}(Y,\rho)
=	\mcal{A}^{(0)}(\rho) + \abar \int_\rho^Y\dif Y_1 \int_{0}^{\rho} \dif \rho_1\,\mcal{A}(Y_1-\rho+\rho_1,\rho_1).\eeq
By continuity, it is easy to understand that its analytic continuation \eqref{Acont} will obey the following integral equation (notice the change in the lower limit of the integral over $Y_1$),
\beq   \mcal{A}(Y,\rho)
=	\mcal{A}(Y=0,\rho) + \abar \int_0^Y\dif Y_1 \int_{0}^{\rho} \dif \rho_1\,\mcal{A}(Y_1-\rho+\rho_1,\rho_1).\eeq
which is an initial-value problem with an initial condition that follows from \eqn{Acont}: $\mcal{A}(Y=0,\rho)=\rmJ_0(2\sqrt{\abar \rho^2})$. Clearly, a similar equation must hold for any choice for the physical tree-level amplitude $\mcal{A}^{(0)}(\rho) $ (the corresponding initial condition will be shortly displayed). Moreover, as
demonstrated in \cite{Iancu:2015vea}, the above equation can be equivalently rewritten in a form which is {\it local} in $Y$:
\begin{equation}
	\label{adlak}
	\mcal{A}(Y,\rho) = \mcal{A}(Y=0,\rho) + 
	\abar \int_0^Y \dif Y_1 \int_0^\rho 
	\dif \rho_1 \mcal{K}_{\rm \scriptscriptstyle DLA}(\rho-\rho_1)
	\mcal{A}(Y_1,\rho_1).
\end{equation}
The kernel in the above equation is given by
\begin{equation}
	\label{kdla}
\mcal{K}_{\rm \scriptscriptstyle DLA}(\rho) \equiv 
\frac{\rmJ_1 \big( 2 \sqrt{\abar \rho^2} \big)}{\sqrt{\abar \rho^2}} = 1 - \frac{\abar \rho^2}{2} + \frac{\abar^2 \rho^4}{12} - \cdots,	
\end{equation} 
with $\rmJ_1$ the respective Bessel function of the first kind. In particular, the first non-trivial contribution to $\mcal{K}_{\rm \scriptscriptstyle DLA}(\rho)$, of $\order{\abar}$, plays the role of a NLO correction to the overall kernel. So, it is reassuring to notice that this correction agrees indeed with the double-logarithmic piece of the full NLO correction to the BK kernel, cf.  \eqn{colleq}.

Given a generic (physical) tree-level amplitude $\mcal{A}^{(0)}(\rho) $, the (unphysical) initial condition $\mcal{A}(Y=0,\rho)$ can be explicitly constructed due to our ability to exactly solve the evolution equation at DLA. This construction involves the following four steps: \texttt{(i)} start with the usual DLA equation in the $\eta$ variable, that is, \eqn{abardla}, and write down the general solution in terms of a Green's function;  \texttt{(ii)} deduce the corresponding solution in the $Y$-representation (which involves TO) via the change of variables $\eta=Y-\rho$;  \texttt{(iii)} use the series expansion of the latter to construct its analytic continuation to $Y<\rho$, and  \texttt{(iv)} 
take the limit $Y=0$ of the last result above. Clearly, steps  \texttt{(ii)}--\texttt{(iv)}  can be short-cut by simply letting $\eta\to -\rho$ in the analytic continuation of the general solution obtained in the first step. Specifically, the  general solution to \eqn{abardla}, can be written as 
\begin{equation}
	\label{abardlasol}
	\bar{\mcal{A}}(\eta,\rho) =\int_0^\rho\dif\rho_1\,\bar f(\eta, \rho-\rho_1)\,
	\mcal{A}^{(0)}(\rho_1),
\end{equation}
where the Green's function $\bar f(\eta, \rho)$ is the solution to  \eqn{abardla} with the initial condition $\mcal{A}^{(0)}(\rho) =\delta(\rho)$. This  Green's function can be easily constructed via iterations or via a Mellin transform \cite{Iancu:2015vea}, and reads 
\beq
 \label{fbess}
\bar f(\eta,\rho) =  \delta(\rho)+ \sqrt{\frac{\abar\eta}{\rho}}\,
 \rmI_1\left(2 \sqrt{\abar\eta\rho}\right),
 \eeq
where $\rmI_1$ is the respective modified Bessel function. The function $\bar f$ is a priori defined for $\eta>0$ but can be extended to negative $\eta$ by using the series expansion of $\rmI_1$. After also taking the limit $\eta\to -\rho$, one eventually finds for $\mcal{A}(Y=0,\rho) =\bar{\mcal{A}}(\eta=-\rho,\rho)$ the following expression\footnote{We would like to stress that this is not the same as the limit $Y=0$ of Eq.~(29) appearing in our earlier work \cite{Iancu:2015vea}: that equation  --- and hence its implication in Eq.~(31) --- were incorrect for the reason already discussed in \footnote~\ref{foot:erratum}.}:
\beq \label{ICcollDLA}
\mcal{A}(Y=0,\rho) = 
\mcal{A}^{(0)}(\rho)- \int_0^{\rho} d\rho_1\,\sqrt\frac{\abar \rho}{\rho-\rho_1} J_1(2 \sqrt{\abar \rho(\rho-\rho_1)})\,\mcal{A}^{(0)}(\rho_1)\,.
\eeq
One can check that when $\mcal{A}^{(0)}(\rho) =1$, the r.h.s. of the above equation reduces indeed to $\rmJ_0(2\sqrt{\abar \rho^2})$. It is interesting to observe how the resummation of the double-collinear logarithms is reorganized at the level of the local evolution equation. Both the  kernel $\mcal{K}_{\rm \scriptscriptstyle DLA}(\rho)$ and the initial condition $\mcal{A}(Y=0,\rho)$ rapidly oscillate for large values of  $\rho$, thus removing the potentially dangerous contributions due to very large daughter dipoles, which would violate the TO constraint. 

So long as we remain at the level of the DLA, the above manipulations may look redundant and not very useful: in order to deduce the local form of the DLA equation in \eqn{adlak}, we used the fact that its solution is {\it a priori} known. But what is interesting about  \eqn{adlak}, is that it can be promoted to full BK accuracy and thus provide an initial-value formulation which is local in $Y$ for the BK evolution with time-ordering. More precisely, as shown  in \cite{Iancu:2015vea}, this extension is possible and also rather unambiguous for the kernel of the equation, but not also for its initial condition.

When written as a differential equation, this collinearly-improved version of the BK equation (collBK) reads
\begin{equation}
	\label{collbk}
\frac{\partial S_{\bx\by}(Y)}{\partial Y} = 
 \frac{\abar}{2 \pi}
 \int
 \frac{\dif^2 \bz \,(\bx\minus\by)^2}{(\bx \minus\bz)^2 (\bz \minus \by)^2}\,
 \mcal{K}_{\rm \scriptscriptstyle DLA}(\rho_{\bx\by\bz})
 \left[S_{\bx\bz}(Y) S_{\bz\by}(Y) - S_{\bx\by}(Y) \right].
\end{equation}
The only non-obvious difference w.r.t.~its DLA counterpart in \eqn{adlak} refers to the argument of the kernel $\mcal{K}_{\rm \scriptscriptstyle DLA}$, which now reads
\begin{equation}
	\label{rhoxyz}
	\rho_{\bx\by\bz}^2 = 
	\ln \frac{(\bx-\bz)^2}{(\bx-\by)^2} 
	\ln \frac{(\by-\bz)^2}{(\bx-\by)^2}.
\end{equation}
This choice is motivated by the matching onto the NLO BK equation: the first non-trivial term in the expansion \eqref{kdla} of $ \mcal{K}_{\rm \scriptscriptstyle DLA}(\rho_{\bx\by\bz})$, that is, $ - \abar \rho_{\bx\by\bz}^2/2$, precisely coincides with the NLO piece involving a double-collinear logarithm in the NLO BK equation \eqref{nlobk}.  Hence, \eqn{collbk} achieves an all-order resummation of the double-collinear logs while precisely including the respective piece from the NLO BK kernel (and nothing more than that). This makes it easy to extend \eqn{collbk} to full NLO accuracy by simply adding the remaining $\abar^2$ pieces in the NLO kernel in  \eqn{nlobk}. As at DLA, the solution $\partial S_{\bx\by}(Y)$ to \eqn{collbk} represents the  physical $S$-matrix only for $Y \geq \rho$, with $\rho=\ln(1/r^2Q_0^2)$ and $r=|\bx-\by|$.

In order to solve  \eqn{collbk}, one also needs its initial condition at $Y=0$ and this turns out to be difficult to construct beyond DLA. Indeed, the function $S_{\bx\by}(Y=0)$ must be chosen in such a way that its evolution  from $Y=0$ up to $Y=\rho$ according to \eqn{collbk} reproduces the desired physical $S$-matrix at tree-level: $S_{\bx\by}(Y=\rho)=S_{\bx\by}^{(0)}$. Hence, in order to obtain  $S_{\bx\by}(Y=0)$, one must solve a boundary-value problem with the boundary condition at the upper limit $Y=\rho$. As already mentioned, we do not know how to solve this problem in general. Instead of that, one can try and use the DLA version of the initial condition, say as obtained by exponentiating $T_0(\rho)\equiv \rme^{-\rho}\mcal{A}(Y=0,\rho) $ with the function $\mcal{A}(Y=0,\rho)$ given by \eqn{ICcollDLA}. Such an approximation would entail some loss of accuracy in the calculation of the amplitude itself, but it should not affect the calculation of its asymptotic properties at large $Y$, like the asymptotic value of the saturation exponent, which is sensitive only to the kernel.

Motivated by this, we have solved   \eqn{collbk} with two different initial conditions, namely the standard GBW model $S_{\bx\by}(Y=0)=\exp(-r^2Q_0^2/4)$ and the collinearly-improved version of this model, with the collinear resummation performed at the level of DLA:  $S_{\bx\by}(Y=0)=\exp[-T_0(r)]$ with $T_0(r)= r^2Q_0^2\rmJ_0(2\sqrt{\abar \rho^2})/4$. The corresponding results for the dipole amplitude are of course very different --- in particular, the solution corresponding to the resummed initial condition shows oscillations in the unphysical domain at large $\rho>Y$, which however become less and less important with increasing $Y$ ---, but the corresponding predictions for the asymptotic value of $\lambda_s$ agree very well with each other, as expected. These predictions are shown too in Fig.~\ref{SresumY}, as the curve ``collBK''. As before, the most interesting plot is the one on the right, which refers to the saturation fronts in $\eta$.  One sees that, except for very small $\abar\lesssim 0.15$, the predictions of collBK strongly deviate from those of the non-local equation in $Y$ that we discussed in the previous section. Furthermore, for values $\abar \gtrsim 0.2$, which are moderately small, the extracted $\bar{\lambda}_s$ is unphysical since it overshoots the LO result $\lambda_0$.

The strong dispersion in the ``collinearly improved'' results that is manifest in Fig.~\ref{SresumY} (right) strongly suggests a failure of the resummation program for the radiative corrections associated with TO: the resummed evolution is indeed stable, but it lacks predictive power. In our opinion, this is related to the fact that the double collinear logarithms are typically very large, $\abar\rho^2\gg 1$, so the higher-order contributions generated by the interference between these very large corrections and the formally subleading ones, of order $\abar\rho$ or $\abar$, are numerically important as well. This problem cannot be cured by extending the resummation program to full NLO accuracy (i.e.~by adding the missing NLO corrections from \eqn{nlobk}). Indeed, two prescriptions for ``collinear improvement''  which are equivalent to NLO accuracy would differently treat  the large higher-order corrections and most likely result in different physical predictions. In the next section, we shall demonstrate that the $\eta$-representation --- i.e., the {\it projectile} evolution with increasing {\it target} rapidity --- offers a better framework for collinear resummations.

\section{Dipole evolution in $\eta$ at NLO}

Given the difficulties with constructing a meaningful perturbative formulation for the dipole evolution with $Y$ and the fact that most of the complications can be attributed to the perturbative treatment of the time-ordering condition, it looks natural to try and reformulate the problem as an evolution in which the successive emissions are directly ordered according to their lifetimes. The relevant evolution rapidity is then $\eta=\ln(\tau_k/\tau_0)$ and is formally the same as the rapidity of the target. 

We should emphasize from the beginning that we will not attempt to follow the evolution of the target. That would be a very difficult problem to study, since our target is a nucleus and in general its wavefunction is saturated for modes softer than the saturation scale.  We will just use ``mixed'' variables to describe the evolution of the projectile: the transverse coordinates will still correspond to the transverse momentum of a gluon emission, while for the longitudinal variable, we shall use the lifetime $\tau_k=2k^+/\kt^2$ of the gluon fluctuation instead of $k^+$.  Also, we shall not propose to compute the Feynman graphs directly in terms of $\tau_k$ (or $\eta$) instead of $k^+$ (or $Y$): $k^+$ is still the best variable for that purpose, since it is not modified by the multiple scattering off the nuclear target. Rather, we shall use the change of variables $\eta\equiv Y-\rho$ to transform the results of (strict) perturbation theory from the $Y$-representation to the $\eta$-representation. 

Strictly speaking, such a change of variables is a non-perturbative operation --- it mixes terms of all orders in the weak coupling expansion, as we have seen in the previous section ---, but its effects can be formally expanded in powers of $\alpha_s$ in order to construct the NLO BK kernel in $\eta$ from the corresponding kernel in $Y$. This construction will be performed in the first part of this section.

In the second part, we shall study the NLO BK evolution in $\eta$ and notably its linearized (BFKL) version w.r.t.~stability issues. Since the evolution in $\eta$ is properly time-ordered by construction, one may not expect any such an issue --- that is, one may expect the strict weak coupling expansion to be well behaved. Somewhat surprisingly though, we shall discover that this is not the case: the NLO corrections in $\eta$ include a double transverse logarithm of a different kind: a genuinely {\it collinear} double-log, associated with emissions where one of the daughter dipoles is much smaller than the parent one. Such ``soft-to-hard'' emissions are atypical in the physical problem at hand: they do not exist at the level of DLA, but in the general case they are allowed by the non-locality of the BFKL (dipole) kernel. They are responsible for the phenomenon known as ``BFKL diffusion'' --- a random walk in $\rho$ occurring on top of the typical ``hard-to-soft'' evolution. Albeit less troublesome than the anti-collinear double logs which appear in the $Y$-representation, these collinear double logs eventually entail a failure of the weak coupling expansion, that we shall analyse via both analytical and numerical studies in this section. This in turn calls for resummations to be discussed in Sect.~\ref{sec:shift}.

\subsection{Building the NLO BK equation in $\eta$}
\label{sec:NLOeta}

Our starting point is the NLO BK evolution in $Y$, that is \eqn{nlobk}, which we succinctly recall here by highlighting only those terms which are relevant to our presents purposes:    
\begin{align}
 \label{nlobky}
 \hspace*{0cm}
 \frac{\partial S_{\bx\by}(Y)}{\partial Y} = &\,
 \frac{\abar}{2 \pi}
 \int 
 \frac{\dif^2 \bz \,(\bx\minus\by)^2}{(\bx \minus\bz)^2 (\bz \minus \by)^2}\,
 \left[S_{\bx\bz}(Y) S_{\bz\by}(Y) - S_{\bx\by}(Y) \right]
 \nn*[0.2cm]
 &-\frac{\abar^2}{4\pi}
 \int
 \frac{\dif^2 \bz \,(\bx\minus\by)^2}{(\bx \minus\bz)^2 (\bz \minus \by)^2}\,
 \ln \frac{(\bx \minus\bz)^2}{(\bx \minus\by)^2} \ln \frac{(\by \minus\bz)^2}{(\bx \minus\by)^2}
 \left[S_{\bx\bz}(Y) S_{\bz\by}(Y) - S_{\bx\by}(Y) \right]
 \nn*[0.2cm]
 &+\,\abar^2 \times \text{``regular''}.
 \end{align}
In writing the r.h.s. we have separated the LO term from the NLO ones and we have explicitly displayed only the NLO piece containing the double anti-collinear logarithm.  All the other NLO terms (including the running coupling corrections and the single transverse logarithms expressing  DGLAP physics) have been collectively denoted as ``regular''. 
We now change variables according to \eqn{abarred}, that is, $\eta = Y - \rho$ with $\rho=\ln(1/r^2 Q_0^2)$, and rewrite the various $S$-matrices in the $\eta$-representation, as
\begin{align}
	\label{sbarred1}
	& S_{\bx\by}(Y) = S_{\bx\by}(\eta+\rho) 
	\equiv \bar{S}_{\bx\by}(\eta)
	\\
	\label{sbarred2}
	& S_{\bx\bz}(Y) = S_{\bx\bz}(\eta+\rho) = 
	S_{\bx\bz}\left(\eta + \rho_{\bx\bz} +\ln\frac{(\bx-\bz)^2}{(\bx-\by)^2}\right) = 
	\bar{S}_{\bx\bz} \left(\eta +\ln\frac{(\bx-\bz)^2}{(\bx-\by)^2}\right),
\end{align}
with the obvious notation $\rho_{\bx\bz} \equiv \ln[1/(\bx-\bz)^2 Q_0^2]$; clearly, a similar rewriting holds for $S_{\bz\by}$. Upon substitution of the above into \eqn{nlobky} we would get an equation non-local in $\eta$. Nevertheless, when working strictly at NLO in $\abar$, one can treat  the rapidity  shift in the argument of $\bar{S}_{\bx\bz}$ as a quantity of order $\mcal{O}(1)$, which is typically much smaller than the rapidity $\eta$ itself (recall that we are eventually interested in large values of $\eta$ such that $\abar\eta\gtrsim 1$) and hence can be expanded out in a Taylor series. Recognizing the fact that each rapidity-derivative of the $S$-matrix like ${\partial \bar{S}_{\bx\bz}}/{\partial \eta}$ is formally  suppressed by a power of $\abar$ --- since one derivative expresses the effect of one step in the $\eta$ evolution (per unit $\eta$) ---, it becomes clear that to the desired order of accuracy, it is enough to keep the first non-trivial term in this expansion, which is linear in the shift:
\begin{align}
	\label{shiftexpand}
	\hspace*{-0.7cm}
	\bar{S}_{\bx\bz} \left(\eta +\ln\frac{(\bx-\bz)^2}{(\bx-\by)^2}\right)
	&\simeq
	\bar{S}_{\bx\bz}(\eta)
	+ \ln\frac{(\bx-\bz)^2}{(\bx-\by)^2}
	\frac{\partial \bar{S}_{\bx\bz}(\eta)}{\partial \eta}
	\nn*[0.2cm]
	\simeq\, &
	\bar{S}_{\bx\bz}(\eta)
	+ 
	\frac{\abar}{2\pi}
	\int
	\frac{ \dif^2 \bu\,(\bx\minus\bz)^2}{(\bx \minus\bu)^2 (\bu \minus \bz)^2}\, 
	\ln\frac{(\bx-\bz)^2}{(\bx-\by)^2}
	\left[\bar{S}_{\bx\bu}(\eta) \bar{S}_{\bu\bz}(\eta) - \bar{S}_{\bx\bz}(\eta) \right],
\end{align} 
where in evaluating the derivative term it was sufficient to use the LO BK equation in $\eta$. Thus a shift in the rapidity argument, which originates from the change of the rapidity variable for the daughter dipoles, is equivalent to adding a term of order $\mcal{O}(\abar)$.  Within the strict perturbative logic that we are temporarily pursuing, the rapidity shift is therefore important only in the LO piece in the r.h.s. of \eqn{nlobky}, in which case it can be expanded out as in \eqn{shiftexpand}. In all the NLO terms, the rapidity shift can be safely neglected, so one can replace e.g.~$S_{\bx\bz}(Y)\to \bar S_{\bx\bz}(\eta)$. Using also the property that the LO term is invariant under $\bx\minus\bz \to \bz\minus\by$ in order to combine some terms, we arrive at
\begin{align}
 \label{nlobketa}
 \hspace*{0cm}
 \frac{\partial \bar{S}_{\bx\by}(\eta)}{\partial \eta} = &\,
 \frac{\abar}{2 \pi}
 \int 
 \frac{\dif^2 \bz \,(\bx\minus\by)^2}{(\bx \minus\bz)^2 (\bz \minus \by)^2}\,
 \left[\bar{S}_{\bx\bz}(\eta) \bar{S}_{\bz\by}(\eta) - \bar{S}_{\bx\by}(\eta) \right]
 \nn*[0.2cm]
 &-\frac{\abar^2}{4\pi}
 \int 
 \frac{\dif^2 \bz \,(\bx\minus\by)^2}{(\bx \minus\bz)^2 (\bz \minus \by)^2}\,
 \ln \frac{(\bx \minus\bz)^2}{(\bx \minus\by)^2} \ln \frac{(\by \minus\bz)^2}{(\bx \minus\by)^2}
 \left[\bar{S}_{\bx\bz}(\eta) \bar{S}_{\bz\by}(\eta) - \bar{S}_{\bx\by}(\eta) \right]
 \nn*[0.2cm]
 &+ \frac{\abar^2}{2\pi^2}
 \int \frac{\dif^2 \bz\, \dif^2 \bu\, (\bx \minus \by)^2 }{(\bx \minus \bu)^2 (\bu \minus \bz)^2 (\bz \minus \by)^2} 
\ln\frac{(\bu-\by)^2}{(\bx-\by)^2}\,
  \bar{S}_{\bx\bu}(\eta) 
 \left[\bar{S}_{\bu\bz}(\eta) \bar{S}_{\bz\by}(\eta) - \bar{S}_{\bu\by}(\eta) \right] 
 \nn*[0.2cm]
 &+\,\abar^2 \times \text{``regular''}.
 \end{align} 
 The third term in the r.h.s. has been generated by expanding out the rapidity shift within the LO term, according to \eqn{shiftexpand}. (In writing this term we  relabelled the integration variables according to $\bu \leftrightarrow \bz$, in order to conform with the notation used in earlier sections.) Remarkably, the $S$-matrix structure of this last term is identical to the one appearing in the double-integration term in \eqn{nlobk}. 
 
 \eqn{nlobketa} is the NLO BK equation for the evolution in $\eta$. It is a local equation in rapidity and differs (functionally) from the corresponding evolution in $Y$ given in \eqn{nlobky} only by an extra term, the one appearing in the third line in \eqn{nlobketa}. 
 
From the discussion in the previous section, we expect the main effect of the change of rapidity variable from $Y$ to $\eta$ to be the elimination of the double anti-collinear logarithms from the perturbative expansion. It is instructive to explicitly check this property at NLO, on the basis of \eqn{nlobketa}. To that aim, we need to compare to the terms of $\order{\abar^2}$ that are explicitly shown in the r.h.s. of \eqn{nlobketa}, in the weak scattering regime where these terms can be linearized. The transverse integrations in these terms look very different from each other, but after linearization they can both be diagonalized via a Mellin transform. Hence, it is convenient to compare the contributions brought by these two terms to the BFKL characteristic function $\omega(\gamma)$ (the Mellin transform of the BFKL kernel). 

Consider therefore the term introduced by the change of variables and which involves a double integration over transverse coordinates. After linearization, this term (to be denoted as $\Delta \bar{T}_{\bx\by}$) becomes
\begin{equation}
 	\label{deltatxy}
 	\Delta \bar{T}_{\bx\by} 
 	= \frac{\abar^2}{2\pi^2}
 \int \frac{\dif^2 \bz\, \dif^2 \bu\, (\bx \minus \by)^2 }{(\bx \minus \bu)^2 (\bu \minus \bz)^2 (\bz \minus \by)^2} 
\ln\frac{(\bu-\by)^2}{(\bx-\by)^2}\,
 \left(\bar{T}_{\bu\bz} + \bar{T}_{\bz\by} - \bar{T}_{\bu\by} \right), 
 \end{equation}
where  the rapidity argument $\eta$ is implicit. 
To compute the respective contribution, denoted as $\Delta\omega(\gamma)$, to the characteristic function, one must insert a power-like {\it Ansatz} for the dipole amplitude within the r.h.s. of  \eqn{deltatxy}:  
$\bar{T}_{\bu\bz}=  |\bu-\bz|^{2\gamma}$ with $0\le \Re(\gamma) \le 1$. One finds
\begin{align}
 \label{domega}
 \Delta\omega(\gamma) &= 
 \frac{1}{r^{2\gamma}}\,
 \frac{\abar^2}{\pi}	
 \int 
 \frac{ \dif^2 \bu \, r^2}{u^2 |\br \minus \bu|^2}\,
 \ln\frac{(\br-\bu)^2}{r^2}\,
 \underbrace{\frac{1}{2\pi}
 \int 
 \frac{ \dif^2 \bz \, u^2}{z^2 |\bu \minus \bz|^2}\,
 \left( z^{2\gamma} + |\bu-\bz|^{2\gamma}-u^{2\gamma} \right)}_{\displaystyle \chi_0(\gamma) u^{2\gamma}}
 \nn
 &= 2 \abar^2 \chi_0(\gamma) \frac{\dif}{\dif \gamma}
 \left\{\frac{1}{2\pi}
 \int
 \frac{ \dif^2 \bu \, r^2}{u^2 |\br \minus \bu|^2}\,
 \left[\frac{(\br-\bu)^2}{r^2} \right]^\gamma
 \right\} 
 = \abar^2 \chi_0(\gamma) \chi_0'(\gamma),
\end{align}
with $\chi_0(\gamma)$ the LO characteristic function. This result is indeed consistent with the expected form for the change in $\omega(\gamma)$ due to a change in the energy (rapidity) scale in the BFKL evolution \cite{Ciafaloni:1998gs,Fadin:1998py}. It is important to study the behavior near the collinear pole at $\gamma = 0$ and, respectively, the anti-collinear  one at $\gamma =1$:
\begin{align}
	\label{chichip}
	\chi_0(\gamma) \chi_0'(\gamma) = 
	\begin{cases}
		{\displaystyle -\frac{1}{\gamma^3}} + 2 \zeta_3 + \mcal{O}(\gamma^2)
		\quad &\text{when} \quad
		\gamma \to 0,
		\\*[0.4cm]
		{\displaystyle -\frac{1}{(1-\gamma)^3}} + 2 \zeta_3 + \mcal{O}\big[(1-\gamma)^2\big]
		\quad &\text{when} \quad
		\gamma \to 1.
	\end{cases}
\end{align}
As expected, the triple pole at $\gamma=1$ which is introduced by the change of variable is such that it precisely cancels the respective pole associated with the double anti-collinear logarithm, i.e.~the second term in the r.h.s. of \eqn{nlobketa}. Accordingly, the NLO BFKL kernel for the evolution in $\eta$ has no triple pole at $\gamma=1$, but it exhibits a triple pole at $\gamma=0$ (cf.~also below \eqn{omegabarser}). We also notice that the change of variables introduces no additional poles (neither double, or single) at  $\gamma=0$ or  $\gamma=1$. 

It is also instructive to see the cancellation of the double anti-collinear log directly in coordinate space. This is indeed possible in the linear approximation, since one of the two integrations in \eqn{deltatxy} for $\Delta \bar{T}_{\bx\by}$ can be explicitly performed. In Appendix \ref{app:nlobfkl}, we find
\begin{align}
 	\label{deltatxyfin}
 	\hspace*{-0.8cm}
 	\Delta \bar{T}_{\bx\by} 
 	=\ &\frac{\abar^2}{4\pi}
 \int \frac{\dif^2 \bz\, (\bx \minus \by)^2 }{(\bx \minus \bz)^2 
 (\bz \minus \by)^2} 
\ln\frac{(\bx-\bz)^2}{(\bx-\by)^2} 
\ln\frac{(\bz-\by)^2}{(\bx-\by)^2}\,
 \left(\bar{T}_{\bx\bz} + \bar{T}_{\bz\by} \right)
 \nn*[0.2cm]
 &- \frac{\abar^2}{4\pi}
  \int \frac{\dif^2 \bz\, (\bx \minus \by)^2 }{(\bx \minus \bz)^2 
 (\bz \minus \by)^2}
 \left[
\ln\frac{(\bx-\by)^2}{(\bx-\bz)^2} 
\ln\frac{(\bz-\by)^2}{(\bx-\bz)^2}\, \bar{T}_{\bx\bz}
+ 
\ln\frac{(\bx-\by)^2}{(\bz-\by)^2} 
\ln\frac{(\bx-\bz)^2}{(\bz-\by)^2}\, \bar{T}_{\bz\by} 
\right]. 
 \end{align}
The first term in the above r.h.s. exactly cancels the ``real'' piece (which contains the double anti-collinear logarithm) in the linearized version of the second term in the r.h.s. of \eqn{nlobketa}.  The second term in \eqn{deltatxyfin} vanishes for large daughter dipoles, but its kernel develops a large double logarithm when either $\bz\to \bx$ or $\bz\to\by$, i.e.~when one of the daughter dipoles is much smaller than their parent (equivalently, for very disparate in size daughter dipoles). This {\em collinear} double logarithm corresponds in Mellin space to the triple pole at $\gamma=0$ exhibited in \eqn{chichip}. 

To summarize, the NLO BFKL equation for the evolution with the target rapidity $\eta$  can be written in the coordinate representation as
\begin{align}
 \label{nlobfkleta}
 \hspace*{-0.9cm}
 \frac{\partial \bar{T}_{\bx\by}(\eta)}{\partial \eta} = &\,
 \frac{\abar}{2 \pi}
 \int 
 \frac{\dif^2 \bz \,(\bx\minus\by)^2}{(\bx \minus\bz)^2 (\bz \minus \by)^2}\,
 \left[\bar{T}_{\bx\bz}(\eta) + \bar{T}_{\bz\by}(\eta) - \bar{T}_{\bx\by}(\eta) \right]
 \nn*[0.2cm]
 &- \frac{\abar^2}{4\pi}
  \int \frac{\dif^2 \bz\, (\bx \minus \by)^2 }{(\bx \minus \bz)^2 
 (\bz \minus \by)^2}
 \left[\ln\frac{(\bx-\by)^2}{(\bx-\bz)^2} 
 \ln\frac{(\bz-\by)^2}{(\bx-\bz)^2}\, \bar{T}_{\bx\bz}(\eta)
 + 
 \ln\frac{(\bx-\by)^2}{(\bz-\by)^2} 
 \ln\frac{(\bx-\bz)^2}{(\bz-\by)^2}\, \bar{T}_{\bz\by}(\eta) \right]
 \nn*[0.2cm]
 &+\frac{\abar^2}{4\pi}
 \int \frac{\dif^2 \bz \,(\bx\minus\by)^2}{(\bx \minus\bz)^2 (\bz \minus \by)^2}\,
 \ln\frac{(\bx-\bz)^2}{(\bx-\by)^2} 
 \ln\frac{(\bz-\by)^2}{(\bx-\by)^2}\,
 \bar{T}_{\bx\by}(\eta)
 \nn*[0.2cm]
 &+\,\abar^2 \times \text{``regular''}.
 \end{align} 
 The NLO corrections explicitly shown in the r.h.s. include two ``real'' terms, proportional to the dipole amplitudes $\bar{T}_{\bx\bz}$ and respectively $ \bar{T}_{\bz\by}$ for the daughter dipoles, and one ``virtual'' term, which involves the amplitude $ \bar{T}_{\bx\by}$ of the parent dipole. In the ``real'' terms, the amplitudes are multiplied by double transverse logarithms which become large in the {\it collinear} regime, i.e.~when one of the daughter dipoles is much smaller than the other one (and than its parent). Interestingly though, this is only the case for the double logarithm multiplying the scattering amplitude of the {\it smallest} dipole. For instance, when $|\bx\!-\!\bz|\to 0$, the double-logarithmic  factor in front of $\bar{T}_{\bx\bz}$ becomes large, $\sim\ln^2[r^2/ (\bx-\bz)^2]$, while that in front of $\bar{T}_{\bz\by}$ approaches to zero. This is likely related to the fact that there is a large phase-space for transverse evolution between the parent dipole and the {\it small} daughter dipole alone. 
 
The  ``virtual'' term involves a double {\it anti-}collinear logarithm, which by itself becomes large for large daughter dipoles. Yet, the corresponding integral over  $\bz$ brings no specially large contribution, because it is not amplified by the dipole amplitude  (in contrast to what was happening for the ``real'' terms in the first line of \eqn{deltatxyfin}). Indeed, this integral yields a pure number,
 \begin{equation}
\label{zetaint}
\frac{\abar^2}{4\pi}
\int \frac{\dif^2 \bz \,(\bx\minus\by)^2}{(\bx \minus\bz)^2 
(\bz \minus \by)^2}\,
 \ln\frac{(\bx-\bz)^2}{(\bx-\by)^2} 
 \ln\frac{(\bz-\by)^2}{(\bx-\by)^2}\,
 \bar{T}_{\bx\by} = \abar^2 \zeta_3 \bar{T}_{\bx\by},
\end{equation}
which confirms that this virtual term is a ``regular'' NLO piece, on the same footing as the other NLO corrections that have been omitted in writing \eqn{nlobfkleta}.  Accordingly, in what follows, we shall mainly focus on the ``real'' NLO corrections in \eqn{nlobfkleta} and the associated double {\it collinear} logarithms. 

\subsection{An instability in the evolution in $\eta$ at NLO}
\label{sec:nloinst}

The fact that the NLO corrections to the BFKL kernel for the evolution in $\eta$ include a piece enhanced by a double collinear logarithm may look at a first sight disturbing: the corresponding piece for the evolution in $Y$ represents a source of instabilities associated with violations of time-ordering, but such instabilities should not exist in the evolution with $\eta$, where the proper time-ordered is {\it a priori} guaranteed. But a bit of thinking reveals that the consequences of the double transverse logarithms are indeed very different in the two cases. In the evolution with $Y$, the {\em anti-collinear} logarithms become large in the {\em typical}, hard-to-soft, evolution of the dipole amplitude: their contribution is enhanced by the fact that large daughter dipoles scatter stronger than their parent. On the contrary, in the evolution with $\eta$, the effects of the {\em collinear} logarithms are suppressed by the scattering, which strongly disfavors the soft-to-hard evolution, i.e.~the emission of very small dipoles. 

Let us present a simple calculation supporting the above arguments. Consider the emission of a very small daughter dipole with transverse size much smaller than $r$ and keep only the dominant contributions to \eqn{nlobfkleta} in the limit where either $\bz\to\bx$, or $\bz\to\by$; one finds
\begin{equation}
	\label{bfkllog}
	\frac{\partial \bar{T}(\eta,r^2)}{\partial \eta} \simeq 
	\abar \int_0^{r^2} \frac{\dif z^2}{z^2}
	\left(1 - \frac{\abar}{2} \ln^2 \frac{r^2}{z^2} \right) 
	\bar{T}(\eta,z^2)\,,
\end{equation}
where $z$ denotes the size of the smaller daughter dipole (either $|\bz\minus\bx|$, or  $|\bz\minus\by|$)  and $\bar{T}(\eta,z^2)$  is the corresponding scattering amplitude. We shall estimate the integral in the r.h.s. with the power-like  {\it Ansatz}  $\bar{T}(z^2)\propto z^{2\gamma}$, with $0 <\gamma\le 1$ to account for color transparency  and also for a possible BFKL ``anomalous dimension''. A simple calculation yields
\begin{equation}
	\label{bfkllog2}
	\frac{\Delta \bar{T}(\eta,r^2)}{\bar{T}(\eta,r^2)} \simeq \abar\eta\left(\frac{1}{\gamma}-\frac{\abar}{\gamma^3}\right),
	\end{equation}
a result that could have been anticipated on the basis of Eqs.~\eqref{deltatxy}--\eqref{chichip}. Unlike for the corresponding analysis in $Y$, cf. Eqs.~\eqref{colleq}--\eqref{colleqiter}, in the present case the presence of double transverse logarithms in the NLO kernel does {\it not} entail similar double-logarithmic corrections in the evolution $\Delta \bar{T}$ of the dipole amplitude. Rather, the NLO piece in the r.h.s. of \eqn{bfkllog2} is a pure-$\order{\abar}$ correction.

From the above discussion, one may conclude that the double collinear logarithms  visible in the r.h.s. of \eqn{nlobfkleta} should be innocuous in practice. However, this conclusion is too strong and must be nuanced, as we shall explain in the remaining part of this section: these collinear logs are troublesome too, in the sense of triggering an instability which calls for all-order resummations. 

With reference to \eqn{bfkllog2}, the emergence of such an instability can be understood as follows: although the evolution in $\eta$ does not give rise to large double logarithms in the solution, it will generate an anomalous dimension $\gamma < 1$ after some steps, typically when $\eta \gtrsim 1/\abar$. We shall later find that this effect is substantial: the value of $\gamma$ which is relevant for the saturation fronts in $\eta$ is in fact not far from 1/2. Hence, the NLO piece in the r.h.s. of \eqn{bfkllog2} is a correction of relative order $\abar/\gamma^2 \sim 4\abar$, which is of $\order{1}$ for the relevant values of $\abar$. Since moreover this correction is negative, it is clear that it will have a large impact on the solution and it has indeed the potential to trigger an instability.

In what follows, we shall first study the emergence of this instability (via analytic methods) at the level of the linear (BFKL) equation, where its consequences turn out to be dramatic:  they prevent the construction of meaningful numerical solutions. Later on, we shall argue that this instability is  somewhat tamed, albeit not fully washed out, by the non-linear effects encoded in the BK equation. Our proof in that sense will not be rigorous, since we shall not attempt to solve the full NLO BK equation in $\eta$. Rather, we shall numerically solve a simplified version of this equation, in which the NLO double collinear logarithm is simply added to the LO kernel (see below for details).

\begin{figure}[t]
\centerline{\hspace*{-.3cm}
  \includegraphics[width=0.5\textwidth]{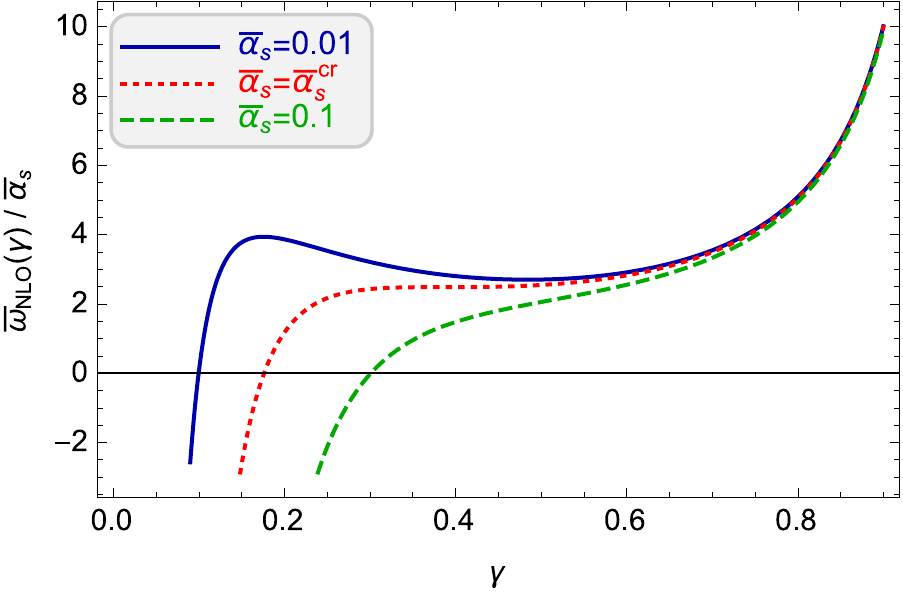}\qquad
\includegraphics[width=0.5\textwidth]{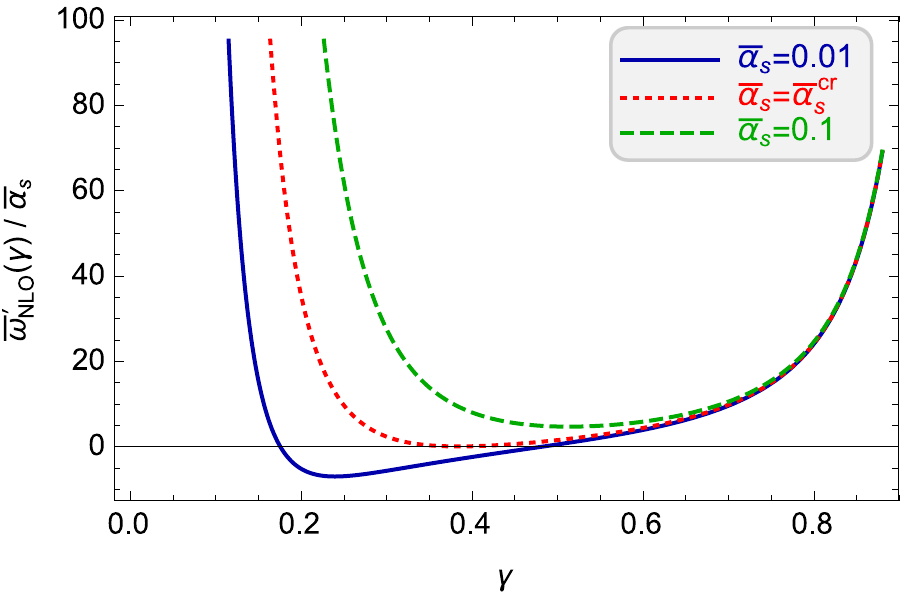}}
\caption{\small Left: The NLO characteristic function $\bar{\omega}(\gamma)$ computed according to \eqn{omegabar} for various values of the coupling $\abar$. Right: Its first derivative $\bar{\omega}'(\gamma)$.}
 \label{fig:alpha} \end{figure}

To start with, we shall construct an explicit solution to the NLO BFKL equation in $\eta$ using the standard technique of the Mellin transform. For the present purposes, it is enough to consider a truncation of this equation which includes only the NLO terms that are explicitly written in the r.h.s. of \eqn{nlobfkleta}. The characteristic function  associated with this truncated equation, as obtained by 
acting on an amplitude which is a pure power and using Eqs.~\eqref{zetaint} and \eqref{domega2},
reads
\begin{equation}
	\label{omegabar}
	\bar{\omega}(\gamma) = 
	\abar \chi_0(\gamma)
	+ \frac{\abar^2}{4}\left[ 
	4 \zeta_3 + 2 \chi_0(\gamma) \chi_0'(\gamma) - \chi_0''(\gamma)
	\right],
\end{equation}
and is displayed (together with its derivative) in Fig.~\ref{fig:alpha}, for three interesting values of $\abar$ (see the discussion below). It admits the following expansions near the two poles:
\begin{align}
\label{omegabarser}
	\bar{\omega}(\gamma) = 
	\begin{cases}
		{\displaystyle \frac{\abar}{\gamma} -\frac{\abar^2}{\gamma^3}} + \abar^2 \zeta_3 + \mcal{O}(\gamma^2)
		\quad &\text{when} \quad
		\gamma \to 0,
		\\*[0.4cm]
		{\displaystyle \frac{\abar}{1-\gamma}} -\abar^2 \zeta_3 + \mcal{O}\big[(1-\gamma)^2\big]
		\quad &\text{when} \quad
		\gamma \to 1.
	\end{cases}
 \end{align}
Assuming as usual that $\bar{T}_{\bx\by}$ depends only on $r$, the general solution to \eqn{nlobfkleta} reads
\begin{equation}
	\label{tbfklsol}
	\bar{T}(\eta,\rho) = 
	\int_{c - \rmi \infty}^{c + \rmi \infty} 
	\frac{\dif \gamma}{2\pi \rmi}\,
	\bar{T}_0(\gamma)
	\exp \left[ \bar{\omega}(\gamma) \eta - \gamma \rho \right],
\end{equation}
with $0<c<1$ and where we recall $\rho=\ln(1/r^2Q_0^2)$. In \eqn{tbfklsol}, $\bar{T}_0(\gamma)$ is the Mellin transform of the initial condition at $\eta=0$, that is
\begin{equation}
	\label{t0gamma}
	\bar{T}_0(\gamma) \equiv \bar{T}(\eta=0,\gamma) = 
	\int_{-\infty}^{\infty} \dif \rho\, \exp(\gamma \rho)\,	
	\bar{T}(\eta=0,\rho),
\end{equation}
but its precise form is not important for what follows. Since we are interested in the solution for $r^2 Q_0^2 \ll 1$ and hence  $\rho > 1$, one must close the integration contour on the right hemisphere. In particular this means that only the single pole at $\gamma=1$ is enclosed by the contour and there will be no large logarithms in the solution due to the cubic pole at $\gamma=0$.  This is of course in agreement with our previous study of \eqn{bfkllog}, but the present analysis will allow us to be more precise.

To that aim, we shall assume that $\eta$ and/or $\rho$ are large enough for the saddle point  method to be a good approximation.  First let us denote by $\mcal{E}(\gamma)$ the exponent in \eqn{tbfklsol}, that is 
\begin{equation}
	\label{epsilon}
	\mcal{E}(\gamma) \equiv \bar{\omega} (\gamma) \eta - \gamma \rho = 
	\left(\frac{\abar}{\gamma} + \frac{\abar}{1-\gamma} - 
	\frac{\abar^2}{\gamma^3} + \text{regular}\right) \eta-\gamma \rho, 
\end{equation}
where we displayed again the pole structure of $\bar{\omega} (\gamma)$, as it will be useful for the subsequent discussion.  A saddle point represents an extremum of this function, that is, a solution to the following equation
\begin{equation}
\label{esaddle}	
\mcal{E}'(\gamma)\equiv \bar{\omega}'(\gamma) \eta - \rho =0.
\end{equation}
For any such a solution  $\gamma_*$, the  saddle point approximation to the amplitude is obtained by expanding the exponent $\mcal{E}(\gamma)$ to quadratic order around $\gamma_*$ and performing the ensuing Gaussian integral over $\gamma$. (Within the initial condition $\bar{T}_0(\gamma)$, one can simply replace $\gamma\to \gamma_*$.) 

This method provides a sensible (stable and physically acceptable) approximation to the asymptotic amplitude at large $\eta$ and generic values of $\rho$ provided the saddle point  lies  on the real axis,  $0<\gamma_* <1$. To study the existence of such a saddle point, it is useful to consider the shape of the derivative  $\bar{\omega}'(\gamma)$ of the characteristic function, that is 
\begin{equation}
\label{omegabarprime}
	\bar{\omega}'(\gamma) = 
	 - \frac{\abar}{\gamma^2} + \frac{\abar}{(1-\gamma)^2} + 
	\frac{3 \abar^2}{\gamma^4} + \text{regular}.
\end{equation}
As visible from the plot in Fig.~\ref{fig:alpha} (right), this function has a unique minimum in the interval $0<\gamma<1$, to be denoted as $\gamma_{\rm c}$, which depends upon $\abar$. So long as the value $\bar{\omega}'(\gamma_{\rm c})$ of the function at its minimum is negative, the saddle point condition \eqref{esaddle}	admits two real solutions  for any $\rho>0$. It is the rightmost one which is physically acceptable since it continuously reduces to the LO saddle point when $\abar \to 0$. As also visible in Fig.~\ref{fig:alpha} (right), this situation occurs so long as  $\abar$ is small enough --- smaller than a critical value $\abar^{\rm cr}$ for which  $\bar{\omega}'(\gamma_{\rm c})=0$. Hence, $\abar^{\rm cr}$ and the associated value $\gamma_{\rm cr}=\gamma_{\rm c}(\abar^{\rm cr})$ of  $\gamma_{\rm c}$ are simultaneously determined by the conditions
\begin{equation}
	\label{abarcrit}
	\bar{\omega}'(\gamma)=0\quad\&\quad\bar{\omega}''(\gamma)=0\ \ \Longrightarrow \ \
	\abar^{\rm cr} \simeq 0.032.
\end{equation}
This critical value $\abar^{\rm cr}$ turns out to be extremely  small\footnote{This value will of course change after including the remaining  NLO corrections, i.e.~the  ``regular'' terms of $\mcal{O}(\abar^2)$ in \eqn{nlobfkleta}.  However, it is very unlikely that it will change a lot, since these regular terms (modulo the DGLAP corrections) do not vary much in the interval $[0,1]$ and thus their first and second derivatives, which control the value of $\abar^{\rm cr}$, cannot give a substantial contribution.}, so for all the physically relevant values of $\abar$ we are in the opposite situation, where the function $\bar{\omega}'(\gamma)$ is positive at its minimum, $\bar{\omega}'(\gamma_{\rm c}) > 0$. In that case, the saddle point condition \eqref{esaddle} admits real solutions only when $\rho$ is larger than the minimum value of the function
$\bar{\omega}'(\gamma) \eta$. This requires  $\rho > \hat{\rho}(\eta)$, where 
\begin{equation}
 \label{rhohatas}
 \hat{\rho}(\eta)\equiv \bar{\omega}'(\gamma_{\rm c}) \eta
 \quad \text{with} \quad \bar{\omega}''(\gamma_{\rm c}) = 0 \quad\&\quad \abar > \abar^{\rm cr}.
\end{equation}
(More precisely, this is just the leading term in the asymptotic expansion of $\hat{\rho}(\eta)$ for large $\eta$; subleading corrections will be computed later.) Specifically, so long as $\rho > \hat{\rho}(\eta)$, there are two real saddle points and the physically acceptable one is the largest one. But for $\rho < \hat{\rho}(\eta)$, \eqn{esaddle} has only complex-valued solutions and the corresponding approximation to the amplitude will develop an oscillating behavior, both as a function of $\eta$ and as a function of $\rho$.

To gain more insight in this behavior, let us display here some numerical values corresponding to $\abar=0.2$, which is not far from the typical value for the physical problem at hand:  solving $\bar{\omega}''(\gamma_\rmc)=0$ for  $\abar=0.2$, we find $\gamma_\rmc \simeq 0.576$ and $\bar{\omega}'(\gamma_\rmc) \simeq 1.57 \simeq 7.85 \abar$, which leads to an extremely fast growth of the scale $\hat{\rho}(\eta)$. Indeed, such an intercept is significantly larger than the intercept $\lambda_0 \simeq 4.88 \abar$ controlling the LO growth of the saturation momentum.  This argument also suggests that this instability cannot be cured by saturation (although it is alleviated by it, as we shall see): for large enough $\eta$, the region of instability extends in the region of linear evolution at $\rho > \lambda_s\eta$.

Although the saddle point approximation gives a meaningful (positive and monotonous)  result for $\bar{T}(\eta,\rho) $ for large enough values of $\rho$, the fact that this result shows oscillations (and in particular it takes negative values) at $\rho < \hat{\rho}(\eta)$ means that this solution is not physically acceptable. In other terms, the NLO BFKL evolution in $\eta$ turns out to be unstable for the physically interesting values of the (fixed) coupling $\abar$. Even though we have used the saddle point approximation, this conclusion remains true for the exact solution, as we have checked via the numerical calculation of the Mellin transform in \eqn{tbfklsol}. 

\begin{figure}[t]
\centerline{\hspace*{-.3cm}
  \includegraphics[width=0.5\textwidth]{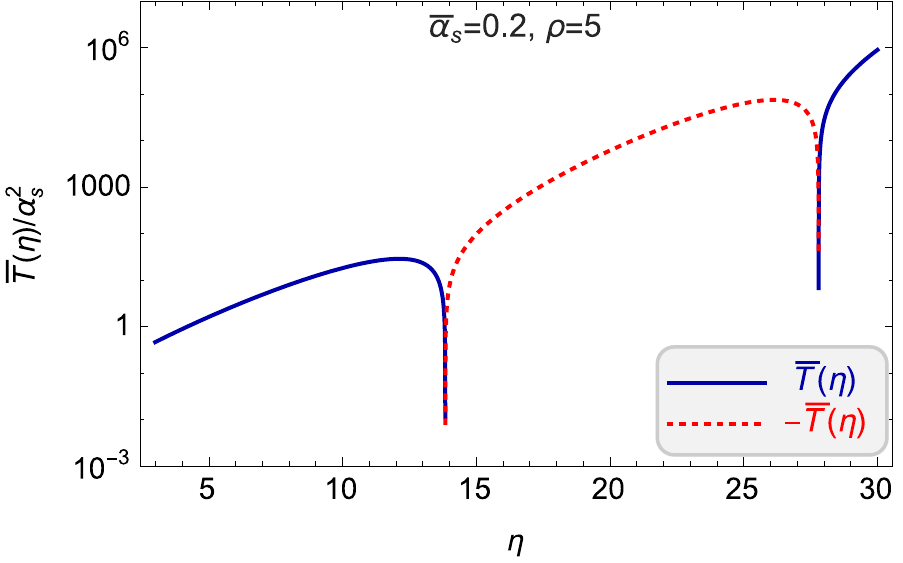}\qquad
\includegraphics[width=0.5\textwidth]{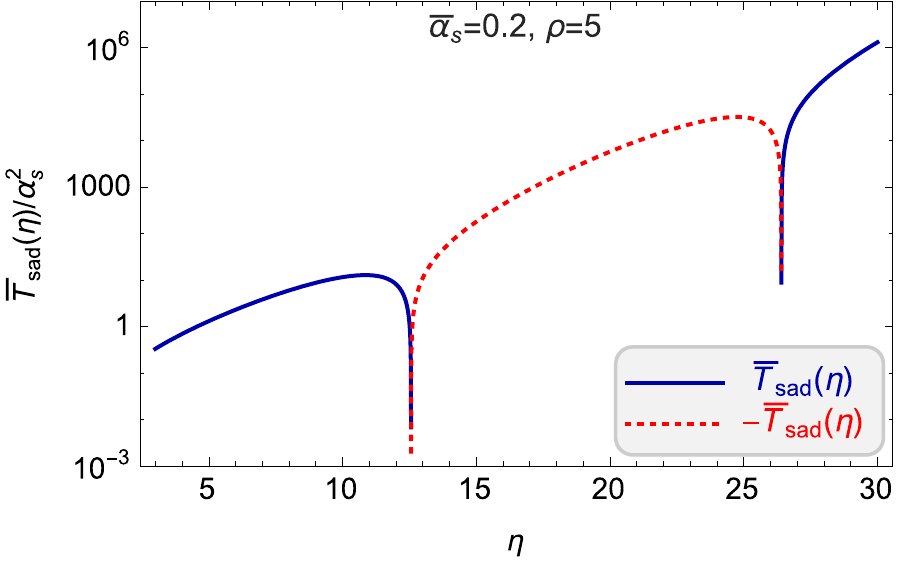}}
\caption{\label{fig:pomsol} \small The amplitude (divided by $\alpha_s^2$) as a function of the target rapidity $\eta$ and for fixed $\rho>0$, as determined from Eqs.~\eqref{tbfklsol} and \eqref{epsilon}. Left: Exact numerical solution. Right: Saddle-point solution corresponding to Eq.~\eqref{app:tsaddlep}.}
 \label{fig:logT} \end{figure}

\begin{figure}[t]
\centerline{\hspace*{-.3cm}
  \includegraphics[width=0.5\textwidth]{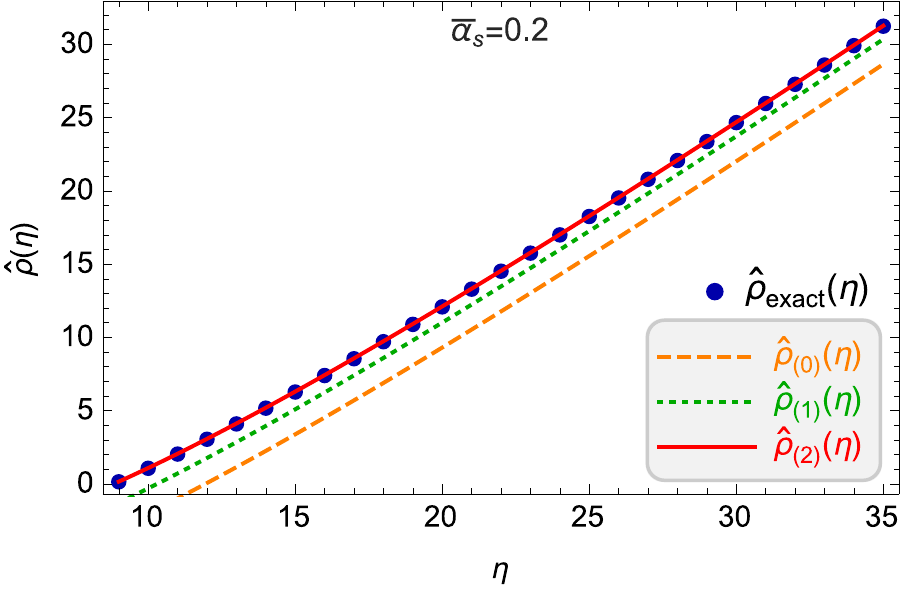}\qquad
\includegraphics[width=0.5\textwidth]{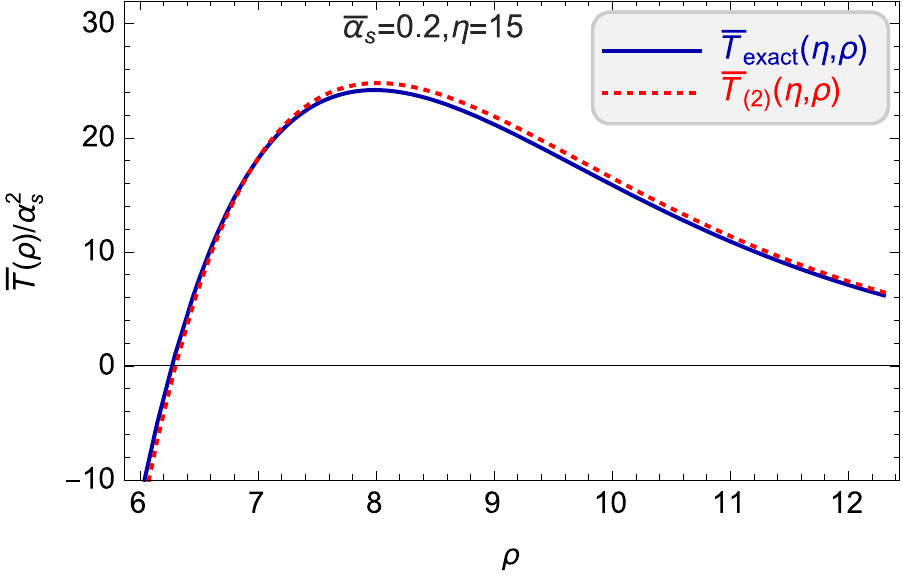}}
 \caption{\small Left: The growth of $\hat{\rho}(\eta)$. The index in the parenthesis refers to the number of included preasympotic terms when expanding in inverse powers of $\eta^{1/3}$, cf.~Eq.~\eqref{app:rhohatnnlo}. Right: The amplitude (divided by $\alpha_s^2$) as a function of $\rho$, in the vicinity of $\hat{\rho}(\eta)$ and for fixed target rapidity $\eta$. Numerically (denoted by exact) determined from Eqs.~\eqref{tbfklsol} and \eqref{epsilon} while the approximate analytical expression is given in Eq.~\eqref{app:tbfklsol}.}
  \label{fig:hatrho} \end{figure}

First, we show in Fig.~\ref{fig:logT} the logarithm of the modulus of $\bar{T}(\eta,\rho) $ as a function of $\eta$ for fixed $\rho=5$ and $\abar=0.2$, as obtained via two calculations: the numerical evaluation of  \eqn{tbfklsol} (left figure) and the saddle point approximation (right figure)\footnote{As mentioned also in Appendix \ref{app:osc}, just for the convenience of the numerics, we take into account only the pole structure of the characteristic function in Eq.~\eqref{epsilon} and for a proper comparison we use the same simplification for the saddle point solution. The same applies to the solutions shown later on in Fig.~\ref{fig:hatrho}.}. The saddle point analysis relevant for large $\eta$ and fixed $\rho$ is given in full detail in Appendix \ref{app:osc}. The two sets of results are qualitatively similar: they both show spikes corresponding to the zeros of  $\bar{T}(\eta,\rho) $, where the amplitude changes from positive values (represented in blue) to negative ones (represented in red). The precise positions of these spikes as well as the absolute values of the amplitude are somewhat different in the calculations.

Second, when $\rho$ is in the vicinity of $\hat{\rho}(\eta)$ (which means that $\rho$ is not fixed any more as it grows with $\eta$), it is useful to expand the characteristic function $\bar{\omega}(\gamma)$ around its inflexion point at $\gamma_\rmc$: by truncating this series, one obtains an approximation to the exponent $\mcal{E}(\gamma)$. The general expansion will be discussed in  Appendix \ref{app:airy}. Here, we shall restrict ourselves to the third order in this expansion (the second order term vanishes by definition), i.e. 
\begin{equation}
	\label{omegaexpand}
	\bar{\omega}(\gamma) = \bar{\omega}(\gamma_{\rm c}) +
	(\gamma - \gamma_{\rm c}) \bar{\omega}'(\gamma_{\rm c}) 
	 + \frac{1}{6}\, 
	 (\gamma - \gamma_{\rm c})^3 \bar{\omega}'''(\gamma_{\rm c}) 
	 + \cdots.
\end{equation}
In this approximation, the exponent in \eqn{tbfklsol} becomes
\begin{equation}
	\label{eapprox}
	\mcal{E}(\gamma) \simeq 
	\bar{\omega}(\gamma_{\rm c})\eta - \gamma_\rmc \rho 
	-(\gamma - \gamma_\rmc) 
	[\rho - \bar{\omega}'(\gamma_\rmc) \eta]
	+\frac{1}{6}\, 
	(\gamma - \gamma_\rmc)^3 \bar{\omega}'''(\gamma_\rmc) \eta.
\end{equation}
At this point, it is convenient to define the diffusion constant
\begin{equation}
	\label{dc}
	D_\rmc \equiv \left[\frac{2}{\bar{\omega}'''(\gamma_\rmc)} \right]^{1/3},
\end{equation}
and to change the integration variable from $\gamma$ to $t$, according to 
 \begin{equation}
 	\label{tdef}
 	\gamma - \gamma_{\rmc} \equiv \frac{D_\rmc t}{\eta^{1/3}}.
 \end{equation}
Then the amplitude in \eqn{tbfklsol} becomes
\begin{equation}
\label{tbfkl1}
	\bar{T}(\eta,\rho) = 
	\frac{\bar{T}_0(\gamma_c) D_\rmc}{\eta^{1/3}}\,
	\exp \left[ \bar{\omega}(\gamma_{\rm c})\eta - \gamma_\rmc \rho \right]
	\int \frac{\dif t}{2 \pi \rmi}\,
	\exp \bigg[- \bigg\{\frac{D_\rmc [\rho - \bar{\omega}'(\gamma_\rmc) \eta]}{\eta^{1/3}} \bigg\} t + \frac{t^3}{3} \bigg].
\end{equation}
The above integration can be performed exactly since it is recognized as a representation of the Airy function, namely
\begin{equation}
\label{tbfklai}
	\bar{T}(\eta,\rho) = 
	\frac{\bar{T}_0(\gamma_c) D_\rmc}{\eta^{1/3}}\,
	\exp \left[ \bar{\omega}(\gamma_{\rm c})\eta - \gamma_\rmc \rho \right]
	{\rm Ai}\left( \frac{D_\rmc [\rho - \bar{\omega}'(\gamma_\rmc) \eta]}{\eta^{1/3}} \right).
\end{equation} 
This expression for the amplitude allows us to deduce a better estimate for  $\hat{\rho}(\eta)$ as compared to the one in Eq.~\eqref{rhohatas}: this is conveniently defined as the largest value of $\rho$ at which the amplitude vanishes. By inspection of \eqn{tbfklai}, it is clear that to the present accuracy $\hat{\rho}(\eta)$ is determined by the rightmost zero of the Airy function, that is $a_1 = -2.338 \dots$; this yields
\begin{equation}
	\label{rhohat}
	\hat{\rho}(\eta) = \bar{\omega}'(\gamma_\rmc) \eta + 
	\frac{a_1}{D_\rmc}\, \eta^{1/3}.
\end{equation}
For  $\rho$ sufficiently close to $\hat{\rho}(\eta)$,  \eqn{tbfklai} can be further simplified:
defining the deviation $\xi \equiv\rho - \hat{\rho}(\eta)$, then for $\xi \ll \eta^{1/3}$ one can expand the Airy function as follows
\begin{equation}
	{\rm Ai}\left( \frac{D_\rmc [\rho - \bar{\omega}'(\gamma_\rmc) \eta]}{\eta^{1/3}} \right) = 
	{\rm Ai}\left( a_1 + \frac{D_\rmc \xi}{\eta^{1/3}} \right)
	\simeq \frac{{\rm Ai}'(a_1) D_\rmc \xi}{\eta^{1/3}},
\end{equation}
so that we are finally led to
\begin{equation}
	\label{tbfklfinal}
	\bar{T}(\eta,\rho) = 
	\frac{\bar{T}_0(\gamma_c) D_{\rmc}^2 {\rm Ai}'(a_1) \xi}{\eta^{2/3}}\,
	\exp \left[ \bar{\omega}(\gamma_{\rm c})\eta - \gamma_\rmc \hat{\rho}(\eta) - \gamma_\rmc \xi \right].
\end{equation}
At this point, it is interesting to notice that the appearance of the Airy function in solutions to the BFKL dynamics is quite unusual in the context of a {\it fixed} coupling --- in general, it rather appears when the coupling is running. Moreover, the $\eta$-dependence of the saturation momentum in BK evolution with running coupling is identical to the one of $\hat{\rho}(\eta)$ if in the latter we let $\eta \to \sqrt{\eta}$.

The results in  Eqs.~\eqref{rhohat} and \eqref{tbfklfinal} provide a rather good approximation for the asymptotic value of the scale  $\hat{\rho}(\eta)$, but they are still not accurate enough to be compared to the numerical results for the amplitude. For the purpose of such a comparison, in Appendix \ref{app:airy} we calculate two successive preasymptotic corrections to Eqs.~\eqref{rhohat} and \eqref{tbfklfinal}, with the results exhibited in Fig.~\ref{fig:hatrho}: the curves denoted as $\hat{\rho}_{(0)}(\eta)$,  $\hat{\rho}_{(1)}(\eta)$ and  $\hat{\rho}_{(2)}(\eta)$ in Fig.~\ref{fig:hatrho} (left) refer respectively to the asymptotic expression in \eqref{rhohat} and the two successive improvements of it as obtained by adding only one or both of the subasymptotic corrections shown in \eqn{app:rhohatnnlo}. Similarly, in Fig.~\ref{fig:hatrho} (right) we compare the exact results for $\bar{T}(\eta,\rho) $ (as a function of $\rho$ for $\eta=15$ and $\abar=0.2$) with our analytic approximation shown  in \eqn{app:tbfklsol}. On both figures, we observe an excellent agreement between the exact results and our best analytic estimates.

We conclude this section with a model-dependent numerical study of the instability in the non-linear evolution in $\eta$ at NLO. This is ``model-dependent'' because we shall not use the full NLO BK equation \eqref{nlobketa}, but only a drastic approximation to it in which, on top of the LO BK terms, we shall keep only a particular NLO correction to the BK kernel which is enhanced by a double-collinear logarithm. Specifically, we shall use the following equation
\begin{equation}
	\label{collbkexp}
\frac{\partial \bar{S}_{\bx\by}(\eta)}{\partial \eta} = 
 \frac{\abar}{2 \pi}
 \int
 \frac{\dif^2 \bz \,(\bx\minus\by)^2}{(\bx \minus\bz)^2 (\bz \minus \by)^2}\,\left[1-\frac{\abar}{2}
	\ln^2 \frac{(\bx-\bz)^2}{(\bz-\by)^2}\right]
 \left[\bar{S}_{\bx\bz}(\eta) \bar{S}_{\bz\by}(\eta) - \bar{S}_{\bx\by}(\eta) \right],
\end{equation}
which in fact represents the expansion to NLO of an equation to be motivated later on, in Sect.~\ref{sec:localeta} (see \eqn{collbk0}). The linearized version of this equation properly includes the double-collinear logarithms in the relevant limit, that is, when one of the daughter dipoles is much smaller than the other one. This may not be obvious when comparing with \eqn{nlobfkleta}: in the latter, the double-collinear logs are important only when they multiply the scattering amplitude of the smallest dipole; for instance, when  $|\bx-\bz|\ll |\by-\bz|\simeq r$, then there is a large double-collinear log only for the ``real'' term proportional to $\bar{T}_{\bx\bz}$ in the second line of  \eqn{nlobfkleta}. On the contrary, in \eqn{collbkexp}, the NLO term from the square brackets {\it a priori} multiplies all the $S$-matrices from the integrand. Yet, after linearization and in the collinear limit where, say, $|\bx-\bz|\to 0$, the contribution to scattering due to the large daughter dipole cancels against the virtual term, $\bar{T}_{\bz\by} - \bar{T}_{\bx\by} \to 0$,  so we are only left with a contribution from the small daughter dipole, multiplied by the same double-collinear log as in \eqn{nlobfkleta}.

\begin{figure}[t]
\centerline{\hspace*{-.3cm}
  \includegraphics[width=0.46\textwidth]{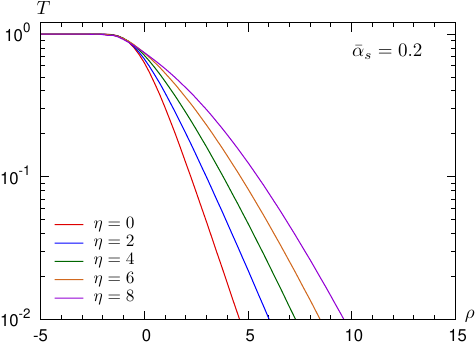}\qquad
\includegraphics[width=0.46\textwidth]{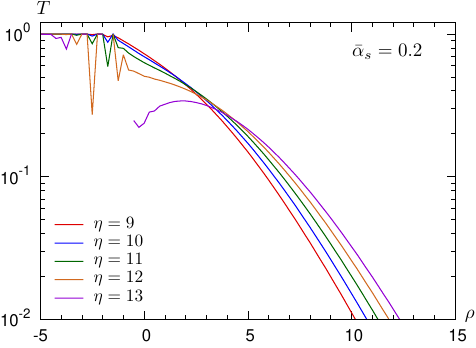}}
\caption{\small Left: The front for small values $\eta$ as obtained by the numerical solution to Eq.~\eqref{collbkexp}. As in the case of BFKL evolution, the solution seems to be physically meaningful in this rapidity regime. Right: The front for larger values of $\eta$ as obtained from the same equation. Again, as in the BFKL case, an instability is formed and renders the solution unphysical.}
 \label{CollEtaExp}
\end{figure}
 
This confirms that \eqn{collbkexp} is indeed correct in the collinear limit and in the weak scattering regime. In Sect.~\ref{sec:localeta}  we shall argue that this equation is not right anymore in the approach towards saturation, so it is not a reliable approximation to the actual non-linear dynamics in $\eta$ at NLO. Yet, by lack of a better equation which is tractable, we have numerically solved  \eqn{collbkexp} and searched for potential instabilities. The results for the saturation fronts, as displayed in Fig.~\ref{CollEtaExp}, show indeed a clear instability, which needs a few units of rapidity (about 8 units for $\abar=0.2$, while it would be $5\div 6$ units for $\abar =0.3$) in order to develop. Interestingly, the instability first manifests itself in the vicinity of saturation, i.e.~in the region where $\bar T\simeq 1$. Even though, as just mentioned, \eqn{collbkexp}  is not really trustable in that region, we believe that this instability is a real feature of the NLO BK equation in $\eta$ and moreover it should indeed be triggered by configurations involving relatively large dipoles, near saturation, as suggested by the numerical results  in Fig.~\ref{CollEtaExp}. Indeed, this is the region in phase-space where the (effective) anomalous dimension $\gamma $ of the BK solution becomes very small (recall the discussion after \eqn{bfkllog2}). A further argument towards instability in the  NLO  BK evolution in $\eta$ will be presented in Sect.~\ref{sec:lt}.

\section{Non-local BK evolution in $\eta$}
\label{sec:shift}

In the previous section, we have seen that the BK evolution in $\eta$ also receives NLO corrections enhanced by a double collinear logarithm which lead to instabilities --- albeit not as severe as for the corresponding problem in $Y$. We have not yet discussed the physical origin of these corrections (this will be done shortly), but given the experience with the $Y$-evolution,  one may anticipate that similar corrections will appear in higher orders and that they need to be resummed in order to obtain a stable evolution. This will lead us to a resummation scheme in $\eta$ which is formally similar to the one developed in Sect.~\ref{sec:BKnlY} for the evolution in $Y$ --- that is, a version of the BK equation which is non-local in rapidity  --- but which differs from the latter in that it is significantly less sensitive to changes in the resummation prescription. 

As we now explain, the double collinear logarithms in the evolution with $\eta$ are related to an ordering problem as well --- not to time-ordering (which is satisfied by construction in this framework), but rather to the ordering of the successive emissions in their longitudinal momentum $k^+$.  Indeed, the physical constraints on the high-energy evolution of the right-moving projectile are the same in both representations ($Y$ and $\eta$): the successive gluon emissions must be simultaneously ordered in longitudinal momenta and in lifetimes. These constraints, that were originally written for the evolution with $Y$ (or $k^+$) in \eqn{TO1}, must be now rewritten for the evolution with $\eta$ (or $\tau_k$); using $\eta=\ln(\tau_k/\tau_0)$ with $\tau_k=2k^+/\kt^2$, one easily finds
\beq\label{TO2}
\tau_q Q^2\gg \tau_k \kt^2\gg \tau_0 Q^2_0\,,\qquad
\tau_q \gg \tau_k \gg \tau_0\,.\eeq
The second condition above is automatically satisfied by the evolution with $\eta$; but the first condition, which represents the ordering in longitudinal momenta, might be violated when the emitted gluon is either too hard ($\kt^2 \gg Q^2$), or too soft ($\kt^2 \ll Q^2_0$). This argument sheds light on the typical configurations which violate these constraints and hence lead to instabilities (in the context of the linear, BFKL, evolution, at least): these are evolutions in which one first creates a very large daughter dipole, with  size $r\gg 1/Q_0$, which then evolves by radiating much smaller dipoles. This also explains why the non-linear physics of saturation reduces this particular instability: saturation suppresses the large dipole with sizes $r\gg 1/Q_s(Y)\ge 1/Q_0$ and thus the phase-space for soft-to-hard evolution decreases.

The integral version of the BK equation obeying the constraints in \eqn{TO2} can now be constructed along the same lines as the corresponding equation for the evolution in $Y$, that is,  \eqn{BKTO} . First, we notice that the inequalities in  \eqn{TO2} imply the following integration range in the rapidity $\eta_1\equiv
\ln(\tau_k/\tau_0)$ of the fluctuation:
\beq
{\rm min}\left\{\eta, \eta-\ln\frac{\kt^2}{Q^2}\right\} \, >\,\eta_1\,>\,
{\rm max}\left\{ 0, \ln\frac{Q_0^2}{\kt^2}\right\},
\eeq
where we kept the notation $\eta\equiv \ln(\tau_q/\tau_0)$ for the rapidity of the incoming dipole (i.e.~the overall range for the evolution with $\eta_1$). Moving to transverse coordinates as appropriate for  the dipole picture and recalling the relation $\kt^2= 1/r_<^2$ with $r_<$ the size of the smallest daughter dipole, cf.  \eqn{ktr<}, we deduce the following properly-ordered version of the BK equation in $\eta$:
\begin{align}
	\label{BKeta}
	 \bar S_{\bx\by}(\eta)= S^{(0)}_{\bx\by}+
	\frac{\abar}{2\pi}
	\int \frac{\dif^2 \bz \,(\bx\minus\by)^2}{(\bx \minus\bz)^2 (\bz \minus \by)^2}\int\limits_
	{\Theta(- \rho_1) |\rho_1|}^{\eta-\Theta(\rho_1- \rho)(\rho_1- \rho)}\dif \eta_1
	\big[ \bar S_{\bx\bz}(\eta_1)
	 \bar S_{\bz\by}(\eta_1) \minus  \bar S_{\bx\by}(\eta_1) \big],
\end{align}
where   $\rho_1=\ln(1/r_<^2Q_0^2)$ and $S^{(0)}_{\bx\by}$ denotes, as before, the tree-level estimate for the amplitude. By taking a derivative w.r.t.~$\eta$, one finally obtains a differential equation which is non-local in $\eta$: 
\begin{align}
	\label{BKetadif}
	\frac{\del \bar{S}_{\bx\by}(\eta)}{\del \eta} = 
	\frac{\abar}{2\pi}
	& \int \frac{\dif^2 \bz \,(\bx\minus\by)^2}{(\bx \minus\bz)^2 (\bz \minus \by)^2}\,
	\Theta\big(\eta \minus \delta_{\bx\by\bz}\big)\,\Theta\big(\eta \minus \Theta(-\rho_1) |\rho_1|\big)
	\nonumber\\*[0.2cm]
	& \qquad \times \big[\bar{S}_{\bx\bz}(\eta \minus \delta_{\bx\by\bz})\bar{S}_{\bz\by}(\eta \minus \delta_{\bx\by\bz}) 
	\minus \bar{S}_{\bx\by}(\eta \minus \delta_{\bx\by\bz}) \big],
\end{align}
with the new rapidity shift $\delta_{\bx\by\bz}$ defined as
\begin{align}
\label{delta}
\delta_{\bx\by\bz} \equiv \Theta(\rho_1- \rho)(\rho_1- \rho) =\Theta(r -r_<)\ln\frac{r^2}{r_<^2}
= {\rm max}\left\{ 0,\ln \frac{(\bx\!-\!\by)^2} {{\rm min}
\{(\bx\!-\!\bz)^2,(\bz\!-\!\by)^2\}}\right\}.	
\end{align}
The two step-functions within the integrand in \eqn{BKetadif} express the condition that the upper limit in the integral over $\eta_1$ in \eqn{BKeta} be larger than the respective lower limit for any value of $\rho_1$.

 As expected, the rapidity shift $\delta_{\bx\by\bz}$ as well as the first step-function introduce constraints on the soft-to-hard evolution: they are effective only when the smallest daughter dipole is (much) smaller than the parent one. In particular, the step-function $\Theta(\eta \minus \delta_{\bx\by\bz})$ effectively acts as an ``ultraviolet cutoff'': it implies a lower limit on $r_<$, namely $r_<^2 \ge r^2\rme^{-\eta}$. (But the ultraviolet divergences also cancel in the standard way, between the ``real'' and ``virtual'' pieces within the square brackets.) The second step-function $\Theta\big(\eta \minus \Theta(-\rho_1) |\rho_1|\big)$ forbids the emission of {\it very} large dipoles, with sizes $r_< \ge \rme^{\eta}/Q_0^2 \gg 1/Q_0^2$. This step function is rather superfluous in the context of the BK equation  \eqref{BKetadif}, where large dipoles are already cut off at the shorter scale $1/Q_s(\eta)$ introduced by saturation. But it might be important in relation with the linearized version of  \eqn{BKetadif}, i.e.~the non-local BFKL equation.

\eqn{BKetadif} represents the counterpart of \eqn{BKTOdiff} for the evolution in $\eta$. Unlike the latter, this is not a boundary-value problem anymore, but an initial-value problem with initial condition $\bar S_{\bx\by}(\eta=0)=S^{(0)}_{\bx\by}$.  Indeed, due to the step-function inside the integrand, the r.h.s. of this equation involves the function  $\bar S_{\bx\by}(\eta)$ only for positive values $\eta>0$ of the rapidity. This being said, the initial-value formulation is complicated by the non-locality in $\eta$: strictly speaking, \eqn{BKetadif} is  a {\it delay} differential equation, due to the shift in the  rapidity arguments and to the step-functions which limit the evolution at small $\eta$. As we shall explain in Sect.~\ref{sec:IC}, some care must be taken when trying to start the evolution with an initial condition formulated at some different rapidity $\eta_0 >0$.



At this stage, the relation between the non-local equation  \eqref{BKetadif} and the NLO corrections containing the double logarithms in \eqn{nlobfkleta} is not yet obvious.  In the next section, we shall show that these corrections are indeed correctly encoded into   \eqn{BKetadif}, in the ``collinear'' limit of interest, i.e.~when one of the daughter dipoles is much smaller than the parent one. This argument also shows that the detailed structure of the equation is not unique (which should be expected, given our previous experience with the non-local evolution in $Y$): any rapidity shift which coincides with $\delta_{\bx\by\bz}$ in the collinear limit and drops to zero when none of the daughter dipoles is small is equally acceptable to the accuracy of interest.

It is one of our main points in this paper to demonstrate that the resummed evolution in $\eta$ shows relatively little scheme dependence --- considerably less than the corresponding evolution in $Y$ and at the level of the expected accuracy (given the approximations), i.e.~$\order{\abar^2}$. To that aim, we will show results obtained with two other prescriptions for the rapidity shift. (We have also considered other choices and found similar results.)  The first one is obtained by replacing the $S$-matrices in the r.h.s. of \eqn{BKetadif} by
\beq
\label{BKeta1} 
\bar{S}_{\bx\bz}(\eta \minus \delta_{\bx\by\bz})\bar{S}_{\bz\by}(\eta \minus \delta_{\bx\by\bz}) 
	\minus \bar{S}_{\bx\by}(\eta \minus \delta_{\bx\by\bz})\,\longrightarrow\,
	\bar{S}_{\bx\bz}(\eta \minus \delta_{\bx\bz; r})\bar{S}_{\bz\by}(\eta \minus \delta_{\bz\by;r}) 
	\minus \bar{S}_{\bx\by}(\eta),
\eeq	
where
\beq
\label{delta1}
\delta_{\bx\bz;r}\equiv {\rm max}\left\{ 0,\ln \frac{r^2}{(\bx\!-\!\bz)^2}\right\}
\eeq
and similarly for $ \delta_{\bz\by;r}$. There are two differences w.r.t.~\eqn{BKetadif}: \texttt{(i)} the rapidity shifts in the ``real'' $S$-matrices are different for the two daughter dipoles and \texttt{(ii)}  there is no such a shift in the ``virtual'' term. These two modifications are consistent with  \eqn{BKetadif} to the accuracy of interest and also with each other. Indeed, with the new prescription, the shift in the ``real'' $S$-matrices is important only for the smallest daughter dipole; e.g., if one has $|\bx\!-\!\bz|\ll r \simeq |\bz\!-\!\by|$, then $\delta_{\bz\by;r}\simeq 0$. Similarly, with the original prescription in \eqn{BKetadif}, the effect of the shift cancels out  (in this limit $|\bx\!-\!\bz|\ll r$) between the ``real'' term associated with the largest daughter dipole and the ``virtual'' term. 

This being said, the new prescription appearing in Eqs.~\eqref{BKeta1}--\eqref{delta1}, which distinguishes between the two daughter dipoles, appears to be favoured by the structure of the NLO corrections, as discussed after \eqn{nlobfkleta}: e.g., when $|\bx\!-\!\bz|\to 0$, the double-logarithmic  factors multiplying the scattering amplitudes $\bar{T}_{\bz\by}$ of the large daughter dipole and $\bar{T}_{\bx\by}$ of the parent dipole do {\it separately} cancel --- similarly to what happens with the rapidity shifts in \eqn{BKeta1}. 


\begin{figure}[t]
\centerline{\hspace*{-.3cm}
  \includegraphics[width=0.55\textwidth]{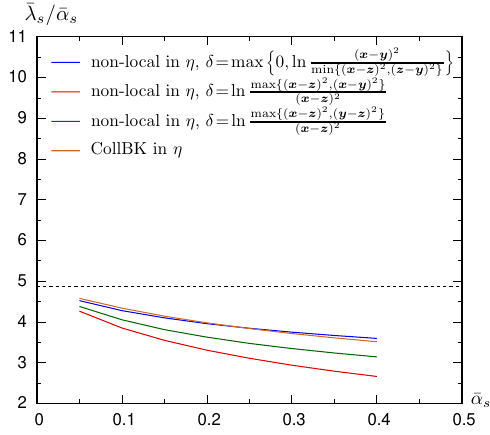}}
\caption{\small The asymptotic speed of the front (divided by $\abar$) in the $\eta$-representation as a function of $\abar$ and for different evolution schemes. All lines are physically acceptable and the correction is of order $\abar$ when compared to the LO result.}
 \label{fig:nloeta} 
 \end{figure}

The final prescription for the rapidity shift in $\eta$ that we shall explicitly consider is obtained via
\beq
\label{BKeta2} 
\bar{S}_{\bx\bz}(\eta \minus \delta_{\bx\by\bz})\bar{S}_{\bz\by}(\eta \minus \delta_{\bx\by\bz}) 
	\minus \bar{S}_{\bx\by}(\eta \minus \delta_{\bx\by\bz})\,\longrightarrow\,
	\bar{S}_{\bx\bz}\left(\eta \minus \ln\frac{r^2_>}{(\bx\!-\!\bz)^2}\right)\bar{S}_{\bz\by}\left(\eta \minus \ln\frac{r^2_>}{(\bz\!-\!\by)^2}\right) 
	\minus \bar{S}_{\bx\by}(\eta),\nn
\eeq	
where $r^2_>\equiv {\rm max}\{(\bx\!-\!\bz)^2,\,(\bz\!-\!\by)^2\}$ is the {\it largest} of the  two daughter dipoles.  Clearly, this prescription has the same collinear limits as that in  \eqn{BKeta1}, whereas for larger daughter dipoles $r_>^2\gtrsim r^2$, the rapidity shifts in \eqref{BKeta2}  are smoothly switched off. An overall step-function $\Theta\big(\eta-\ln({r_>^2}/{r_<^2})\big)$ (ensuring that the rapidity arguments of all the $S$-matrices remain positive) is now implicitly understood.

Using these three prescriptions for the rapidity shift, we have solved the non-local equation up to very high values of $\eta$ and extracted the asymptotic value of the saturation exponent $\bar{\lambda}_s $, with the results displayed in Fig.~\ref{fig:nloeta}. These results should be compared to those in Fig.~\ref{SresumY} (right) and notably to the curves denoted there as ``non-local'' (which, we recall, have been obtained by solving the non-local equation in $Y$ with two different prescriptions and then reinterpreting the results in terms of $\eta$).  Clearly, the non-local evolution in $\eta$ shows much less scheme dependence than the one in $Y$: the various curves shown  in Fig.~\ref{fig:nloeta} do all show the same trend and they stay relatively close to each other, within the limits expected in view of our approximations. E.g., for $\abar=0.3$, the difference between the most ``extreme'' predictions in Fig.~\ref{fig:nloeta}  --- those obtained with the prescriptions in \eqn{BKetadif} and \eqn{delta1}, respectively --- is $\delta\bar\lambda_s\simeq \abar\simeq 3\abar^2$, which is half the corresponding difference in  Fig.~\ref{SresumY} (right). Also, the ratio $\delta\bar\lambda_s/\bar\lambda_s$  between the variance and the average value $\bar\lambda_s\simeq 3\abar$ is  $\delta\bar\lambda_s/\bar\lambda_s\simeq 0.3\simeq \abar$, as expected for a set of calculations which should differ only at NLO. Equally important, the variance $\delta\bar\lambda_s$ grows only slowly with $\abar$ in Fig.~\ref{fig:nloeta}, whereas it grows much faster  in  Fig.~\ref{SresumY} (right). One must also add that none of the curves in Fig.~\ref{fig:nloeta} has an inflection point and that they all stay below the LO result (contrary to what was happening in Fig.~\ref{SresumY} for some of the curves).

The comparison between  Fig.~\ref{fig:nloeta} and Fig.~\ref{SresumY} (right) also reveals that the curves corresponding to the prescription  \eqref{delta1} for the evolution in $\eta$ and respectively \eqref{deltaopt}  for the evolution in $Y$ yield results which look similar by eye and in fact turn out to be {\it identical}. This is not a coincidence: one can check that, with these two particular prescriptions, the non-local evolution in $Y$ can be {\it exactly} mapped onto the corresponding evolution in $\eta$ via our standard change of rapidity variables and functions, $Y=\eta+\rho$ and ${\bar S}_{\bx\by}(\eta)= S_{\bx\by}(Y=\eta+\rho)$. 
But to some extent, the very existence of such an ``ideal'' change of variables is an accident: if one starts with any other of the prescriptions that we have used so far, e.g.~\eqn{BKetadif} or \eqref{BKeta2} for the evolution in $\eta$, and one makes the change of variables $\eta\to Y$, one finds an equation in $Y$ which has some pathologies --- typically the rapidity arguments of the various $S$-matrices can become smaller than $\rho$ in some corners of the phase-space. (Vice-versa, if one starts with a generic prescription in $Y$, the change of variables to $\eta$ produces rapidity arguments which can become negative: $\eta<0$. This is why, for moderately small values of $\abar$, Eq.~\eqref{Delta} was leading to an intercept larger than the LO one after converting the results in terms of $\eta$.). This ultimately reflects the fact that, after the resummation of the double transverse logarithms,
the non-local evolution in  $\eta$ is more stable (in the sense of less  scheme-dependent) than that in $Y$.

In  Sect.~\ref{sec:bcy},  in the context of the evolution with $Y$, we described an alternative strategy for performing resummations where the ``collinearly improved''  equation remains {\it local} in rapidity but the corresponding kernel receives double logarithmic corrections to all orders. It is interesting to study whether a similar construction is also possible for the evolution with $\eta$. This will be discussed in detail  in Sect.~\ref{sec:localeta}, where we shall see that such a local reformulation is ``almost'' possible. That is, one can indeed construct a resummed version of the BFKL kernel which includes the double collinear logs to all orders: this is in fact identical to the corresponding kernel for the evolution in $Y$, cf.  \eqn{collbk}, up to the replacement of the (double) anti-collinear logs by collinear ones (see \eqn{rhobarxyz}). But the natural non-linear completion of this equation, which is shown in \eqn{collbk0} and has the same structure as \eqn{collbk}, does not correctly describe the approach towards saturation (the Levin-Tuchin law) --- as we shall check by comparing with the respective behavior of the  non-local equation \eqref{BKetadif}. 

This being said, the fact that this local equation is correct in the BFKL regime together with the ``pulled'' nature of the saturation front guarantees that this equation is in fact
appropriate for most purposes: it correctly predicts the shape and speed of the saturation front at all points except very close to the unitarity limit $\bar{T}=1$. To check that, we have also plotted in Fig.~\ref{fig:nloeta} the respective prediction for the asymptotic saturation exponent (denoted as ``collBK''); as one can see, this is indeed close to the results obtained from the non-local equation and, perhaps accidentally, almost identical to the corresponding prediction of the original prescription for the rapidity shift in \eqn{BKetadif}. This stability should be contrasted to what happens for the resummed evolution in $Y$, where the ``collBK'' curve in Fig.~\ref{SresumY} (right) rapidly deviates from all the other predictions when increasing $\abar$.

Another drawback of the local resummation as compared to the non-local one is the fact that for the former, unlike for the latter, we do not know how to systematically include all the other NLO corrections, i.e.~the ``regular'' terms of $\order{\abar^2}$ in \eqn{nlobfkleta}. This will be further discussed in the next section.

\section{Matching to NLO BK in $\eta$}
\label{sec:match}

Our purpose in this section is twofold. First, we will explicitly check that the non-local version of the BK equation in $\eta$, as constructed in the previous section, contains indeed the NLO corrections enhanced by double-collinear logarithms (as visible in the second line in the r.h.s. of \eqn{nlobfkleta}). Second, we shall extend this non-local equation to full NLO accuracy, by matching onto the NLO BK equation given in \eqref{nlobketa}. As a result, we shall obtain an equation which is non-local in $\eta$, which is perturbatively correct up to $\order{\abar^2}$ and which performs an all-order resummation of the kernel corrections enhanced by double-collinear logarithms, while missing ``regular'' terms starting with $\order{\abar^3}$.  The subsequent discussion applies to the two prescriptions for the rapidity shift shown in Eqs.~\eqref{BKeta1}-\eqref{delta1}  and  respectively \eqref{BKeta2}, which share with the full NLO result the fact that  the rapidity shift is non-zero only for the scattering amplitude associated with the smallest daughter dipole. For definiteness, we shall write our formulae for the case of the prescription in Eqs.~\eqref{BKeta1}-\eqref{delta1}, to be referred to as the ``canonical'' one.  Thus,  our starting point is the following equation
\begin{align}
        \label{bketato}
        \frac{\del \bar{S}_{\bx\by}(\eta)}{\del \eta} = 
        \frac{\abar}{2\pi}
        \int \frac{\dif^2 \bz \,(\bx\minus\by)^2}{(\bx \minus\bz)^2 (\bz \minus \by)^2}\,
        \Theta\big(\eta \minus \delta_{\bx\by\bz}\big)
        \big[\bar{S}_{\bx\bz}(\eta \minus \delta_{\bx\bz;r})\bar{S}_{\bz\by}(\eta \minus \delta_{\bz\by;r}) 
        \minus \bar{S}_{\bx\by}(\eta) \big].
\end{align}
 We first identify the $\mcal{O}(\abar^2)$ piece in the r.h.s.~of \eqn{bketato}. Similar to what we have done in \eqn{shiftexpand}, we start by expanding the real terms to linear order in the shift and then use the leading-order BK equation in $\eta$ to deduce, e.g.
\begin{align}
        \label{shiftexpand2}
        \bar{S}_{\bx\bz} \left(\eta - \delta_{\bx\bz;r}\right)
        \simeq\,&
        \bar{S}_{\bx\bz}(\eta)
        - \delta_{\bx\bz;r}
        \frac{\partial \bar{S}_{\bx\bz}(\eta)}{\partial \eta}
        \nn*[0.2cm]
        \simeq\, &
        \bar{S}_{\bx\bz}(\eta)
        - 
        \frac{\abar}{2\pi}
        \int
        \frac{ \dif^2 \bu\,(\bx\minus\bz)^2}{(\bx \minus\bu)^2 (\bu \minus \bz)^2}\, 
        \delta_{\bx\bz;r}
        \big[\bar{S}_{\bx\bu}(\eta) \bar{S}_{\bu\bz}(\eta) - \bar{S}_{\bx\bz}(\eta) \big],
\end{align} 
with an analogous expression for $\bar{S}_{\bz\by}$. The step function in \eqn{bketato} does not play any role in perturbation theory for $\eta>0$, so it will be replaced by unity. Then, after some convenient relabelling of the integration variables, we find that the $\mcal{O}(\abar^2)$ contribution contained in the r.h.s.~of \eqn{bketato} reads
\begin{align}
\label{a2shift}
-\frac{\abar^2}{2\pi^2}
 \int \frac{\dif^2 \bz\, \dif^2 \bu\, (\bx \minus \by)^2 }{(\bx \minus \bu)^2 (\bu \minus \bz)^2 (\bz \minus \by)^2}\, 
\delta_{\bu\by;r}\, \bar{S}_{\bx\bu}(\eta) 
 \big[\bar{S}_{\bu\bz}(\eta) \bar{S}_{\bz\by}(\eta) - \bar{S}_{\bu\by}(\eta) \big].
 \end{align}
In the second step we subtract this $\mcal{O}(\abar^2)$-piece from \eqn{bketato}, while in the third step we add all the NLO terms of the BK equation in \eqref{nlobketa} to finally get 
\begin{align}
 \label{nlobkres}
 \hspace*{-0.9cm}
 \frac{\del \bar{S}_{\bx\by}(\eta)}{\del \eta} = &\, 
        \frac{\abar}{2\pi}
        \int \frac{\dif^2 \bz \,(\bx\minus\by)^2}{(\bx \minus\bz)^2 (\bz \minus \by)^2}\,
        \Theta\big(\eta \minus \delta_{\bx\by\bz}\big)
        \big[\bar{S}_{\bx\bz}(\eta \minus \delta_{\bx\bz;r})\bar{S}_{\bz\by}(\eta \minus \delta_{\bz\by;r}) 
        \minus \bar{S}_{\bx\by}(\eta) \big]
 \nn*[0.2cm]
 &-\frac{\abar^2}{4\pi}
 \int 
 \frac{\dif^2 \bz \,(\bx\minus\by)^2}{(\bx \minus\bz)^2 (\bz \minus \by)^2}\,
 \ln \frac{(\bx \minus\bz)^2}{(\bx \minus\by)^2} \ln \frac{(\by \minus\bz)^2}{(\bx \minus\by)^2}
 \left[\bar{S}_{\bx\bz}(\eta) \bar{S}_{\bz\by}(\eta) - \bar{S}_{\bx\by}(\eta) \right]
 \nn*[0.2cm]
 &+ \frac{\abar^2}{2\pi^2}
 \int \frac{\dif^2 \bz\, \dif^2 \bu\, (\bx \minus \by)^2 }{(\bx \minus \bu)^2 (\bu \minus \bz)^2 (\bz \minus \by)^2} 
\bigg[\ln\frac{(\bu \minus\by)^2}{(\bx \minus \by)^2} + \delta_{\bu\by;r} \bigg]
  \bar{S}_{\bx\bu}(\eta) 
 \left[\bar{S}_{\bu\bz}(\eta) \bar{S}_{\bz\by}(\eta) - \bar{S}_{\bu\by}(\eta) \right] 
 \nn*[0.2cm]
 &+\,\abar^2 \times \text{``regular''},
 \end{align} 
which is one of the main results in the current work. Notice that we have combined the term generated by the rapidity shift in \eqn{a2shift} with the third term in \eqn{nlobketa} respectively, and that the sum in the square bracket is just the $Y$-shift $\Delta_{\bu\by;r}$ defined in Eq.~\eqref{deltaopt}. By construction, \eqn{nlobkres} reduces to the full NLO BK equation in $\eta$ when expanded to order $\abar^2$, meaning that the associated error is now of $\order{\abar^3}$. The structure of the equation and the associated error would remain unchanged if one were to use the prescription \eqref{BKeta2} for the rapidity shift  instead of the ``canonical'' one in Eqs.~\eqref{BKeta1}-\eqref{delta1}.

From the discussion after  \eqn{nlobketa}, we already know that the double anti-collinear logarithm visible in the second term in \eqn{nlobkres} is cancelled in the relevant regime (i.e.~for {\it large} daughter dipoles) by a corresponding piece generated by the integral over $\bu$ in the third term. Moreover, given the definition of the shift in Eq.~\eqref{delta1}, it should be clear that the third term in  \eqn{nlobkres} is also free of double {\it collinear} logs (i.e.~those appearing for {\it small} daughter dipoles).  Thus, taken together, the two $\mcal{O}(\abar^2)$-terms which are explicit in the r.h.s. of \eqn{nlobkres} do not contain any large contributions that may cause instabilities. However, one should be careful when numerically integrating this equation, because one expects large cancellations between these two terms in the anti-collinear regime\footnote{We also point out that these two terms are structurally equivalent to the respective terms appearing in \cite{Beuf:2014uia} where one is evolving in $Y$. This is as expected, since ``regular terms'' of order $\abar^2$ have the same functional form in $Y$- and $\eta$-representation. The main difference when compared to the result in \cite{Beuf:2014uia}, as extensively discussed earlier, is in the first term in Eq.~\eqref{nlobkres} and in the fact that it represents an evolution in $\eta$, which is an initial value problem and with a resummation which is under control.}. 

At this point, one may observe that the previous discussion implicitly demonstrates the other point that we would like to make in this section, namely the fact that the non-local equation \eqref{bketato} properly includes the NLO corrections enhanced by double-collinear logarithms. Indeed, we have just shown that \texttt{(i)} Eqs.~\eqref{nlobkres} and  \eqref{nlobketa} are equivalent to $\mcal{O}(\abar^2)$, and \texttt{(ii)}
the NLO corrections which are  explicit in the r.h.s. of  \eqn{nlobkres} do not include  double-collinear logs anymore. It follows that these double logs are fully included in the first term in the r.h.s.  of  \eqn{nlobkres}, which is the non-local BK equation in $\eta$. For completeness, in Appendix~\ref{app:local}
we shall also verify this via an explicit calculation starting with \eqn{a2shift}.

\comment{\section{Speed and slope of the front in $\eta$}
\label{sec:front}

\section{Starting the evolution with shifted arguments}
\label{sec:theta}
}

\section{The effective characteristic function}
\label{sec:BFKLeta}

Although it is well defined and poses no special problems for numerical studies, the equation \eqref{bketato} is quite special in that it is non-local in the evolution ``time'' $\eta$. So, it is perhaps less obvious how to adapt to this equation methods that were traditionally used in the past for the analytic studies of the BFKL or the BK equations. In this section, we shall nevertheless show that this equation has an underlying mathematical structure which is not too far away from that of the more familiar equations aforementioned, which are local in $\eta$. Its linearized version can be studied in Mellin space, similarly to the BFKL equation; this analysis, to be presented in  Sect.~\ref{sec:chi},  features a special characteristic function, which has already appeared in the original studies of the collinear resumations of the NLO BFKL equation  \cite{Salam:1998tj,Ciafaloni:1998iv,Ciafaloni:1999yw,Ciafaloni:2003rd}. Moreover, to the accuracy of interest (i.e.~in so far as the resummation of the double collinear logs is concerned) and with some caveat deep at saturation to be discussed, this non-local equation can also be rewritten in a local form, as already mentioned at the end of Sect.~\ref{sec:shift}; this local version will be constructed in Sect.~\ref{sec:localeta} below.

\subsection{Asymptotic eigenvalue branch}
\label{sec:chi}

We start with the ``canonical'' form of the non-local evolution equation in $\eta$, as given in Eq.~\eqref{bketato}. Linearizing and using the symmetry under $\bx\minus\bz \leftrightarrow \bz\minus\by$, we obtain
\begin{align}
	\label{lineareta}
	\frac{\del \bar{T}_{\bx\by}(\eta)}{\del \eta} = 
	\frac{\abar}{2\pi}
	\int \frac{\dif^2 \bz \,(\bx\minus\by)^2}{(\bx \minus\bz)^2 (\bz \minus \by)^2}\,
	\Theta\big(\eta - \delta_{\bx\by\bz}\big)
	\big[2 \bar{T}_{\bx\bz}(\eta - \delta_{\bx\bz;r}) - 
	\bar{T}_{\bx\by}(\eta) \big].
\end{align}
We define the Laplace transform of the amplitude according to 
\begin{equation}
	\label{laplacedef}
	\bar{T}_{\bx\by}(\omega) = 
	\int_0^{\infty} \dif \eta\, \rme^{-\omega \eta}\, 
	\bar{T}_{\bx\by}(\eta) 
\end{equation}
and by taking the Laplace transform of Eq.~\eqref{lineareta}, we deduce
\begin{equation}
	\label{lap1}
	\omega \bar{T}_{\bx\by}(\omega) - \bar{T}_{\bx\by}(\eta=0) = 
	\frac{\abar}{2\pi}
	\int \frac{\dif^2 \bz \,(\bx\minus\by)^2}{(\bx \minus\bz)^2 (\bz \minus \by)^2}\,
	\int_{\delta_{\bx\by\bz}}^{\infty}
	\dif \eta\,\rme^{-\omega \eta}
	\big[2 \bar{T}_{\bx\bz}(\eta - \delta_{\bx\bz;r}) - 
	\bar{T}_{\bx\by}(\eta) \big].
\end{equation}
Note there is some complication which has prevented us to identify the Laplace transform of the amplitude in the r.h.s. of the above equation. For the real term this due to the fact that the lower limit is different than the shift, while for the virtual one it is simply because this limit is non-zero. Yet, as we shall show in Appendix \ref{sec:rem}, this is irrelevant for the present purposes; that is, one can equally well replace \eqn{lap1} by
\begin{align}
	\label{lap2}
	\omega \bar{T}_{\bx\by}(\omega)\, - &\,  \bar{T}_{\bx\by}(\eta=0) =\frac{\abar}{2\pi}
	\int \frac{\dif^2 \bz \,(\bx\minus\by)^2}{(\bx \minus\bz)^2 (\bz \minus \by)^2}
	\left[2 \rme^{-\omega \delta_{\bx\bz;r}} \bar{T}_{\bx\bz}(\omega)  - \bar{T}_{\bx\by}(\omega) \right].
	\end{align}
We further take a Mellin transform w.r.t.~the dipole size (cf.~Eq.~\eqref{tmellin}) and Eq.~\eqref{lap2} becomes
\begin{align}
	\label{lapmel1}
	\hspace*{-0.3cm}
	\omega \bar{T}(\omega,\gamma)-\bar{T}(\eta=0,\gamma) = 
	\bar{T}(\omega,\gamma)\, \frac{\abar}{2\pi}
	\int \frac{\dif^2 \bz \,(\bx\minus\by)^2}{(\bx \minus\bz)^2 (\bz \minus \by)^2}
	\left\{2 \rme^{-\omega \delta_{\bx\bz;r}} 
	\left[\frac{(\bx\minus\bz)^2}{(\bx\minus\by)^2} \right]^{\gamma} 
	-1 \right\}.
\end{align}
We shall now focus on the integration over the transverse coordinate $\bz$, which simply gives a function of $\gamma$ and $\omega$. Including a factor of $1/2\pi$ for convenience, we call this function $\chi(\gamma,\omega)$ and we split it in two pieces according to
\begin{align}
	\label{chigodef}
	\chi(\gamma,\omega) =\ & 
	\frac{1}{2\pi}
	\int \frac{\dif^2 \bz \,(\bx\minus\by)^2}{(\bx \minus\bz)^2 (\bz \minus \by)^2}
	\left\{2\left[\frac{(\bx\minus\bz)^2}{(\bx\minus\by)^2} \right]^{\gamma} - 1\right\}
	\nn
	& + \frac{1}{\pi}
	\int \frac{\dif^2 \bz \,(\bx\minus\by)^2}{(\bx \minus\bz)^2 (\bz \minus \by)^2}
	\left[\frac{(\bx\minus\bz)^2}{(\bx\minus\by)^2} \right]^{\gamma}
	\left( \rme^{-\omega \delta_{\bx\bz;r}} -1 \right).
\end{align}
The first term in the above is recognized as the LO characteristic function $\chi_0(\gamma)$. Given the definition of the shift in Eq.~\eqref{delta1}, it is clear that the second term, to be called $\delta\chi(\gamma,\omega)$, has support only for $|\bx-\bz|<r$. As usual we first let $\bz \to \bx-\bz$ and then we easily perform the angular integration (which is unrestricted) since the only respective dependence is in the denominator factor $(\br-\bz)^2$. We find  
\begin{equation}
	\label{dchigo}
	\delta\chi(\gamma,\omega) =
	2 \int_0^1 \frac{\dif z}{z}\,\frac{z^{2(\gamma+\omega)}-z^{2\gamma}}{1-z^2} = \psi(\gamma) - \psi(\gamma+\omega),
	\end{equation}
where in writing the first equation we have also made use of the fact that the integral is scale invariant. Notice that the integration is finite, since the numerator vanishes as $z\to 1$. Now it is straightforward to see that
\begin{equation}  
\label{chireal}	
\chi(\gamma,\omega) = 
2 \psi(1) - \psi(\gamma + \omega) - \psi(\gamma).
\end{equation}
Such an expression has been known for a long time, cf.~\cite{Salam:1998tj}. It is the characteristic function of a linear equation for the unintegrated gluon distribution, where one has enforced the appropriate kinematic constraints in the BFKL equation in order to match the correct DGLAP double logarithmic limit for both soft-to-hard and hard-to-soft evolution \cite{Andersson:1995ju}.

One can now solve \eqn{lapmel1} to obtain
\begin{equation}
	\label{tomegagamma}
	\bar{T}(\omega,\gamma) = \frac{\bar{T}(\eta=0,\gamma)}
	{\omega - \abar \chi(\gamma,\omega)}.
\end{equation}
Thus, we can write the solution to Eq.~\eqref{lineareta} as the double integral
\begin{equation}
	\label{linearsol}
	\bar{T}(\eta,r) \simeq 
	 \int_{\omega_0 - \rmi \infty }^{\omega_0+\rmi \infty}
	\frac{\dif \omega}{2\pi\rmi} \,
	\int_{\frac{1}{2} - \rmi \infty }^{\frac{1}{2}+\rmi \infty}
	\frac{\dif \gamma}{2\pi\rmi} \, 
	\frac{\bar{T}(\eta=0,\gamma)}{\omega - \abar 
	\chi(\gamma,\omega)} \,
	\rme^{\omega \eta} \big(r^2 Q_0^2\big)^\gamma,
\end{equation}
where $\omega_0$ is such that all the singularities of the integrand are on the left of the corresponding contour. The approximate equality sign in \eqn{linearsol} is meant to recall that we have made some approximation in going from \eqn{lap1} to \eqn{lap2}, whose consequences will be discussed  in Appendix \ref{sec:rem}.

 The denominator in Eq.~\eqref{linearsol} gives rise to an infinite number of single poles in the $\omega$-plane, however one can show that there is a unique pole on the positive real axis. This is important as it implies that the dominant behavior of the amplitude at large $\eta$ is controlled by this unique pole, that we shall denote as $\bar{\omega}$. This pole is clearly defined by  
\begin{equation}
	\label{omegaplus}
	\bar{\omega} -\abar \chi(\gamma,\bar{\omega}) =0
	\quad \& \quad 
	\bar{\omega}>0,
\end{equation}
and the corresponding approximation to the amplitude, valid at large $\eta$, reads
\begin{equation}
	\label{linearsol2}
	\bar{T}(\eta,r) \approx 
	\int_{\frac{1}{2} - \rmi \infty }^{\frac{1}{2}+\rmi \infty}
	\frac{\dif \gamma}{2\pi\rmi} \, 
	\bar{T}(\eta=0,\gamma)\,
	\rme^{\bar{\omega}(\gamma) \eta} \big(r^2 Q_0^2\big)^\gamma,
\end{equation}
where we have also neglected the residue associated with the pole $\bar{\omega}(\gamma)$.

Now we would like to check the pole structure (in the $\gamma$-plane) of $\bar{\omega}(\gamma)$ order by order in perturbation theory. Solving iteratively Eq.~\eqref{omegaplus} we find
\begin{equation}
	\label{omegaplusser}
	\bar{\omega}(\gamma) = \abar \chi_0(\gamma) 
	-\abar^2 \chi_0(\gamma)\psi'(\gamma) + \abar^3
	\left\{\chi_0(\gamma)\big[\psi'(\gamma)\big]^2 - 
	\frac{\chi_0^2(\gamma) \psi''(\gamma)}{2} \right\} + \cdots.	
\end{equation}
The assumption that the LO term in the above is $\abar \chi_0(\gamma)$, automatically picks the positive $\bar{\omega}(\gamma)$ solution in Eq.~\eqref{omegaplus}. Expanding Eq.~\eqref{omegaplusser} around $\gamma=0$ we get
\begin{equation}
	\label{omegapluspole0}
	\bar{\omega}(\gamma) = \frac{\abar}{\gamma} - 
	\abar^2 \left(\frac{1}{\gamma^3} + \frac{\pi^2}{6\gamma}\right)
	+\abar^3 \left[\frac{2}{\gamma^5} + \mcal{O}\left(\frac{1}{\gamma^3}\right) \right]+ \cdots,
\end{equation}
while around $\gamma=1$ we have
\begin{equation}
	\label{omegapluspole1}
	\bar{\omega}(\gamma) = \frac{\abar}{1-\gamma} -
	\frac{\pi^2}{6} \frac{\abar^2}{(1-\gamma)}
	+\abar^3\, \mcal{O}\left(\frac{1}{(1-\gamma)^2} \right)+ \cdots.
\end{equation}

The main features of this pole structure could have been anticipated in view of our previous discussions in Sects.~\ref{sec:NLOeta} and \ref{sec:nloinst}. The dominant poles at $\gamma=0$, of the type $\sim \abar^{\kappa}/\gamma^{2\kappa-1}$, are associated with the emission of very small daughter dipoles. (In coordinate space, they correspond to kernel corrections enhanced by double collinear logarithms.)
We had already checked that for the cubic pole $\sim \abar^{2}/\gamma^{3}$ occurring at NLO: in \eqn{bfkllog2}, this pole was generated by integrating over small daughter dipoles with size $z\ll r$. There are no corresponding poles at $\gamma=1$ (e.g.~no cubic pole  $\sim \abar^{2}/(1-\gamma)^{3}$), in agreement with the fact that for the evolution in $\eta$ there are no double transverse logarithms associated with the emission of very {\it large} dipoles.

It is furthermore interesting to notice the
absence of {\em sub-leading} poles, of the type $\sim \abar^{\kappa}/\gamma^{2\kappa-2}$ (and similarly $\sim \abar^{\kappa}/(1-\gamma)^{2\kappa-2}$): this guarantees that we are not interfering with the NLO corrections due to DGLAP evolution. For example, that would not have been the case if the subleading term in $\delta_{\bx\bz;r}$ was a constant as $\bz\to\bx$. In this sense, our choice in \eqn{delta1} is indeed optimal. 

The remaining poles of lower orders, and for any power of $\abar$, are not unique; they depend on the specific choice for the rapidity shift, i.e.~upon the prescription used for the resummation. This is in particular the case for all the poles $\gamma=1$, except of course for the leading-order one. In particular, with our ``optimal'' choice $\delta_{\bx\bz;r}$ for the rapidity shift, we still generate single NLO poles at both $\gamma=0$ and $\gamma=1$ (and with residue $-\pi^2/6$ for both of them). This is at variance with the discussion of the NLO approximation in Sect.~ \ref{sec:NLOeta} (notably after \eqn{chichip}), where we have seen that the change of variables $Y\to \eta$ introduces no additional poles (neither double, or single) at  $\gamma=0$ or  $\gamma=1$.  There is of course no contradiction, since the resummation (i.e.~the rapidity shift) is only meant to deal with the {\it dominant} respective poles, i.e.~the cubic poles at NLO. This being said, it is interesting to notice that one can ``fine-tune'' the resummation in such a way to also remove these single poles: this cannot be done by merely readjusting the rapidity shift (the latter is fixed by the condition to avoid the NLO double-poles as alluded to above), but it can be achieved by also introducing a shift in the virtual term, as explained in Appendix \ref{sec:shiftv}. This is of course not compulsory and in what follows we shall stick to our ``canonical'' version of the resummation, cf. \eqn{bketato}.

\subsection{A local equation with a resummed kernel}
\label{sec:localeta}

Comparing Eqs.~\eqref{chireal} and \eqref{omega0}, it is quite clear that the effect of the resummation at BFKL level merely amounts to a shift
\begin{equation}
	\label{replace}
	\psi(\gamma) \to \psi(\gamma+\omega)
\end{equation}
in the argument of one piece from the LO characteristic function $\chi_0(\gamma)$ --- that piece which includes the simple pole at $\gamma=0$. As a matter of facts, one can check that, for the purpose of resumming the double-collinear logarithms in the BFKL kernel, it suffices to perform this shift in the pole term alone \cite{Salam:1998tj}:
\begin{equation}
	\label{replace2}
	\frac{1}{\gamma} \to \frac{1}{\gamma+\omega}.
\end{equation}
That is, instead of \eqn{chireal}, on could as well use the following version for the resummed characteristic function:
\begin{equation}  
\label{chireal1}	
\chi(\gamma,\omega) = \frac{1}{\gamma+\omega} + \left[\chi_0(\gamma)-\frac{1}{\gamma}\right],
\end{equation}
in which there is no shift in the regular (no-pole) piece of the LO characteristic function. Via the positive-pole condition \eqref{omegaplus}, this would generate a solution $\bar{\omega}(\gamma) $ with the same global properties as those discussed in relation with Eqs.~\eqref{omegaplusser}--\eqref{omegapluspole1}:  the dominant poles at $\gamma=0$ are properly included to all orders and there are no subleading (DGLAP-like) poles.

In this section, we shall consider yet another alternative for the resummed characteristic function, which is different from both \eqref{chireal} and \eqref{chireal1} (albeit equivalent to them to the accuracy of interest and in the BFKL regime) and which has the virtue to allow for a reformulation of the evolution equation  which is local in $\eta$ but with a resummed kernel. As we shall shortly see, this is the counterpart in the $\eta$-representation of the ``collBK'' (the collinearly-improved version of the BK equation) discussed in Sect.~\ref{sec:bcy} in the context of the evolution with $Y$.

This alternative version for $\chi(\gamma,\omega) $ will in fact be implicitly defined by the resummed evolution equation that we shall derive in what follows. To that aim, we start with a study of the evolution in the collinear regime, where one of the daughter dipoles is very small. For that purpose, it suffices to keep only the dominant poles at $\gamma=0$, as generated by the would-be pole piece of the characteristic function alone, i.e.~$\chi(\gamma,\omega) = {1}/({\gamma+\omega})$. Then the pole condition  \eqref{omegaplus} reduces to a quadratic equation for $\bar{\omega} $ that can be  easily solved:
\begin{equation}
	\label{omegadlasmall}
	\omega \simeq \frac{\abar}{\gamma + \omega} 
	\, \Longrightarrow \,
	\bar{\omega} \simeq \frac{1}{2}
	\left( - \gamma + \sqrt{\gamma^2 + 4 \abar} \right) = \frac{\abar}{\gamma} - 
	\frac{\abar^2}{\gamma^3} 	+\frac{2\abar^3}{\gamma^5}-\cdots,
\end{equation}
where we have kept only the positive solution. The last equality shows the formal expansion of $\bar{\omega} $ in powers of $\abar$: as anticipated, this correctly reproduces the dominant poles at $\gamma=0$ to all orders. On the other hand, the complete solution has a finite limit when $\gamma\to 0$, namely $\bar{\omega}(0)=\sqrt{\abar}$.

It is a straightforward exercise to show that the above function $\bar{\omega}(\gamma)$  is the exact eigenvalue corresponding to the local evolution equation\footnote{A simple way to see this consists in expanding the kernel  $\mathcal{K}_\sdla$ as shown in \eqn{kdla} and then acting with each term in this expansion on the power-like amplitude $\bar{T}\propto z^{2\gamma}$ to reconstruct the series in the r.h.s. of \eqn{omegadlasmall}.}
\begin{equation}
	\label{tdlalocal}
	\frac{\del \bar{T}(\eta,r^2)}{\del \eta}
	= \abar \int_0^{r^2} \frac{\dif z^2}{z^2}\,
	\mathcal{K}_\sdla \left(\ln\frac{r^2}{z^2} \right)
	\bar{T}(\eta,z^2),
\end{equation}
where $\mathcal{K}_\sdla$ has been defined in \eqn{kdla}.  But even though it involves the same kernel
as the equation \eqref{adlak} describing (resummed) DLA evolution in $Y$,  the above equation differs from 
\eqn{adlak}  in an essential way:  the integral over $z$ in Eq.~\eqref{tdlalocal} is restricted to small daughter dipoles with $z^2< r^2$, whereas the corresponding integral in \eqn{adlak} rather runs over  large dipoles with $z^2> r^2$.

At this level, it is possible to follow a strategy similar to that employed in Sect.~\ref{sec:bcy} in order to extend Eq.~\eqref{tdlalocal} to a local evolution equation which encompasses the LO BK equation (and reduces to it in the absence of the collinear resummation). In fact, simply by analogy with \eqn{collbk}, it is quite clear that the following equation looks like a natural generalization of Eq.~\eqref{tdlalocal} to the full BFKL dynamics and also to the non-linear regime:
\begin{equation}
	\label{collbk0}
\frac{\partial \bar{S}_{\bx\by}}{\partial \eta} = 
 \frac{\abar}{2 \pi}
 \int
 \frac{\dif^2 \bz \,(\bx\minus\by)^2}{(\bx \minus\bz)^2 (\bz \minus \by)^2}\,
 \mcal{K}_{\rm \scriptscriptstyle DLA}(\bar{\rho}_{\bx\by\bz})
 \left(\bar{S}_{\bx\bz} \bar{S}_{\bz\by} - \bar{S}_{\bx\by} \right).
\end{equation}
This looks formally similar to  \eqn{collbk}, but it differs from it in so far as the argument  of  $\mcal{K}_{\rm \scriptscriptstyle DLA}$ is concerned: this is now defined as (compare to \eqn{rhoxyz})
\begin{equation}
	\label{rhobarxyz}
	\bar{\rho}_{\bx\by\bz}^2 \equiv
	\ln^2 \frac{(\bx-\bz)^2}{(\bz-\by)^2}\,.
\end{equation}
Clearly, $\bar{\rho}_{\bx\by\bz}$ reduces to $\ln({r^2}/{z^2})$, as it should, when one of the two daughter dipoles (whose size is denoted as $z$) is much smaller than the other one, while it dies away when the two daughter dipoles have comparable sizes. Hence, as expected, the resummation performed by $\mcal{K}_{\rm \scriptscriptstyle DLA}$ is only effective for the soft-to-hard evolution (in contrast to  \eqn{collbk}, where a similar kernel but with a different argument performs the anti-collinear resummation applicable to the hard-to-soft evolution).

Although it looks appealing --- for the same reasons as its counterpart,  \eqn{collbk}, in the evolution with $Y$, namely, the fact that it has the same non-local and non-linear structure as the LO BK equation --- \eqn{collbk0} is not fully right, not even in the approximations of interest. Specifically, this equation is an acceptable resummation in the linear, BFKL, regime, where it generates a characteristic function with all the good properties previously discussed in relation with Eqs.~\eqref{chireal} and \eqref{chireal1}. (This can be checked order by order in $\abar$, by using the perturbative expansion of  $\mcal{K}_{\rm \scriptscriptstyle DLA}$, cf. \eqn{kdla}.) But it becomes incorrect in the non-linear regime and more precisely in the approach towards saturation, as we now explain. The collinear (small daughter dipole) resummation is indeed relevant in this regime, since the soft-to-hard evolution controls the approach of the dipole $S$-matrix towards the black disk limit $\bar S=0$, known as the Levin-Tuchin formula\footnote{Incidentally, this also explains why the anti-collinear (large  daughter dipoles) resummation is unimportant for that issue (we mean of course the evolution with $Y$):  \eqn{collbk} predicts exactly the same result for the Levin-Tuchin formula as the LO BK equation.}
 \cite{Levin:1999mw,Iancu:2003zr}. Let us briefly review the argument and derive in the process the corresponding prediction of \eqn{collbk0}. 
 
 Assume that the measured dipole with size $r$ is deeply at saturation, $r^2 \bar Q_s^2(\eta) \gg 1$. Then, clearly, its $S$-matrix is only tiny, $\bar{S}(\eta,r^2)\ll 1$, but we would like to know how it approaches to 0 when further increasing $\eta$ and/or $r$. To study this evolution based on \eqn{collbk0}, one can \texttt{(i)}  restrict oneself to smaller daughter dipoles, but which are still at saturation, namely such that $1/\bar Q_s^2(\eta) \ll z^2 \ll r^2$ (indeed, the contributions from even smaller daughter dipoles, with size $z^2\ll 1/\bar Q_s^2$, cancel between real and virtual terms) and \texttt{(ii)} keep just the virtual term, which is linear in $\bar{S}$ (the real term,  involving the product of two $S$-matrices for two large dipoles, is even smaller). This yields
\begin{equation}
	\label{s0local}
	\frac{\del \bar{S}(\eta,r^2)}{\del \eta} 
	= -\abar \bar{S}(\eta,r^2)
	\int_{1/\bar Q_s^2}^{r^2} \frac{\dif z^2}{z^2}
	\mathcal{K}_\sdla\left(\ln\frac{r^2}{z^2} \right)
	\simeq -\sqrt{\abar}\, \bar{S}(\eta,r^2),
\end{equation}
where the final result in the r.h.s. holds up to corrections of relative order $\sqrt{\abar}$ to the prefactor. Note that this final result is independent of the saturation scale $\bar Q_s$: indeed, when the size $z$ of the daughter dipole is small enough, such that $\bar\rho\equiv \ln({r^2}/{z^2})$ becomes of $\order{1/\sqrt{\abar}}$, the function $\mcal{K}_{\rm \scriptscriptstyle DLA}$ --- which we recall is proportional to the Bessel function $\rmJ_1 \big( 2 \sqrt{\abar \bar \rho^2} \big)$  --- oscillates very fast and effectively kills the contribution from even smaller daughter dipoles. Hence the effective range for the integration over $z^2$ is  $r^2 \exp (-1/\sqrt{\abar}) \ll z^2 \ll r^2$, where the lower limit is indeed much larger than $1/\bar Q_s^2(\eta)$ for $\eta$ large enough. \eqn{s0local} is easily seen to imply a power-like behavior
\begin{equation}
	\label{s0power}
	\bar{S}(\eta,r^2) \propto \rme^{-\sqrt{\abar}(\eta-\eta_s)}\propto
	{\left[{r^2 \bar Q_s^2(\eta)}\right]^{-\frac{\sqrt{\abar}}{\bar{\lambda}_s}}},
\end{equation}
where $\eta_s$ is the value of the rapidity at which $\bar Q_s^2(\eta_s) = 1/r^2$ and
we have also used $\bar Q_s^2(\eta)\propto \rme^{\bar{\lambda}_s\eta}$.

The result in \eqn{s0power} is very different (in particular, it has a different functional form) from the LO prediction in \eqn{sbelow} and as a matter of facts it is {\it not} correct: as we shall see  in Sect.~\ref{sec:lt}, the respective prediction of the non-local equation \eqref{bketato} has the same functional structure as the LO result \eqref{sbelow}, but with a modified coefficient in front of the exponent. The above derivation of \eqn{s0power} also gives us a hint about what may have gone wrong: the peculiar power-like behavior visible in  \eqn{s0power} is clearly a consequence of the fact that the virtual term in  \eqn{collbk0} is multiplied by $\mcal{K}_{\rm \scriptscriptstyle DLA}$, that is,  this term too is affected by the resummation. This property contradicts the actual structure of the NLO corrections to the BK equation in $\eta$: as discussed after \eqn{nlobfkleta}, the double collinear logarithms  occurring at NLO are important only for the scattering of the smallest daughter dipole; but they are absent both in the scattering of the other, large, daughter dipole, and in the virtual term.

\section{Saturation fronts in $\eta$}
\label{sec:etafronts}

This section will be devoted to the ``phenomenology'' of the non-local equation \eqref{bketato} for the evolution with $\eta$, that is, to its predictions for interesting quantities like the saturation exponent and the anomalous dimension (including their dependence upon $\eta$ prior to asymptotics) and for phenomena like geometric scaling and the approach towards saturation. For the case of a fixed coupling, we will be able to address all these issues via analytic calculations, with results that will be then confirmed by the numerics. For the case of a running coupling, we shall present only numerical results which confirm that the effects of the resummation (non-locality) are important in that case too.

\subsection{Saturation saddle point and geometric scaling}
\label{sec:sat}

In order to study the speed and the shape of the saturation fronts in $\eta$, we will follow the same strategy as for the LO BK equation in Sect.~\ref{sec:lobk}. This is indeed possible since \texttt{(i)} the analytic study in Sect.~\ref{sec:lobk} relied only on the linearized version of the evolution equation, and \texttt{(ii)} after linearization, the non-local equation \eqref{bketato} admits a similar mathematical treatment (in particular, the construction of explicit solutions via the Mellin representation), as we have seen in Sect.~\ref{sec:BFKLeta}.

\begin{table}
\begin{center}
\begin{tabular}{|c|c|c|c|}
 \hline
         $\abar$ & $\bar{\lambda}_s$ & $\bar{\gamma}_s$ & $\bar{D}_s$         
         \\ \hline\hline
         $ \to 0$ & \hfill $4.88 \abar$ & $0.628 $ & \hfill $97.0 \abar$
         \\ \hline
         $ 0.1$ & $ 0.384 = 3.84\abar$ & $0.589 $ & \hfill $6.18 = 61.8 \abar$         
         \\ \hline
         $ 0.2$ & $ 0.657 = 3.29\abar$ & $0.565 $ & \hfill $9.74 = 48.7 \abar$         
         \\ \hline
         $ 0.3$ & $ 0.876 = 2.92\abar$ & $0.548 $ & \hfill $12.4 = 41.3 \abar$         
         \\ \hline
         $ 0.4$ & $ 1.058 = 2.65\abar$ & $0.535 $ & \hfill $14.6 = 36.4 \abar$         
         \\ \hline
 \end{tabular}
 \end{center}
 \caption{\small Asymptotic speed, slope and diffusion coefficient for the non-local equation \eqref{bketato} for various values of the coupling constant $\abar$.}
 \label{tab:one}         
 \end{table}

\begin{figure}[t]
\centerline{\hspace*{-.3cm}
  \includegraphics[width=0.46\textwidth]{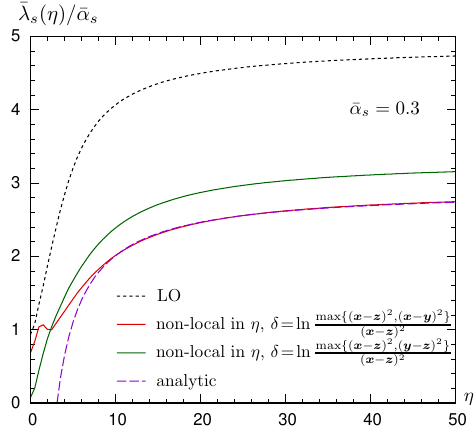}\qquad
\includegraphics[width=0.46\textwidth]{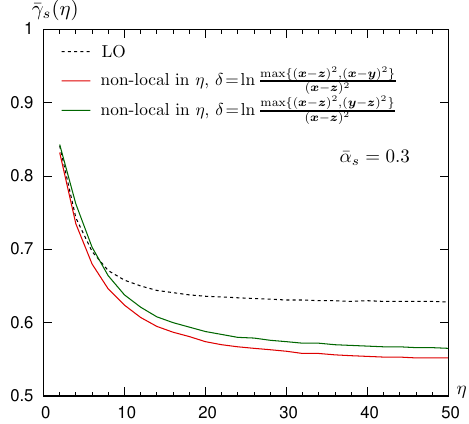}}
\caption{\small Left: The front speed (divided by $\abar$) as a function of $\eta$, obtained from the numerical solution to Eq.~\eqref{bketato} for two different $\delta$-shifts. The analytic asymptotic expansion for the ``canonical'' shift and the LO result are also shown. Right: The slope as a function of $\eta$, obtained from the same equation.}
 \label{fig:etadep}
\end{figure}

Specifically, we start with the BFKL solution expressed as an inverse Mellin transform in \eqn{linearsol2}. This involves the characteristic function $\bar{\omega}(\gamma)$ which is defined by the solution to Eq.~\eqref{omegaplus}. Even though the latter cannot be analytically inverted, it is just an algebraic equation which can be easily solved numerically in order to construct $\bar{\omega}(\gamma)$. For our purposes, even this is not necessary, as we now explain. The saturation saddle point is determined as explained 
in Sect.~\ref{sec:lobk}, that is, this is solution (to be denoted as $\bar{\gamma}_s$) to the following equation:
\begin{equation}
        \label{omegasad}
        \frac{\dif\bar{\omega}(\gamma)}{\dif \gamma} = \frac{\bar{\omega}(\gamma)}{\gamma}.
\end{equation}
Using Eq.~\eqref{omegaplus} and the chain differentiation rule, we immediately deduce
\begin{equation}
        \label{dchidg}
        \frac{\dif \bar{\omega}}{\dif\gamma}
        =\abar\,\frac{\del \chi(\gamma,\bar{\omega})}
        {\del \gamma}
        + \abar\,\frac{\del \chi(\gamma,\bar{\omega})}
        {\del \bar{\omega}}\, 
        \frac{\dif \bar{\omega}}{\dif \gamma}
        \,\Rightarrow\, 
        \frac{\dif \bar{\omega}}{\dif \gamma} = 
        \frac{\abar \del \chi(\gamma,\bar{\omega})/
        \del \gamma}{1 - \abar \del \chi(\gamma,\bar{\omega})/\del \bar{\omega}},
\end{equation}
and we thus see that Eq.~\eqref{omegasad} is equivalent to
\begin{equation}
\label{omegaplussad}
        \frac{\del \chi(\gamma,\bar{\omega})/
        \del \gamma}{1 - \abar \del \chi(\gamma,\bar{\omega})/\del \bar{\omega}} = 
        \frac{\chi(\gamma,\bar{\omega})}{\gamma}.
\end{equation}
With $\chi(\gamma,\omega)$ defined in Eq.~\eqref{chireal} and for a given $\abar$, it is a trivial numerical exercise to solve Eqs.~\eqref{omegaplus} and \eqref{omegaplussad} and determine the asymptotic slope $\bar{\gamma}_s$ and the asymptotic speed $\bar{\lambda}_s =\bar{\omega}(\bar{\gamma}_s)/\bar{\gamma}_s$ of the front. The amplitude above the saturation line is given by an expression analogous to Eq.~\eqref{tabove}, in which we replace $Y\to \eta$, $Q_s \to \bar{Q}_s$, $\gamma_0 \to \bar{\gamma}_s$ and $D_0= 2 \abar \chi''_0(\gamma_0) \to \bar{D}_s \equiv 2 \bar{\omega}''(\bar{\gamma}_s)$, that is  
\begin{equation}
        \label{tnlabove}
        T(\eta,r) = \big(r^2 \bar{Q}_s^2\big)^{\bar{\gamma}_s}
        \bigg(\ln \frac{1}{r^2 \bar{Q}_s^2} + c\bigg)
        \exp \bigg[ - \frac{\ln^2 \big(r^2 \bar{Q}_s^2\big)}{\bar{D}_s \eta} \bigg].
\end{equation}
Notice that the second derivative $\bar{\omega}''(\gamma)$ can be calculated in terms of partial derivatives of $\chi(\gamma,\omega)$, just by extending what we did for the first derivative in Eq.~\eqref{dchidg}. The range of validity for Eq.~\eqref{tnlabove} and the corresponding geometric scaling are as described below  Eq.~\eqref{tabove}. In Table \ref{tab:one} we give the values of $\bar{\lambda}_s$, $\bar{\gamma}_s$ and $\bar{D}_s$ for some representative values of $\abar$.  We have checked that these values are indeed very well reproduced by the numerical solutions to  \eqn{bketato}  (recall in particular the numerical results for
$\bar{\lambda}_s$ in Fig.~\ref{fig:nloeta}); they are furthermore in agreement with the numerical results for $\abar=0.3$ that are displayed in Fig.~\ref{fig:etadep}  and will be discussed shortly.

Regarding the speed of the front, we can also indicate the  $\eta$-dependence in the approach towards the asymptotic behavior at $\eta\to \infty$. This is
given by an expression analogous to Eq.~\eqref{dlogqs} which adapted to the present notations becomes
\begin{equation}
        \label{dlogqsnl}
      \bar{\lambda}_s(\eta)\,\equiv\,  \frac{\dif \ln \bar{Q}_s^2}{\dif \eta} \simeq
        \bar{\lambda}_s
        -\frac{3}{2\bar{\gamma}_s}\,\frac{1}{\eta}.
        \end{equation} 
        
In Fig.~\ref{fig:etadep} we show the  $\eta$-dependence  of both the front speed $\bar{\lambda}_s(\eta)$ and the anomalous dimension $\bar{\gamma}_s(\eta)$, as obtained from  numerical solutions to  \eqn{bketato} with $\abar=0.3$. The function $\bar{\lambda}_s(\eta)$ is defined as in \eqn{dlogqsnl} with the saturation scale $ \bar{Q}_s^2(\eta)$ extracted from the numerical data. As for $\bar{\gamma}_s(\eta)$, this is obtained simultaneously with the diffusion coefficient $\bar{D}_s(\eta)$ by fitting the numerical saturation fronts with the {\em Ansatz} in \eqn{tnlabove}. In Fig.~\ref{fig:etadep} (left) we also show the pre-asymptotic behavior predicted by \eqn{dlogqsnl}, which turns out to provide an excellent fit down to $\eta=7 \div 8$.

\begin{figure}[t]
\centerline{\hspace*{-.3cm}
  \includegraphics[width=0.46\textwidth]{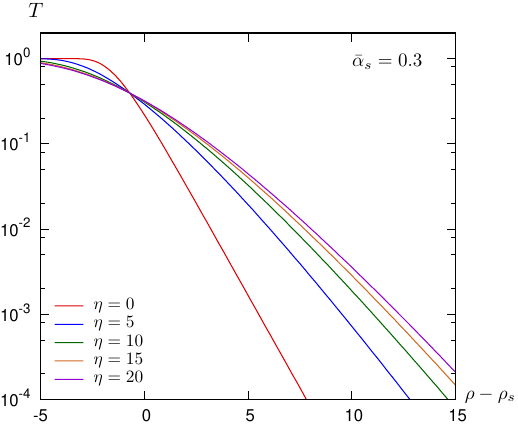}\qquad
\includegraphics[width=0.46\textwidth]{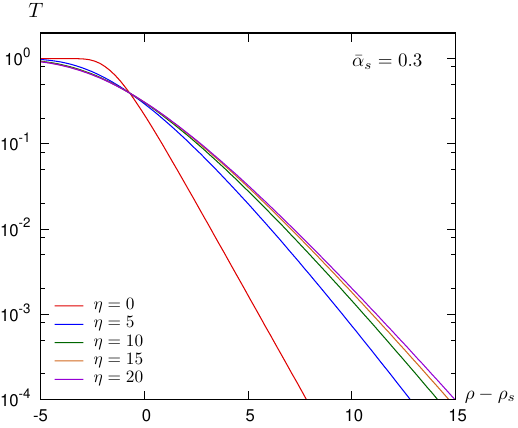}}
\caption{\small Left: The amplitude obtained from the non-local equation \eqref{bketato} plotted as a function of the scaling variable $\rho-\rho_s$.  Right: The same for the amplitude obtained by solving LO BK. In both cases the amplitude exhibits geometric scaling. The non-local equation leads to a front which is a bit less steep.}
 \label{fig:scaling}
\end{figure}

In Fig.~\ref{fig:scaling} (left) we study the quality of geometric scaling for the numerical solutions  to  \eqn{bketato} (for $\abar=0.3$); that is, we plot the amplitude as a function of the difference $\rho-\bar\rho_s(\eta) \equiv  \ln[{1}/{r^2 \bar{Q}_s^2}]$, with  $ \bar{Q}_s^2(\eta)$ itself determined by the numerics. For the sake of comparison, we also show the respective curves for the LO BK equation (in the right panel). As one can see, the scaling provided by the non-local equation is still good, albeit slightly less so than at LO, and it becomes better and better with increasing $\eta$ (the successive curves corresponding to larger and larger values of $\eta$ approach a common shape in a region in $\rho-\bar\rho_s(\eta)$ which extends with $\eta$, via BFKL diffusion).

\begin{figure}[t]
\centerline{\hspace*{-.3cm}
  \includegraphics[width=0.46\textwidth]{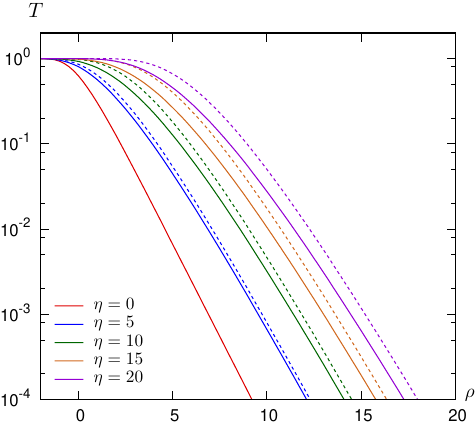}\qquad
\includegraphics[width=0.46\textwidth]{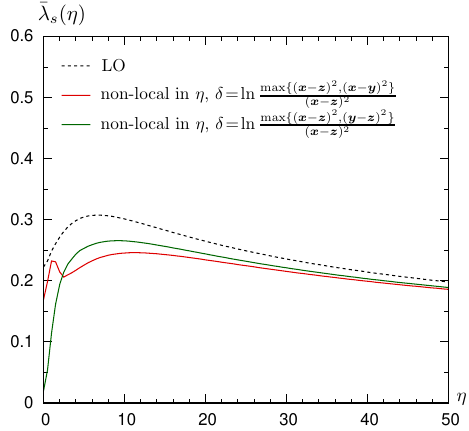}}
 \caption{\small Left: The amplitude as obtained from running coupling evolution (and where the scale in the coupling is set to run with the smallest of the sizes of the three dipoles). Solid lines stand for the solutions to the non-local equation \eqref{bketato}, while dashed stand for those obtained from LO BK. Right: The speed of the fronts as a function of $\eta$, for running coupling evolution according to the non-local equation \eqref{bketato} (for two different $\delta$-shifts) and for LO BK.}
 \label{fig:rc}
\end{figure}

Finally, in Fig.~\ref{fig:rc} we present the predictions of the running coupling version of \eqn{bketato}, with the ``minimal dipole size'' prescription for the running coupling, $\abar(r_{\rm min})$ where $r_{\rm min} = \min\{|\bx \minus\by|,|\bx \minus\bz|,|\by \minus\bz|\}$. In the left panel, we show the saturation fronts for various rapidities (for comparison, the respective fronts at LO are shown too, with dashed lines). In the right panel, we show the $\eta$-dependence of the saturation exponent as predicted by the LO BK equation and by the non-local equation with two prescriptions for the rapidity shift. The first observation is that the running of the coupling is dramatically slowing down the evolution, with or without the collinear resummation: the typical values of the saturation exponent are smaller by roughly a factor of 3 as compared to the case of a fixed coupling $\abar=0.3$. This being said, the reduction in the value of $\bar{\lambda}_s(\eta)$ due to the collinear resummation is still visible:
the results for $\bar{\lambda}_s(\eta)$ corresponding to the two prescriptions for the rapidity shift are rapidly converging to each other with increasing $\eta$, but they remain visibly smaller than the respective LO result, up to rapidities as large as $\eta= 50$.

\subsection{The solution below $Q_s$: the Levin-Tuchin formula with resummation}
\label{sec:lt}

As a final application of the non-local evolution equation \eqref{bketato}, let us use it in order to determine the  limiting form of the dipole $S$-matrix deeply at saturation, i.e.~for very large dipole sizes $r^2 \bar{Q}_s^2(\eta) \gg 1$.  As already stressed in previous discussions, where we have studied the same problem, first, for the LO BK equation, cf. Eq.~\eqref{sbelow}, and then for the local form of the collinear resummation in $\eta$, cf.~\eqn{s0power}, this study allows for two important simplifications (to the leading double-logarithmic accuracy): \texttt{(i)} one can neglect the term quadratic in $\bar{S}$ in the evolution equation  and \texttt{(ii)} the integration over the daughter dipole size $z$ becomes logarithmic when $z$ is much smaller than the parent dipole size $r$. The only change in the argument w.r.t.~the respective discussion of the LO BK equation in Sect.~\ref{sec:lobk} is the non-locality of the ``real'' term (the term quadratic in $\bar{S}$) in rapidity.

In the interesting regime, where one of the two daughter dipoles is much smaller than the other one, this non-locality is important only for the $S$-matrix associated with that smaller dipole. Hence, to the accuracy of interest, \eqn{bketato} reduces to 
\begin{equation}
	\label{app:shiftedlog}
	\frac{\partial \bar{S}(\eta,r^2)}{\partial \eta} \simeq \abar \bar{S}(\eta,r^2)
	\int_0^{r^2} \frac{\dif z^2}{z^2}\, \Theta \left(\eta -\ln\frac{r^2}{z^2} \right)
	\left[\bar{S}\left(\eta -\ln\frac{r^2}{z^2},z^2 \right) -1 \right],
\end{equation}
where the r.h.s. includes a factor of 2 to account for the fact that the small dipole with size $z$ can be any of the two daughter dipoles. As already mentioned, the integration over $z^2$ is logarithmic only for the virtual term. But the range for this integration is controlled by the condition that the ``real'' $S$-matrix $\bar{S}\left(\eta -\ln\frac{r^2}{z^2},z^2 \right) $ (corresponding to the ``large'' daughter dipole) be negligibly small. This happens for $z^2$ larger than a typical value that we shall denote as $1/\widetilde{Q}_s^2$, for which this $S$-matrix is close to the saturation line. Clearly, this scale $\widetilde{Q}_s$ plays the role of the saturation momentum, but evaluated at a special value of the rapidity, which is itself dependent upon $\widetilde{Q}_s$; namely, this saturation condition reads
\begin{equation}
	\label{qstilde}
	\bar{S}\left(\eta - \ln r^2 \widetilde{Q}_s^2, 1/\widetilde{Q}_s^2 \right) \sim \order{1}
	\,\Longrightarrow\, {\widetilde{Q}_s^2}\,\simeq\,\bar{Q}_s^2(\eta -\ln r^2 \widetilde{Q}_s^2)
	\simeq\,
Q_0^2
	\rme^{\bar{\lambda}_s \left(\eta - \ln r^2 \widetilde{Q}_s^2 \right)}
\end{equation}
where we have assumed the asymptotic behavior $\bar{Q}_s^2(\eta) = Q_0^2\, \rme^{\bar{\lambda}_s \eta}$. \eqn{qstilde} is readily solved to give
\beq
	\label{qstildesol}
	r^2 \widetilde{Q}_s^2 = (r^2 \bar{Q}_s^2)^{1/(1+\bar{\lambda}_s)}.
\eeq
The integration in \eqn{app:shiftedlog} can now be easily done to the leading logarithmic accuracy of interest:
\begin{equation}
	\label{app:shiftedlog1}
	\frac{\partial \bar{S}(\eta,r^2)}{\partial \eta} \simeq  - \abar \bar{S}(\eta,r^2)
	\int_{1/\widetilde{Q}_s^2}^{r^2} \frac{\dif z^2}{z^2} = 
	-{\abar}\, \bar{S}(\eta,r^2)  \frac{\ln (r^2 \bar{Q}_s^2)}{1+\bar{\lambda}_s}=
	- \frac{\abar\bar{\lambda}_s}{1+\bar{\lambda}_s}\,(\eta-\eta_s),
\end{equation}
where $\eta_s$ obeys the condition $\bar{Q}_s^2(\eta_s)=1/r^2$. The final integration over $\eta \ge \eta_s$ yields
\begin{equation}
	\label{ltres}
	\bar{S}(\eta,r^2) \simeq \exp \left[ -  \frac{\abar\bar{\lambda}_s}{2(1+\bar{\lambda}_s)}\,(\eta-\eta_s)^2\right]=
 \exp \left[ - \frac{\abar}{2\bar{\lambda}_s(1+ \bar{\lambda}_s)} \,\ln^2\big[r^2 \bar{Q}_s^2(\eta)\big] \right].
\end{equation}
Thus, the functional form of the $S$-matrix is similar to the one obtained at LO,  cf. Eq.~\eqref{sbelow}, except that the coefficient in the exponent gets the extra multiplicative factor  $1/(1+\bar{\lambda}_s)$. Recalling that $\bar{\lambda}_s\sim\order{\abar}$, one concludes that the NLO modification in the exponent of the Levin-Tuchin formula should be a multiplicative factor equal to $1-\bar{\lambda}_s$. This can indeed be checked on the basis of the NLO BK equation in $\eta$, as discussed in Sect.~\ref{sec:NLOeta}. This argument also suggests that the NLO BK equation should develop an instability in the saturation regime if $\abar$ is large enough so that $\bar{\lambda}_s>1$.

Finally, as a consistency check of our work, let us show how  \eqn{ltres} arises from the corresponding result for the evolution in $Y$.  Strictly speaking, one should use a resummed version of the BK equation in $Y$, which includes the effects of time-ordering (e.g.~the non-local equation \eqref{bkyto}). Note however that the TO constraint in $Y$-evolution is relevant only for large daughter dipoles, i.e.~for the hard-to-soft evolution, therefore it has no incidence on the Levin-Tuchin formula, which therefore preserves the same form after resummation as at LO order, that is, Eq.~\eqref{sbelow}, where however the saturation momentum is affected by the resummation: $Q_s^2(Y) = Q_0^2\, \rme^{\lambda_s Y}$ with $\lambda_s$ the asymptotic intercept of the saturation momentum in the presence of TO, as exhibited (as a function of $\abar$)   in Fig.~\ref{SresumY} (left). Using the relation between the two rapidities, that is, $Y = \eta + \ln(1/r^2 Q_0^2)$ and the formula $\lambda_s = \bar{\lambda}_s/(1+\bar{\lambda}_s)$ between the corresponding saturation exponents, cf. \eqn{speedslope}, we get
\begin{equation}
	\label{rqs}
	r^2 Q_s^2(Y) = \big[r^2 \bar{Q}_s^2(\eta)\big]^{1/(1+\bar{\lambda}_s)}\end{equation}
and then Eq.~\eqref{sbelow} leads to \eqn{ltres}, as it should.  Eqs.~\eqref{qstilde} and \eqref{rqs} make clear that the scale $\widetilde{Q}_s^2$ is nothing else but $Q_s^2(Y)$.

\section{More insights on the initial condition and the rapidity shift}
\label{sec:IC}

In the previous sections, we have noticed that the non-local equations in $\eta$ (and unlike the corresponding equations in $Y$) are formally well-defined as initial-value problems. With reference to our ``canonical'' equation  \eqref{bketato}, it is quite clear that, once the initial condition is given at $\eta=0$, that is, $\bar S_{\bx\by}(\eta=0)=S^{(0)}_{\bx\by}$, then this equation allows us to determine the function  $\bar S_{\bx\by}(\eta)$ for all positive values $\eta>0$. This crucially relies on the presence of the step-function in the integrand of  \eqn{bketato}, which ensures that the shifted rapidity arguments remain positive semi-definite even for very small daughter dipoles. In this section, we would like to discuss more general formulations of the initial-value problem and also the physical meaning of this step-function and of the rapidity shift itself.

Notice first that $\eta=0$ corresponds to a value $\xbj=1$ for Bjorken $x$, which is clearly not a good choice for formulating the initial condition for the high-energy evolution. One should rather start at a much lower value  $\xbj^0\ll 1$, but such that $\abar\eta_0\ll 1$ as well, in order for the early evolution, up to $\eta_0\equiv\ln(1/\xbj^0)$, to be indeed negligible and for the small-$x$ approximations to apply at $\eta >\eta_0$. By inspection of \eqn{bketato}, it is quite clear that this equation cannot be solved with the initial value formulated at some generic $\eta_0>0$: indeed, for a sufficiently small daughter dipole, the rapidity argument of the corresponding $S$-matrix, i.e. $\eta-\ln(r^2/r^2_<)$ with $r^2_< \ll r^2$ can become smaller than $\eta_0$. (We recall that $r_<=\mbox{min}(|\bz-\bx|,\,|\bz-\by|)$ and in this section we shall often use the simpler notation $r_<\equiv z$.) This can be avoided by modifying the argument of the step-function as follows:
\begin{align}
        \label{bketao}
        \frac{\del \bar{S}_{\bx\by}(\eta)}{\del \eta} = 
        \frac{\abar}{2\pi}
        \int \frac{\dif^2 \bz \,(\bx\minus\by)^2}{(\bx \minus\bz)^2 (\bz \minus \by)^2}\,
        \Theta\big(\eta\minus \eta_0\minus \delta_{\bx\by\bz}\big)
        \big[\bar{S}_{\bx\bz}(\eta\minus \delta_{\bx\bz;r})\bar{S}_{\bz\by}(\eta \minus \delta_{\bz\by;r}) 
        \minus \bar{S}_{\bx\by}(\eta) \big].
\end{align}
Interestingly though, the equation itself depends upon the initial rapidity $\eta_0$: this is unavoidable when working with a delay equation which is non-local in the evolution ``time''.

At this point, it becomes interesting to gain some more insight into the physical meaning of the rapidity shift and, related to that, of the constraint introduced by the step-function in the above equation. Let us first recall from Sect.~\ref{sec:to} that $\eta$ is a direct measure of the lifetime of a fluctuation: one can write (cf. \eqn{eta1def})
\beq
\eta\equiv \eta_r=  \ln\frac{\tau_r}{\tau_0}\ \Longrightarrow\ \eta -\ln\frac{r^2}{z^2} =\ln\frac{\tau_z}{\tau_0}= \eta_z
 \eeq
where the subscripts, $r$ or $z$, on lifetimes or rapidities refer to the transverse size of the respective dipole fluctuation and we recall that we consider a small daughter dipole with size $z\ll r$ (hence $\eta_z <\eta$). That is, the shifted rapidity is a measure of the actual lifetime of the daughter dipole. This shows that, in general the $S$-matrix does not depend upon the kinematical rapidity $\eta$ of the fluctuation, but rather upon its lifetime $\tau$. When the emitted dipole is much smaller than its parent $z\ll r$, it also has a much shorter lifetime $\tau_z\ll \tau_r$, and  its scattering amplitude must properly be evaluated at the longitudinal scale set by this lifetime $\tau_z$. 

Furthermore, the step-function visible in \eqn{bketao}, that for the present purposes can be rewritten as $
 \Theta\big(\eta - \eta_0 - \ln(r^2/z^2))=\Theta\big(\eta_z - \eta_0)$, has the role to eliminate those small-$z$ fluctuations whose lifetime $\tau_z$ is smaller than some fixed scale introduced by the initial rapidity $\eta_0$. This scale is largely arbitrary --- it is a part of our model for the initial condition --- but a better understanding of it, together with an improved formulation of the initial condition, can be reached for the particular case where the target is a large nucleus described by the MV model.
 
 To that aim, consider the evolution at early stages, say for a (parent-dipole) rapidity $\eta$ such that $\eta-\eta_0\lesssim 1$. Then the constraint introduced by the step-function  in \eqn{bketao} removes from the evolution all the emissions of very small dipoles with $z\ll r$: such dipoles would have lifetimes much smaller than the characteristic scale introduced by $\eta_0$. For a large nucleus, this  characteristic scale is the longitudinal extent of the target, that we shall denote as $L$. (In the target rest frame and in a mean field approximation in which the nucleus is assumed to be homogeneous, $L=R_A\simeq A^{1/3} R_0$, where $R_0$ is the radius of a nucleon and $R_A$ that of the nucleus.) It is then natural to choose $\eta_0=\ln(L/\tau_0)=\ln(R_A/R_0)$, which holds in any frame. 
 
 In the context of the LO BK equation, which is local in rapidity, it seems appropriate to encode the effect of all the fluctuations with lifetimes smaller than $L$ into a model for the initial condition $S^{(0)}(r)$ at $\eta_0$; e.g., in the MV model one simply ignores all such fluctuations. However, beyond LO, the non-local evolution couples fluctuations with widely different lifetimes and the same happens at early stages, where the evolution includes effects from fluctuations with lifetimes smaller than the target width. Such fluctuations do not contribute to leading logarithmic accuracy --- they do not provide contributions of $\order{\abar\eta}$ when $\abar\eta\gtrsim 1$ ---, which explains why they were ignored in writing \eqn{bketao}. Yet, they are a part of the complete physical picture and they might influence the evolution at early stages. It is therefore interesting to estimate their contribution to the early evolution in more detail.  In this study, one can treat the scattering between the short-lived fluctuations and the nucleus within the semi-classical approximation, that is, within the MV model.
   
Let us start by anticipating the final result of this study: this is a modified version of \eqn{bketao}, which reads
 \begin{align}
          \label{bketafin}
        \frac{\del \bar{S}_{\bx\by}(\eta)}{\del \eta} = \,
        \frac{\abar}{2\pi}
        \int \frac{\dif^2 \bz \,(\bx\minus\by)^2}{(\bx \minus\bz)^2 (\bz \minus \by)^2}\,
        \Big[\bar{S}_{\bx\bz}(\eta \minus \delta_{\bx\bz;r})\bar{S}_{\bz\by}(\eta \minus \delta_{\bz\by;r}) 
        \minus \bar{S}_{\bx\by}(\eta) \Big].
\end{align}
As compared to \eqn{bketao} there is no step-function anymore (hence, no explicit dependence upon $\eta_0$) but it is understood that the dipole $S$-matrices in the r.h.s. are given by the MV model whenever their rapidity arguments are smaller than $\eta_0$. With this prescription,  \eqref{bketafin} represents indeed a well-defined initial value problem with the initial condition formulated at $\eta=\eta_0$. For instance, the $S$-matrix  $\bar{S}_{\bx\bz}(\eta \minus \delta_{\bx\bz;r})$ is given by the solution to this equation when $\eta \minus \delta_{\bx\bz;r} > \eta_0$ and by the MV-model estimate $S^{(0)}_{\bx\bz}$ whenever $\eta \minus \delta_{\bx\bz;r} \le \eta_0$. In other terms, the MV model is not used {\it locally} in $\eta$, at $\eta=\eta_0$, as in a traditional initial value problem; it is also used at smaller rapidities $\eta <\eta_0$, i.e. for gluon fluctuations whose lifetimes are much smaller than $L$. This goes beyond the usual validity range of this model --- which, we recall, is intended for a color dipole whose coherence length is comparable to $L$ \cite{McLerran:1993ni,McLerran:1993ka}  --- and requires some explanation.

\comment{\begin{align}
	\label{Sglob}
	\bar{S}_{\bx\bz}(\eta \minus \delta_{\bx\bz;r}) = 
	\begin{cases}
		{\displaystyle	 } \text{from the solution to \eqn{bketafin} when} \ 
		&\eta \minus \delta_{\bx\bz;r} > \eta_0,
		\\*[0.4cm]
		{\displaystyle S^{(0)}_{\bx\bz}} \quad \text{(MV Model) when} \ 
		&\eta \minus \delta_{\bx\bz;r} \le \eta_0.
	\end{cases}
\end{align}
}

Before developing our argument, let us observe that the difference between Eqs.~\eqref{bketafin}  and \eqref{bketao} is important only at early rapidities $\eta\gtrsim\eta_0$, as expected on physical grounds. Indeed, the conditions $\eta\minus\ln(r^2/z^2)<\eta_0 <\eta$ impose a strong limitation on the size of the smallest daughter dipole,  $z^2 \ll r^2 \exp[-(\eta\minus\eta_0)]$; for such small values of $z$,
the combination of $S$-matrices inside the integrand of \eqref{bketafin} is rapidly vanishing, due to color transparency and to real vs. virtual cancellations. 
 
 A rigorous derivation for \eqn{bketafin} can be given by using the generalisation of the BK equation to the case of an extended target, as presented in the context of jet quenching
 \cite{Liou:2013qya,Iancu:2014kga}. Here however we shall present a simplified argument, which replaces most of the formal manipulations in \cite{Liou:2013qya,Iancu:2014kga} by heuristic considerations.
 
The question that we would like to address is as follows: given a dipole with size $r$ and rapidity $\eta\gtrsim\eta_0$, how is its scattering modified by the emission of soft gluons with very small sizes $z\ll r$ and hence short lifetimes $\tau_z\ll L$ ? The interesting case is when the parent dipole is not too large, $r\lesssim 1/Q_s$, where $Q_s=Q_s(\eta_0)$ is the saturation momentum in the MV model. Indeed, larger dipoles with $r\gg 1/Q_s$ are in the black disk regime already at tree-level and this cannot change after including the effects of the radiation. The small daughter dipole with $z\ll r \lesssim 1/Q_s$ will be in the color transparency regime and its $S$-matrix can be computed in the single scattering approximation. This is an important simplification. Based on it, we would like to argue that the contribution of such small fluctuations to the change of the $S$-matrix for the parent dipole can be evaluated as
 \begin{align}
        \label{bksmall}
        \hspace*{-0.5cm}
        \frac{\del \bar{S}(\eta, r)}{\del \eta}\bigg |_{\eta\gtrsim\eta_0} = 
       - 2 \, \frac{\abar}{2\pi}
        \int \frac{\dif^2 \bz \,r^2}{z^2 (\br \minus \bz)^2}\,
        \Theta\Big(\eta_0 \minus \eta + \ln\frac{r^2}{z^2} \Big)
        \Big[{S}^{(0)}(r)T^{(0)}(z)\Big],
\end{align}
where the overall factor of 2 accounts for the fact that the ``small dipole'' with size $z$ can be any of the two daughter dipoles (in the notations of \eqn{bketafin}, one either  has $z= |\bx-\bz|$, or $z= |\bz-\by|$). The step-function limits the integration to the short-lived fluctuations of interest for the problem at hand.
 The $S$-matrix ${S}^{(0)}(r)$ and the single-scattering amplitude $T^{(0)}(z)$ which occur in the r.h.s. are computed according to the MV model, as appropriate for $\eta\sim\eta_0$.

The $S$-matrix ${S}^{(0)}(r)$ is the probability for the parent dipole to survive the medium in a color singlet state and without radiating; this is shown in \eqn{SMV} that we conveniently rewrite here as
 \beq\label{S0}
 {S}^{(0)}(r)\,=\,\rme^{-\frac{1}{4}L \hat q r^2}\,. 
 \eeq
Our present notation is meant to emphasise  that the saturation scale in the exponent is proportional to the width $L$ of the medium: $Q_s^2=\hat q L$. (The proportionality coefficient $\hat q$ is logarithmically dependent upon $r$, as shown in \eqn{SMV}, but this dependence is inessential for the present purposes.) Furthermore, $T^{(0)}(z)$ is the scattering amplitude for a small dipole with size $z\ll 1/Q_s$ which propagates throughout the {\em whole} medium:
 \beq\label{T0}
 T^{(0)}(z) \,=\,\frac{1}{4}L \hat q z^2\,.
 \eeq
 The emergence of this global amplitude may look surprising in view of the fact that the fluctuation has only a very short lifetime $\tau_z\simeq 2q^+z^2\ll L$. But in the context of \eqn{bksmall}, the quantity $T^{(0)}(z) $ is not meant to represent the scattering amplitude of just {\it one} fluctuation, but the cumulated effect of an arbitrary number of such fluctuations which can occur anywhere inside the medium.

\begin{figure}[t] 
\centerline{\includegraphics[width=0.75\textwidth]{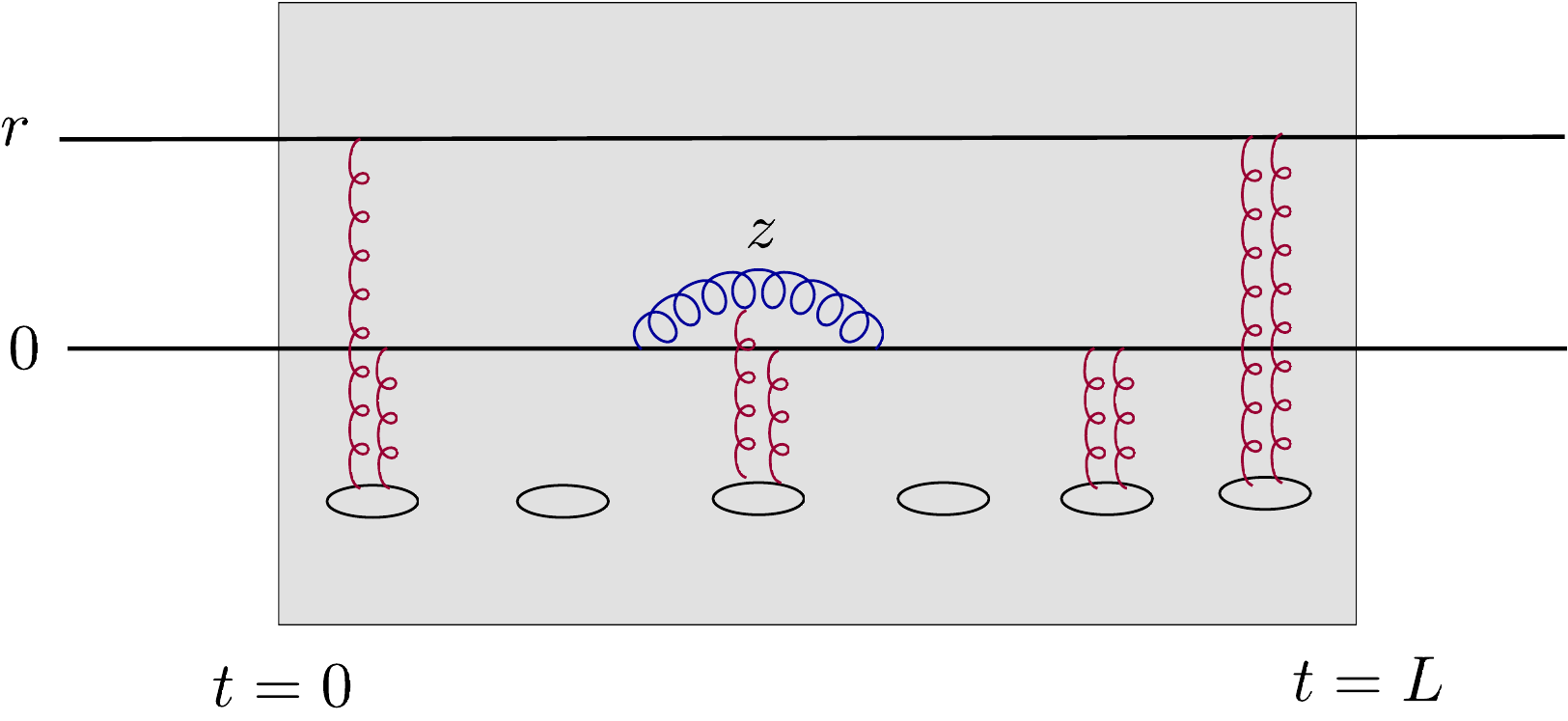}}
 \caption{\small Pictorial representation of the quantum evolution of the dipole scattering
 amplitude via the emission of a small-size, short-lived, soft gluon fluctuation. The dipole with size
 $r$ propagates through the target with longitudinal extent $L$ and scatters off the valence quarks via
 2-gluon exchanges (in the spirit of the MV model). The gluon fluctuation with size $z\ll r$ undergoes a single scattering off the valence quark that it overlaps with in time.}
 \label{fig:dip}
\end{figure}

To better understand that, it might be useful to have a glance at Fig.~\ref{fig:dip} which illustrates
 the dynamics under consideration: the dipole with size $r$ undergoes successive independent scatterings off the color sources (``valence quarks'') in the nucleus and also radiates a very short-lived
gluon fluctuation, which can overlap in the longitudinal direction (and interact) with only one such a color source. But this fluctuation can be emitted anywhere inside $L$, that is, it can renormalize the scattering between the dipole and any of the color sources.  The change in the dipole $S$-matrix associated with a fluctuation localized around the color source $i$ can be estimated as  
\begin{align}
\label{Schain}
\Delta S(r) &\,= S_{Lt_2}(r)S_{t_2t_1}(z)S_{t_2t_1}(|\br-\bz|)S_{t_10}(r) - S_{L0}(r)\,\simeq\,
S_{L0}(r)\big[S_{t_2t_1}(z)-1\big]\nn &\,\simeq\,-S(r)\frac{1}{4} \tau_z \hat q z^2
\end{align}
in rather schematic notations where $t_1$ ($t_2$) is the time\footnote{These ``time'' variables are truly values of the light-cone coordinate $x^+$, which plays the role of a time for the right-moving projectile.}
when the fluctuation is emitted (reabsorbed), with $t_1 < t_i < t_2$, $t_2\minus t_1=\tau_z$ is the gluon lifetime, $S_{t_10}(r)$ is the $S$-matrix accummulated by the dipole of size $r$ during its propagation from $t=0$ up to $t=t_1$, etc; so, in particular $S_{L0}(r)=S(r)$. (We have omitted the upper script $(0)$ on the MV-model $S$-matrices,  to simplify notations.) Furthermore, we have used the fact that, in the interval $t_1 < t < t_2$, the original dipole is replaced by the two daughter dipoles, which however are very asymmetric: $|\br-\bz|\simeq r\gg z$. Hence, $S_{t_2t_1}(|\br-\bz|)\simeq S_{t_2t_1}(r)$, which enabled us to reconstruct $S_{L0}(r)$ as $S_{L0}(r)=S_{Lt_2}(r)S_{t_2t_1}(r)S_{t_10}(r)$ in the first, ``real'', term in the r.h.s. of \eqn{Schain}. The final estimate in \eqn{Schain} follows after using the single-scattering approximation for $S_{t_2t_1}(z)$. This final result is independent of the position $t_i$ of the source $i$ and can be used to deduce the rate for the change in the $S$-matrix: 
\beq
\frac{\rmd S}{\rmd t}\,\simeq\,\frac{\Delta S}{\tau_z} = -S(r)\frac{1}{4} \hat q z^2
\eeq
The global change integrated over all times $0 < t < L$ is obtained by multiplying this rate by a factor of $L$ and thus yields the product $S(r)T(z)$ visible in the integrand of \eqn{bksmall}.

To finally justify \eqn{bketafin}, we observe that for early rapidities and short-lived fluctuations, such that $\eta -\ln(r^2/z^2)<\eta_0 \lesssim \eta$, the combination of $S$-matrices in the integrand of  \eqn{bketafin} is indeed equivalent to that in \eqn{bksmall} to the accuracy of interest.

\eqn{bketafin} is one of our main results in this paper: this is the physically most complete version of our resummation of double-collinear logarithms in the BK evolution with respect to the target rapidity $\eta$. Clearly, the extension of this new equation to full NLO accuracy is the same as presented in Sect.~\ref{sec:match}: it suffices to replace the ``BK-like'' equation  appearing in the first line of \eqn{nlobkres} by \eqn{bketafin}.

\begin{figure}[t]
\centerline{\hspace*{-.3cm}
  \includegraphics[width=0.46\textwidth]{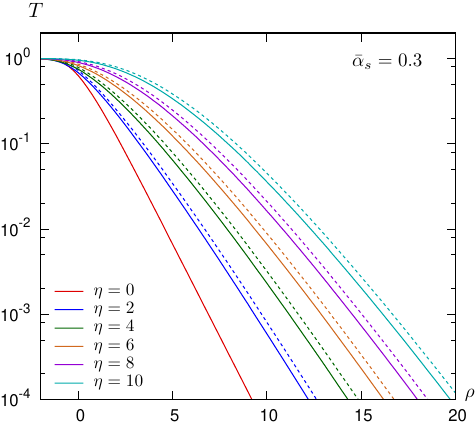}\qquad
\includegraphics[width=0.46\textwidth]{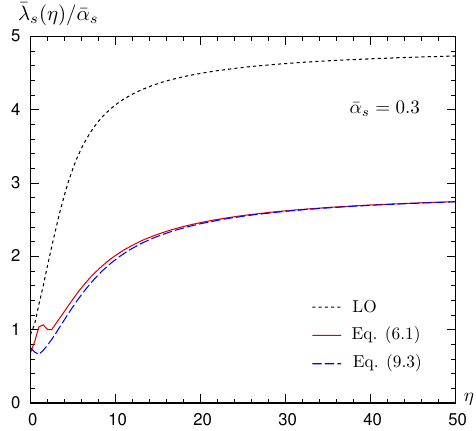}}
 \caption{\small Comparing the numerical solutions to the ``canonical'' non-local equation  \eqref{bketato} (the same as \eqref{bketao} with $\eta_0=0$) and the new \eqn{bketafin}, for 
 $\abar=0.3$.
  Left: The respective predictions for the dipole amplitude: continuum lines correspond to \eqn{bketafin} and dashed lines to \eqn{bketato}.  Right: The speed of the fronts as a function of $\eta$; for illustration, we also show the respective prediction of the LO BK equation.}
 \label{fig:newIC}
\end{figure}

In practice though, we do not expect significant changes when using \eqn{bketafin}  instead of \eqref{bketao} (including at NLO accuracy). Indeed, as already stressed, the difference refers at most to the very early evolution. To demonstrate that, we compare the numerical solutions to  Eqs.~\eqref{bketafin}  and \eqref{bketao} in Fig.~\ref{fig:newIC}. One can see a small difference  in the speed of the saturation fronts at early rapidities $\eta\lesssim 4$, but the asymptotic predictions for
$\bar\lambda_s$ are indeed the same, as expected. In Appendix \ref{app:delay} we study a 0-dimensional toy-model with delay and show indeed that the asymptotic speed does not depend on the detailed way that one starts the evolution.

\section{Conclusions and perspectives}
\label{sec:conc}

Our main observation in this paper is that, when studying the high-energy evolution in pQCD beyond leading-order, the BK equation for dilute-dense scattering must be formulated as an evolution with respect to the rapidity $\eta$ of the dense target, and not with the rapidity $Y$ of the dilute projectile. 

This is first of all needed for the physical interpretation of the results. For instance, the structure functions for deep inelastic scattering must be computed in terms of Bjorken $x$ to admit their standard interpretation in terms of parton distributions in the target. Similarly, the variable $x$ for the evolution of the saturation momentum $Q_s(x)$ must refer to the longitudinal momentum of the target in order to have a meaningful interpretation in terms of non-linear effects in the gluon distribution of the target.

However, our present analysis shows that the choice of a rapidity variable goes beyond such issues of physical interpretation: the use of the target rapidity $\eta$ is already compulsory in the formulation of the evolution equation at NLO and beyond. This is due to the fact that the perturbative expansion for 
the evolution with $Y$ is afflicted with severe instabilities, whose consequences remain out of control  --- in the sense of showing a strong scheme-dependence after translating the results in terms of $\eta$ --- even after the resummation of the dominant radiative corrections --- those enhanced by  anti-collinear double logarithms.  The ultimate reason why perturbation theory in $Y$ is so ill-behaved is because the effects of the anti-collinear logarithmic corrections to the BK kernel are amplified by the typical, ``hard-to-soft'', evolution for dilute-dense scattering. 

Besides this fundamental problem, we have identified additional difficulties with the anti-collinear resummations in $Y$, which refer to the formulation of the initial-value problem in the presence of the constraint of time-ordering. Albeit of more ``technical'' nature, these difficulties have a serious impact on the applications to the phenomenology (they affect the results at low and intermediate rapidities) and we have not been able to offer solutions to them in practice.

We have shown that all these difficulties can be circumvented by working directly with the rapidity $\eta$ of the dense target. The general idea is indeed natural, given that the evolution in $\eta$ guarantees the proper time-ordering of the soft gluon emissions and thus avoids the emergence of anti-collinear double-logarithmic corrections. But albeit natural, this strategy is still highly non-trivial, as it requires several clarifications and developments that we have successively addressed in this paper.

First, one needs the NLO BK equation for the evolution in $\eta$; in Sect.~\ref{sec:NLOeta}, we have shown that this can be easily obtained via a change of variables from the corresponding equation in $Y$, which is known. Second, albeit void of double {\it anti}-collinear logarithms, the NLO corrections to the BK equation in $\eta$ contain double {\it collinear} logs, which matter for the atypical,  ``soft-to-hard'', emissions permitted by the BFKL diffusion. Their effects accumulate with increasing $\eta$
and eventually lead to instabilities, which are however milder than for the evolution with $Y$ (cf. the discussion in Sect.~\ref{sec:nloinst}). In Sect.~\ref{sec:shift}, we have demonstrated that the evolution in $\eta$ stabilises after all-order collinear resummations, which are formally similar to the anti-collinear  ones in $Y$, but differ from the latter in two important aspects: \texttt{(i)} they lead to well-posed initial-value problems (up to minor subtleties related to the non-locality in rapidity, which are clarified in Sect.~\ref{sec:IC}), and  \texttt{(ii)} their predictions show a reasonably small scheme-dependence, of the order of the expected perturbative accuracy for the resummations. 

Interestingly, we found that the collinear resummations in $\eta$ affect not only the linear dynamics in the BFKL (or weak scattering) regime, but also the non-linear dynamics in the vicinity of unitarity/saturation. This observation allowed us to discard one of the resummed equations --- the only one to be  {\it local} in $\eta$, cf. \eqn{collbk0} ---,  which is not correct in the approach to saturation. Furthermore, among the {\it non-local} (in $\eta$) versions of the resummation that we proposed, and which differ from each other in the precise structure of the rapidity shift, we have selected only two: those which match better the detailed structure of the NLO corrections in $\eta$ (cf. Sect.~\ref{sec:shift} and the beginning of Sect.~\ref{sec:match}). For these two prescriptions, we have also presented the extension of the corresponding resummed equation to full NLO accuracy: this is \eqn{nlobkres}, which may be seen as our main new result in this paper.

In order to gain more insight in the physical consequences of the resummation, we performed rather exhaustive analytic and numerical studies using one of the non-local evolution equations aforementioned, namely the ``canonical'' equation displayed in Eqs.~\eqref{bketato} or \eqref{bketafin}. These two equations differ from each other only in the formulation of the initial value problem, which is indeed subtle in the presence of a non-locality in the evolution ``time'' $\eta$. \eqn{bketafin} is more complete on physical grounds (see Sect.~\ref{sec:IC} for details), but in practice the differences w.r.t. \eqn{bketato} are only tiny,  as demonstrated by the numerical comparison in Fig.~\ref{fig:newIC}.

Via analytic studies of \eqn{bketato} in the BFKL regime, in Sect.~\ref{sec:BFKLeta}, we have clarified the relation between our present approach and the collinear resummations performed earlier, in the context of the NLO BFKL equation.  In Sect.~\ref{sec:etafronts} we have studied the solution to \eqn{bketato}, via both analytic and numerical methods, and found that the resummation has indeed important effects in slowing down the evolution and also changing the shape of the saturation front (the ``anomalous dimension at saturation''), while keeping the property of geometric scaling. These effects remain sizeable in the presence of a running coupling and contribute to producing an effective ($\eta$-dependent) saturation exponent which is consistent with the phenomenology (see  Fig.~\ref{fig:rc}. (right)).

To summarize, our preliminary studies based on the collinearly-improved version of the BK equation alone --- where by ``collinear improvement'' we now mean the collinear resummations in $\eta$ leading to Eqs.~\eqref{bketato} or \eqref{bketafin} --- already look promising for the phenomenology. It would be of course very interesting to confirm such good expectations via explicit applications of this equation to the phenomenology of deep inelastic scattering at HERA and that of particle production at forward rapidities in proton-proton and proton-nucleus collisions at the LHC. The results thus obtained (say, in terms of the parametrisation of the initial condition for the resummed equation) could then serve as a basis for predictions for future studies of semi-hard processes at the Electron-Ion Collider.

But the ultimate test of the usefulness of these resummations and, more generally, of our current understanding of the high-energy evolution in perturbative QCD would consist in solving the resummed equation with full NLO accuracy, that is, \eqn{nlobkres}, for which we expect a very good accuracy and hence highly reliable predictions: for a fixed coupling, one expects an error of $\order{\alpha_s^3}$ in the prediction for the saturation exponent, meaning that the relative error $\delta\bar\lambda_s/\bar\lambda_s\simeq  \abar^2$ should be only $\sim 10\% $ for $\abar=0.3$. Of course, solving this full NLO equation would be a much more challenging task than our current solutions to the collinearly-improved BK equation \eqref{bketato}, due to the complicated non-linear and non-local (notably in the transverse plane) structure of  \eqn{nlobkres}. Yet, this should be doable in practice, since  \eqn{nlobkres} does not look more complicated than the respective equation for the evolution with $Y$
 \cite{Iancu:2015joa} and which has been numerically solved in Ref.~\cite{Lappi:2016fmu}.

\section*{Acknowledgments} 

While working on this project, we have benefited from useful discussions with many of our colleagues. In particular we would like to express our special thanks to Giovanni Chirili, Yuri Kovchegov and Tuomas Lappi.  A.H.M.~and D.N.T.~would like to acknowledge l'Institut de Physique Th\'eorique de Saclay for hospitality during the early stages of this work. E.I, A.H.M. and D.N.T. would also like to thank the Institute for Nuclear Theory at the University of Washington and the organizers of the program ``Probing Nucleons and Nuclei in High Energy Collisions (INT-18-3)'' for their hospitality and the Department of Energy for partial support while this work was close to completion. The work of B.D. and part of the work of E.I. are supported by the Agence Nationale de la Recherche project ANR-16-CE31-0019-01. The work of A.H.M.~is supported in part by the U.S. Department of Energy Grant \# DE-FG02-92ER40699. 

\appendix
 
\appendix
 
\section{Integrals for the BFKL evolution at NLO in the $\eta$-representation}
\label{app:tint}

\subsection{NLO BFKL}
\label{app:nlobfkl}

In the first part of this Appendix we show how one goes from \eqn{deltatxy} to \eqn{deltatxyfin} (the latter being necessary to obtain the NLO BFKL equation in the $\eta$-representation), by performing one of the two 2-dimensional integrations. 

The key point is to make a suitable change of variables, so that all three terms in \eqn{deltatxy} involve, for example, $\bar{T}_{\bz\by}$. To this end, in the first term we let $\bz \to \bu+\by-\bz$, in the third one we let $\bz \leftrightarrow \bu$, while we leave the second one as it is. Then one sees that the first two terms become identical and putting all three terms together we have
\begin{align}
	\label{deltatxyapp1}
	\hspace*{-0.6cm}
	\Delta \bar{T}_{\bx\by} 
 	= \frac{\abar^2}{2\pi^2}
 	\int \frac{\dif^2 \bz\, (\bx \minus \by)^2 }{(\bx \minus
 	\bz)^2  (\bz \minus \by)^2}\, 
 	\bar{T}_{\bz\by}
 	\int \frac{\dif^2 \bu}{(\bu\minus\bz)^2}
 	\bigg[
 	\frac{(\bx \minus \bz)^2 }{(\bx \minus
 	\bu)^2}\,
 	\ln\frac{(\bu\minus\by)^4}{(\bx\minus\by)^4}
 	-\frac{(\bz \minus \by)^2 }{(\bu \minus \by)^2}\,
 	\ln\frac{(\bz\minus\by)^2}{(\bx\minus\by)^2}
 	\bigg]. 
\end{align}
Each of the two terms in the above integration over $\bu$ is singular, thus we will try to reshuffle the terms in order to get integrations that individually converge. To this end, we decompose the logarithm appearing in the first term in \eqn{deltatxyapp1} as
\begin{equation}
	\label{logdec}
	\ln\frac{(\bu\minus\by)^4}{(\bx\minus\by)^4}
	=\ln\frac{(\bu\minus\by)^4}{(\bx\minus\by)^2 (\bz\minus\by)^2}
	+ \ln\frac{(\bz\minus\by)^2}{(\bx\minus\by)^2}. 
\end{equation}
Employing Eq.~(143) in \cite{Balitsky:2009xg}, or equivalently Eq.~(A.26) in \cite{Hatta:2017fwr}, we get 
\begin{equation}
	\label{uint1}
	\int \frac{\dif^2 \bu\, (\bx \minus \bz)^2 }{(\bx \minus
 	\bu)^2  (\bu \minus \bz)^2}\,
 	\ln\frac{(\bu\minus\by)^4}{(\bx\minus\by)^2 (\bz\minus\by)^2}
 	 = \pi \ln^2 \frac{(\bz \minus \by)^2}{(\bx \minus \by)^2}.
\end{equation}
Notice that although the integration gives two individual singularities originating from $\bu=\bx$ and $\bu=\bz$, they eventually cancel each other leaving a finite integral. One could see this directly at the level of the integrand, by splitting the logarithm into two equal parts and making the change of variable $\bu \to \bx+\bz-\bu$ in one of the two parts. Since the dipole kernel is invariant under this change, one finds
\begin{equation}
	\label{uint1b}
	\int \frac{\dif^2 \bu\, (\bx \minus \bz)^2 }{(\bx \minus
 	\bu)^2  (\bu \minus \bz)^2}\,
 	\ln\frac{(\bu\minus\by)^4}{(\bx\minus\by)^2 (\bz\minus\by)^2}
 	 = \int \frac{\dif^2 \bu\, (\bx \minus \bz)^2 }{(\bx \minus
 	\bu)^2  (\bu \minus \bz)^2}\,
 	\ln\frac{(\bu\minus\by)^2 (\bx\minus\by+\bz\minus\bu)^2}{(\bx\minus\by)^2 (\bz\minus\by)^2},
\end{equation}
in which it is clear that the logarithm cancels the singularities at $\bu=\bx$ and $\bu=\bz$. Now, by using Eq.~(A.22) in \cite{Hatta:2017fwr} we have
\begin{equation}
\label{uint2}
	\ln\frac{(\bz\minus\by)^2}{(\bx\minus\by)^2}
 	\int  \dif^2 \bu 
 	\bigg[\frac{(\bx \minus \bz)^2 }{(\bx \minus
 	\bu)^2  (\bu \minus \bz)^2} - 
 	\frac{(\bz \minus \by)^2 }{(\bz \minus
 	\bu)^2  (\bu \minus \by)^2}\bigg] = 
 	-2 \pi \ln \frac{(\bz \minus \by)^2}{(\bx \minus \by)^2}
 	\ln \frac{(\bz \minus \by)^2}{(\bx \minus \bz)^2}.
\end{equation}
The two terms in Eqs.~\eqref{uint1} and \eqref{uint2} are combined together to give for the $\bu$-integration in Eq.~\eqref{deltatxyapp1}
\begin{equation}
	\label{uint}
	\int  \dif^2 \bu\,  \cdots
	= \pi  \ln \frac{(\bz \minus \by)^2}{(\bx \minus \by)^2}
 	\ln \frac{(\bx \minus \bz)^2}{(\bx \minus \by)^2}
 	- \pi \ln \frac{(\bz \minus \by)^2}{(\bx \minus \by)^2}
 	\ln \frac{(\bz \minus \by)^2}{(\bx \minus \bz)^2}.	
\end{equation}
By further symmetrizing the $\bz$-integrand in \eqn{deltatxyapp1} in $\bx-\bz$ and $\bz-\by$, we finally arrive at \eqn{deltatxyfin}. One can calculate the characteristic function of each of the two terms in \eqn{deltatxyfin} to find
\begin{align}
	\label{domega1}
	&\Delta \omega^{(1)} = \frac{\abar^2}{2}\,\chi_0(\gamma)\chi_0'(\gamma)+
	\frac{\abar^2}{4}\,\chi_0''(\gamma),
	\\*[0.2cm]
	\label{domega2}
	&\Delta \omega^{(2)} = \frac{\abar^2}{2}\,\chi_0(\gamma)\chi_0'(\gamma)-
	\frac{\abar^2}{4}\,\chi_0''(\gamma),
\end{align}
and their sum agrees with \eqn{chichip} as it should.

\subsection{NLO piece of the local equation \eqref{bketato}}
\label{app:local}

In the second part of this Appendix we uncover the double logarithm for small daughter dipoles which is contained in the non-local equation Eq.~\eqref{bketato} at order $\abar^2$. 

We start from Eq.~\eqref{a2shift}, we linearize to get the equation for the amplitude and then we follow the exact same steps described above Eq.~\eqref{deltatxyapp1}. It becomes obvious that we arrive at an equation very similar to Eq.~\eqref{deltatxyapp1}, except that the logarithms are replaced by $\delta$-shifts, more precisely we have
\begin{align}
    \label{deltatshifta2}
	\Delta \bar{T}_{\bx\by} 
 	= \frac{\abar^2}{2\pi^2}
 	\int \frac{\dif^2 \bz\, (\bx \minus \by)^2 }{(\bx \minus
 	\bz)^2  (\bz \minus \by)^2}\, 
 	\bar{T}_{\bz\by}
 	\int \frac{\dif^2 \bu}{(\bu\minus\bz)^2}
 	\bigg[-2\,
 	\frac{(\bx \minus \bz)^2 }{(\bx \minus
 	\bu)^2}\,
 	\delta_{\bu\by;r}
 	+\frac{(\bz \minus \by)^2 }{(\bu \minus \by)^2}\,
 	\delta_{\bz\by;r}
 	\bigg]. 
\end{align}
In the regime where the shifts are non-zero, the above is identical to Eq.~\eqref{deltatxyapp1}. However the two equations are very different in the regime of large dipoles. Since the shift simply vanishes, Eq.~\eqref{deltatshifta2} does not lead to any large double logarithms. On the contrary, Eq.~\eqref{deltatxyapp1} does contain such logs, which in fact are necessary to cancel those in the second term in Eq.~\eqref{nlobketa} (when linearized).

Let us consider the integration over $\bu$ in Eq.~\eqref{deltatshifta2}. Each of the two terms there is singular, the first when $\bu=\bz$ and the second when $\bu=\bz$ or $\bu=\by$. Nevertheless, it is rather easy to see that these singularities cancel when summed, and therefore the $\bu$-integration leads to a finite function of $\bx$, $\by$ and $\bz$. Although it appears difficult to perform exactly the integration with the shift in Eq.~\eqref{delta1}, we can make progress by looking at small daughter dipoles, and after all this is the regime that we are interested in.

\begin{figure}[t]
\begin{center}
\includegraphics[width=0.5\textwidth]{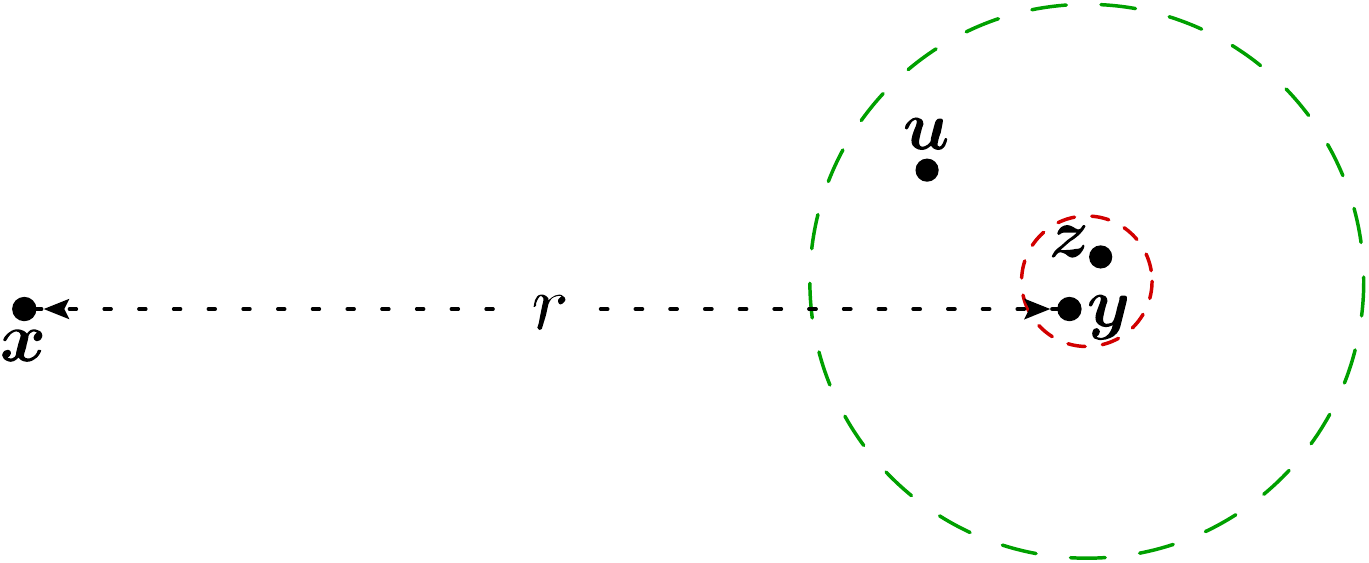}
\end{center}
\caption{\small The transverse coordinates in the strongly ordered regime defined in Eq.~\eqref{uorder}, which in the end leads to the double logarithmic contribution. The points $\bx$, $\by$ and $\bz$ are fixed and are such that $|\bz-\by| \ll r$. The point $\bu$ should be well outside the small (red) circle with radius $\sim |\bz-\by|$. At the same time it should stay inside the larger (green) circle whose radius is much smaller than $r$ but much larger than $|\bz-\by|$.}
\label{fig:uorder} 
\end{figure}

Thus, since $\bar{T}_{\bz\by}$ appears on the r.h.s.~of Eq.~\eqref{deltatshifta2}, we assume $|\bz-\by| \ll r$. Then the dominant contribution to the $\bu$-integration arises from the strongly ordered regime
\begin{equation}
	\label{uorder}
	|\bz-\by| \ll |\bu-\by| \ll r,
\end{equation}
and in particular this leads to the approximate equalities 
\begin{equation}
	\label{approxsizes}
	|\bx-\bz| \simeq |\bx-\bu| \simeq r
	\quad \& \quad 
	|\bu-\bz| \simeq |\bu - \by|.
\end{equation}
An example of such a configuration and the allowed integration region are shown in Fig.~\ref{fig:uorder}. We now see that only the first term in the square bracket in Eq.~\eqref{deltatshifta2} leads to a logarithmic integration and we have
\begin{equation}
	\label{uint3}
	\int \dif^2\bu\, \cdots \simeq 
	- 2 \pi \int_{|\bz-\by|^2}^{r^2}
	\frac{\dif |\bu-\by|^2}{|\bu-\by|^2} \ln \frac{r^2}{|\bu-\by|^2} = 
	-\pi \ln^2 \frac{r^2}{|\bz-\by|^2},  
\end{equation}
which is in agreement with the $\bz \to \by$ limit of \eqn{uint} as it should. Finally, using the above, Eq.~\eqref{deltatshifta2} leads to
\begin{equation}
	\label{deltatorder}
	\Delta\bar{T}(r^2) \simeq - \frac{\abar^2}{2}
	\int_0^{r^2} \frac{\dif z^2}{z^2}\,\ln^2\frac{r^2}{z^2}\, 
	\bar{T}(z^2),
\end{equation}
where in the integrand we have let $|\bz-\by| \to z$. This is indeed the double logarithmic NLO correction to the kernel in the $\eta$-representation as exhibited in Eq.~\eqref{bfkllog}.

\section{Oscillations of the BFKL solution at large $\eta$}
\label{app:osc}

In this Appendix we start from the general BFKL solution given in \eqn{tbfklsol} and we perform a saddle-point integration in order to find an analytical expression valid in the regime $0 <\rho \ll \eta$.

This is the regime of the Pomeron intercept problem and we will see that the solution is unstable in the sense that it develops oscillations in $\eta$ (and also in $\rho$), except in the case that $\abar$ is extremely small\footnote{One may argue that neither $Y$, nor $\eta$ are the correct rapidity variables. In fact the correct choice should be $(Y+\eta)/2 = \ln(s/QQ_0)$, with $Q\sim 1/r$. With such a choice $\gammap$ is complex, but $\omegap$ is real due to the symmetry of the kernel under $\gamma \leftrightarrow 1-\gamma $. Thus, there are no oscillations in $Y$, but only in $\rho$ (since $\gammap$ is still complex) \cite{Ross:1998xw}. In fact this happens only for large values of $\rho$, so the issue is kind of milder, but in principle the oscillation is not a nice feature anyway. Still, in the saturation problem, one has to solve $\bar{\omega}'(\gamma) = \bar{\omega}(\gamma)/\gamma$ which is asymmetric in $\gamma \leftrightarrow 1-\gamma $, no matter what the choice of the energy scale is.}.
In the asymptotic regime of interest, one must solve the saddle point condition
\begin{equation}
	\label{omegap}
   \bar{\omega}'(\gamma_{\mathbb{P}}) =0. 
\end{equation}
At leading order $\gamma_{\mathbb{P}}=1/2$ with $\bar{\omega}(\gamma_{\mathbb{P}}) = 4(\ln2) \abar$. At NLO, at the level of \eqn{omegabar}, and given the structure in \eqn{omegabarprime}, one readily sees that there is a real solution only for very small values of $\abar$. More precisely, when $\abar$ is smaller than the critical value $\abar^{\rm cr} \simeq 0.032$ in \eqn{abarcrit}, then \eqn{tbfklsol} admits a well-defined solution. When $\abar > \abar^{\rm cr}$, the asymptotic dynamics is dominated by two complex solutions to \eqn{omegap} which are conjugate to each other and thus guarantee that the solution is real. It suffices to consider only one of the two saddle points, calculate the respective contribution and then add its complex conjugate. We approximate the exponent in \eqn{tbfklsol} as
\begin{equation}
	\label{epsilonp}
	\mcal{E}(\gamma) \simeq \bar{\omega}(\gammap) \eta
	+\frac{\bar{\omega}''(\gammap)\eta}{2} (\gamma - \gammap)^2 - \gamma\rho.
\end{equation}
We follow the integration contour
\begin{equation}
	\label{pomcontour}
	\gamma = \gammap + r \rme^{\rmi \thetap}\, 
	\Rightarrow\,\dif \gamma = \rme^{\rmi \thetap} \dif r
\end{equation}
where $\thetap$ is a fixed angle to be shortly determined. In order to have a compact notation let us also define
\begin{align}
	\label{app:omegap}
	&\bar{\omega}(\gammap) \equiv \omegap
	\\
	&\bar{\omega}''(\gammap) = |\bar{\omega}''(\gammap)| 
	\rme^{\rmi \betap} \equiv \Dp \rme^{\rmi \betap}.
\end{align}
To avoid any confusion we point out that $\gammap$ and $\omegap$ are complex, while $\Dp$ and $\betap$ are real. These constants are fixed by the saddle point condition and they will all appear in the solution. Now \eqn{epsilonp} reads
\begin{equation}
	\label{epsilonp2}
	\mcal{E}(\gamma) \simeq \omegap \eta - \gammap \rho
	+\frac{\Dp \eta}{2}\, \rme^{\rmi(\betap+2\thetap
	)} r^2 - 
	\rho \rme^{\rmi \thetap} r.
\end{equation}
The steepest descent method requires that the amplitude decreases as fast as possible as we move away from the saddle point (while obviously staying on the contour, here defined in \eqn{pomcontour}). This condition uniquely selects $\thetap$ as
\begin{equation}
	\label{app:thetap}
	\betap +2 \thetap = \pi\, \Rightarrow \,
	\thetap = \frac{\pi}{2} - \frac{\betap}{2} 
	\quad \& \quad
	\rme^{\rmi \thetap} = \rmi \rme^{-\rmi \betap/2} 
\end{equation}
and \eqn{epsilonp2} becomes
\begin{equation}
\label{epsilonp3}
	\mcal{E}(\gamma) \simeq \omegap \eta - \gammap \rho
	-\frac{\Dp \eta}{2}\,  r^2 - 
	\rmi  \rme^{-\rmi \betap/2}  \rho r.
\end{equation}	
The next step is to shift the integration variable $r$ in such a way that the linear term in \eqn{epsilonp3} vanishes. This can be achieved by letting $r\to r+\rmi \rme^{-\rmi\betap 2}\rho/
(\Dp \eta)$ and the exponent becomes
\begin{equation}
\label{epsilonp4}
	\mcal{E}(\gamma) \to \omegap \eta - \gammap \rho
	-\frac{\rme^{\rmi \betap}\rho^2}{2 \Dp \eta}
	-\frac{\Dp \eta}{2}\,  r^2.
\end{equation}
We are finally able to perform the integration in \eqn{tbfklsol}. Putting everything together, including the prefactor evaluated at $\gammap$, we arrive at 
\begin{equation}
	\label{app:tsaddlep}
	\bar{T}(\eta,\rho) = \bar{T}_0(\gammap) \rme^{-\rmi\betap/2}
	\frac{\rme^{-\textstyle\frac{\rme^{\rmi \betap}\rho^2}{2\Dp\eta}}}{\sqrt{2\pi\Dp\eta}}\,\rme^{\omegap \eta- \gammap \rho}
	\,+\, \textrm{c.c.}
\end{equation}
where c.c.~stands for complex conjugate. 

For simplicity, let us assume the initial condition $\bar{T}_0(\gamma) = \alpha_s^2/[\gamma^2 (1-\gamma)^2]$, which corresponds to dipole-dipole scattering averaged over the target dipole area. For the convenience of the numerics when solving Eq.~\eqref{tbfklsol}, we retain only the pole structure of the characteristic function, that is, we shall neglect the regular pieces in Eq.~\eqref{epsilon}. Then, using the same simplified characteristic function for a proper comparison, we readily obtain the saddle point solution from Eq.~\eqref{app:tsaddlep}. As we have already shown in Fig.~\ref{fig:pomsol} the agreement of the two oscillating solutions is excellent in the regime of validity. Notably, let us take (for example) $\abar=0.2$ for which we find $\omegap = 0.591 + 0.225\rmi$. From the dominant exponential in \eqn{app:tsaddlep} we easily determine the half period of the oscillation in $\eta$ which is $\tau_{\eta} = \pi/\Im \omegap = 13.9$ which is clearly seen in both solutions in Fig.~\ref{fig:pomsol}.

\section{The BFKL solution around $\hat{\rho}(\eta)$}
\label{app:airy}

In this Appendix we would like to improve our BFKL solution in the regime $\rho > \hat{\rho}(\eta)$. To this end, we shall extent the calculation done in Sect.~\ref{sec:nloinst} and obtain two subleading terms to be added on the r.h.s. of Eqs.~\eqref{rhohat} and \eqref{tbfklfinal}.

Starting from \eqn{tbfklsol} we now expand $\bar{\omega}(\gamma)$ around $\gamma_\rmc$ to all orders, i.e.
\begin{equation}
	\label{app:omegaall}
	\bar{\omega}(\gamma) = \bar{\omega}(\gamma_{\rm c}) +
	(\gamma - \gamma_{\rm c}) \bar{\omega}'(\gamma_{\rm c}) 
	 + \frac{1}{6}\, 
	 (\gamma - \gamma_{\rm c})^3 \bar{\omega}'''(\gamma_{\rm c}) 
	 +\sum_{m=4}^{\infty} \frac{(\gamma - \gamma_{\rm c})^m \bar{\omega}^{(m)}(\gamma_{\rm c})}{m!}.
\end{equation}
Similarly we expand the initial condition $\bar{T}_0(\gamma)$ as
\begin{equation}
	\label{app:tzeroall}
	\bar{T}_0(\gamma) = \bar{T}_0(\gamma_\rmc) + 
	\sum_{l=1}^{\infty}\frac{(\gamma - \gamma_{\rm c})^m 
	\bar{T}_0^{(l)}(\gamma_\rmc)}{l!}.
\end{equation}
We furthermore introduce for our convenience the variable
\begin{equation}
	\label{app:z}
	z \equiv  \frac{D_\rmc [\rho -\bar{\omega}'(\gamma_{\rm c} \eta]}{\eta^{1/3}},
\end{equation}
and by using the integration variable $t$ defined in \eqn{tdef} in the main text we write the amplitude as
\begin{equation}
	\label{app:tbfklser}
	\hspace*{-0.3cm}
	\bar{T}_0(\eta,\rho) = \frac{\bar{T}_0(\gamma_\rmc)D_\rmc}{\eta^{1/3}} 
	\exp \left[ \bar{\omega}(\gamma_{\rm c})\eta \!-\! \gamma_\rmc \rho \right]
	\int \frac{\dif t}{2 \pi \rmi}\,
	\left(1 \!+\! \sum_{l=1}^{\infty} \frac{b_l t^l}{\eta^{l/3}}\right)
	\exp\left(\!-z t +\!\frac{t^3}{3} \right)
	\exp\left(\sum_{m=4}^{\infty} \frac{\kappa_m t^m	}{\eta^{\frac{m-1}{3}}} \right).
\end{equation}
The expansion coefficients appearing in the above are defined as
\begin{equation}
	\label{app:bkappa}
	b_l = \frac{\bar{T}_0^{(l)}(\gamma_\rmc) D_\rmc^l}{\bar{T}_0 (\gamma_\rmc) l!}
	\quad \text{and} \quad
	\kappa_m = \frac{\bar{\omega}^{(m)}(\gamma_\rmc) D_\rmc^m}{m!}.
\end{equation}
No approximation has been done so far, that is, \eqn{app:tbfklser} is the exact generalization of \eqn{tbfkl1} to all orders in the relevant Taylor expansion. Now from the two series in \eqn{app:tbfklser} we neglect terms which fall faster than $1/\eta^{2/3}$ as $\eta$ becomes large, so we have for these two factors
\begin{equation}
	\label{app:etaexpand}
	\left(1 + \frac{b_1 t}{\eta^{1/3}} + \frac{b_2 t^2}{\eta^{2/3}}
	\right)
	\exp \left( 
	\frac{\kappa_4 t^4}{\eta^{1/3}} + \frac{\kappa_5 t^8}{\eta^{2/3}}
	\right)
	\simeq 1 + \frac{b_1 t +\kappa_4 t^4}{\eta^{1/3}}
	+ 
	\frac{1}{\eta^{2/3}}
	\left[ 
	b_2 t^2 + (b_1 \kappa_4 +\kappa_5)t^5 + \frac{\kappa_4^2}{2}\, t^8
	\right].
\end{equation}
Using the identity
 \begin{equation}
 	\label{app:airyder}
 	\int \frac{\dif t}{2 \pi \rmi}\, t^n 
 	\exp\left(-z t +\frac{t^3}{3} \right)
 	 = (-1)^n \frac{\dif^n {\rm Ai}(z)}{\dif z^n}
 \end{equation}
and the fact that all derivatives of the Airy function can be expressed in terms of ${\rm Ai}(z)$ and ${\rm Ai}'(z)$, since ${\rm Ai}''(z) = z {\rm Ai}(z)$, it is not difficult to show that the amplitude reads 
\begin{align}
	\label{app:tbfklsol1}
	\hspace*{-0.9cm}
	\bar{T}_0(\eta,\rho) = &\ \frac{\bar{T}_0(\gamma_\rmc)D_\rmc}{\eta^{1/3}} 
	\exp \left[ \bar{\omega}(\gamma_{\rm c})\eta - \gamma_\rmc \rho \right]
	\bigg\{
	{\rm Ai}(z) + \frac{1}{\eta^{1/3}} \left[\kappa_4 z^2 {\rm Ai}(z) +(2\kappa_4 - b_1) {\rm Ai}'(z) \right]
	\nn
	& \hspace*{-1.1cm}
	+\frac{1}{\eta^{2/3}}
	\left[
	b_2 z {\rm Ai}(z) \!-\! (b_1 \kappa_4+\kappa_5)
	\left[4 z {\rm Ai}(z) \!+\! z^2  {\rm Ai}'(z)\right]
	+\frac{\kappa_4^2}{2}
	\left[ 
	\left(28 z \!+\! z^4 \right) {\rm Ai}(z) +12 z^2 {\rm Ai}'(z)
	\right]
	\right] 
	\bigg\}.
\end{align}
Now it is possible to find at which $z$ the above vanishes. In turn, this determines $\hat{\rho}(\eta)$ up to order $1/\eta^{1/3}$, and we find
\begin{equation}
	\label{app:rhohatnnlo}
	\hat{\rho}(\eta) = \bar{\omega}'(\gamma_\rmc) \eta + 
	\frac{a_1}{D_\rmc}\, \eta^{1/3} + \frac{b_1 - 2\kappa_4}{D_{\rmc}} + \frac{(\kappa_5 - 4 \kappa_4^2) a_1^2}{D_{\rmc}}\, \frac{1}{\eta^{1/3}}. 
\end{equation}
It is a matter of tedious, but straightforward, algebra to expand the solution in \eqn{app:tbfklsol1} around its zero, and in terms of $\xi = \rho - \hat{\rho}(\eta)$ we finally arrive at
\begin{align}
	\label{app:tbfklsol}
	\bar{T}(\eta,\rho) = &\
	\frac{\bar{T}_0(\gamma_c) D_{\rmc}^2 {\rm Ai}'(a_1) \xi}{\eta^{2/3}}\,
	\exp \left[ \bar{\omega}(\gamma_{\rm c})\eta - \gamma_\rmc \hat{\rho}(\eta) - \gamma_\rmc \xi \right]
	\bigg[1+ \frac{\kappa_4 a_1^2}{\eta^{1/3}} 
	\nn
	+ &\frac{a_1}{2 \eta^{2/3}} 
	\big(-b_1^2+ 2 b_2 +32 k_4^2 + a_1^3 \kappa_4^2 -12\kappa_5 \big) +\frac{2 a_1 \kappa_4 D_\rmc \xi}{\eta^{2/3}}
	+\frac{a_1 D_{\rmc}^2 \xi^2}{6 \eta^{2/3}}
	\bigg].
\end{align}
\comment{As in the previous Appendix, assuming the initial condition $\bar{T}_0(\gamma) = \alpha_s^2/[\gamma^2 (1-\gamma)^2]$, which corresponds to dipole-dipole scattering averaged over the target dipole area, we can construct the numerical solution to \eqn{tbfklsol} (for our purposes it is enough to consider only the pole structure of $\bar{\omega}(\gamma)$ as given in \eqn{epsilon}).} 

It is also worthwhile to note in \eqn{app:rhohatnnlo} for $\hat{\rho}(\eta)$ that the coefficient of the $1/\eta^{1/3}$ is universal, since it depends only on the constants $\kappa_m$ which are related to the kernel, but not on the constants $b_l$ which are fixed by the initial condition. On the contrary the constant term is not universal. All this is natural since universal terms should not be affected by an arbitrary shift $\eta \to \eta + \eta_0$. Since there is no $\eta^{2/3}$ term in the expansion in \eqn{app:rhohatnnlo}, a shift cannot induce a change in the $1/\eta^{1/3}$ term. On the contrary, a shift in the very leading term proportional to $\eta$ gives rise to a change in the constant term. Using the same reasoning, one does not expect to find more universal terms in \eqn{app:rhohatnnlo}. The next term would be $ \sim 1/\eta^{2/3}$ and it would be affected by a shift in the $\eta^{1/3}$ term.

\section{The remainder in Eq.~\eqref{lapmel1}}
\label{sec:rem}

In this Appendix we would like to show that the approximations involved in going from \eqn{lap1} to \eqn{lap2} are indeed harmless for the purposes of the discussion in Sect.~\ref{sec:chi}. To be more precise, we will show that the terms neglected when writing Eq.~\eqref{lap2} do not modify the asymptotic eigenvalue branch of the non-local equation with the ``canonical'' shift and thus they do not affect the (exponent of the) asymptotic behavior of the solution. 

To that aim, we return to  \eqn{lap1}, and evaluate the r.h.s. without any approximation: we first shift the rapidity integration variable in the real term and then we isolate an integration from 0 to $\infty$, both in the real and in the virtual term, to reconstruct the Laplace transform of the amplitude; we thus find
\begin{align}
	\label{lap3}
	\hspace*{-0.8cm}
	\omega \bar{T}_{\bx\by}(\omega)\, - &\,  \bar{T}_{\bx\by}(\eta=0) =\frac{\abar}{2\pi}
	\int \frac{\dif^2 \bz \,(\bx\minus\by)^2}{(\bx \minus\bz)^2 (\bz \minus \by)^2}
	\left[2 \rme^{-\omega \delta_{\bx\bz;r}} \bar{T}_{\bx\bz}(\omega)  - \bar{T}_{\bx\by}(\omega) \right]
	\nn
	- & \,\frac{\abar}{2\pi} \int \frac{\dif^2 \bz \,(\bx\minus\by)^2}{(\bx \minus\bz)^2 (\bz \minus \by)^2}
	\left[2 \int_{0}^{\delta_{\bx\by\bz} - \delta_{\bx\bz;r}}
	\!\! \dif \eta\, \rme^{-\omega (\eta +\delta_{\bx\bz;r}) } \bar{T}_{\bx\bz}(\eta)  
	- \int_{0}^{\delta_{\bx\by\bz}}
	\!\! \dif \eta\, \rme^{-\omega \eta} \bar{T}_{\bx\by}(\eta)
	\right].
\end{align}
The second term in this equation has been neglected in writing \eqn{lap2}; we shall refer to it as the ``remainder''. After taking a Mellin transform of Eq.~\eqref{lap3}, the remainder to be added to the r.h.s.~of Eq.~\eqref{lapmel1} is
\begin{equation}
\label{remainder}
\hspace*{-0.4cm}
- \frac{\abar}{2\pi}
	\int\frac{\dif^2 \bz \,(\bx\minus\by)^2}{(\bx \minus\bz)^2 (\bz \minus \by)^2}
	\left\{
	2\left[\,\frac{(\bx\minus\bz)^2}{(\bx\minus\by)^2} \right]^{\gamma}\!
	\int_{0}^{\delta_{\bx\by\bz} - \delta_{\bx\bz;r}}
	\!\! \dif \eta\, \rme^{-\omega (\eta+\delta_{\bx\bz;r})} \bar{T}(\eta,\gamma)
	- \int_{0}^{\delta_{\bx\by\bz}}
	\!\! \dif \eta\, \rme^{-\omega \eta} \bar{T}(\eta,\gamma)
	\right\}.
\end{equation}	
One can already suspect that the above cannot affect the asymptotics, since the $\eta$-integration does not extend all the way to infinity. For simplicity, we shall work to lowest order in $\abar$, which means we can set $\omega=0$. By further expanding $\bar{T}(\eta,\gamma)$ around $\eta=0$, the remainder becomes
\begin{align}
	\label{remainder1}
	& - \bar{T}(\eta=0,\gamma)\,
	\frac{\abar}{2\pi}
	\int \frac{\dif^2 \bz \,(\bx\minus\by)^2}{(\bx \minus\bz)^2 (\bz \minus \by)^2}
	\left\{
	\left[2\,\frac{(\bx\minus\bz)^2}{(\bx\minus\by)^2} \right]^{\gamma} 
	\big(\delta_{\bx\by\bz} - \delta_{\bx\bz;r}\big)
	- \delta_{\bx\by\bz}
	\right\}
	\nn
	& -\bar{T}'(\eta=0,\gamma)\,
	\frac{\abar}{4\pi}
	\int \frac{\dif^2 \bz \,(\bx\minus\by)^2}{(\bx \minus\bz)^2 (\bz \minus \by)^2}
	\left\{
	\left[2\,\frac{(\bx\minus\bz)^2}{(\bx\minus\by)^2} \right]^{\gamma} 
	\big(\delta_{\bx\by\bz} - \delta_{\bx\bz;r}\big)^2
	- \delta_{\bx\by\bz}^2
	\right\} - \cdots.
\end{align}
In a given line in the above, the real and the virtual terms are individually divergent, however their sum is finite and the remainder to lowest order in $\abar$ gives
\begin{equation}
	\label{remainder2}
	\abar f_0(\gamma)\bar{T}(\eta=0,\gamma)
	+\abar f_1(\gamma)\bar{T}'(\eta=0,\gamma) + \cdots ...
\end{equation}
where $f_i(\gamma)$ are regular functions of $\gamma$ in the interval $[0,1]$. Similarly for higher orders in $\abar$. Thus, the remainder can only modify the numerator in Eq.~\eqref{linearsol} by changing the coefficient of $\bar{T}(\eta=0,\gamma)$ and adding terms proportional to the derivatives $\bar{T}^{(n)}(\eta=0,\gamma)$, but it does not modify the asymptotics determined by the positive eigenvalue $\bar{\omega}(\gamma)$ defined in Eq.~\eqref{omegaplus}.

\section{Shifting the virtual term}
\label{sec:shiftv}

In this Appendix we shall study a modification to our non-local equation with the ``canonical'' shift given in \eqref{bketato} which leads to certain attractive features. 

Focusing on the NLO terms in Eqs.~\eqref{omegapluspole0} and \eqref{omegapluspole1}, we see that there is a single pole with residue $-\pi^2/6$ both at $\gamma=0$ and at $\gamma=1$. Now recall that when we match to NLO BK, we must subtract the $\mcal{O}(\abar^2)$ of the shift to avoid double counting, cf.~the piece proportional to $\delta_{\bu\by;r}$ in the third term in Eq.~\eqref{nlobkres}. Hence such poles, now with a positive residue, remain as part of the total eigenvalue. They are presumably very weak to cause any potential problems, nevertheless, since they do not exist in NLO BFKL, it may be desirable to construct a scheme in which they are absent. This can be done, for example, by changing the virtual term in Eq.~\eqref{bketato} as follows 
\begin{equation}
	\label{shiftvirt}
	\bar{S}_{\bx\by}(\eta) \to 
	\Theta(\eta - 2\Delta_{\bx\bz;r})
	\bar{S}_{\bx\by}(\eta - 2\Delta_{\bx\bz;r}),
\end{equation}
where $\Delta_{\bx\bz;r}$ is defined in Eq.~\eqref{deltaopt} and is significant only for large daughter dipoles. In order to match to NLO BK in the $\eta$-representation, we must add to the r.h.s.~of Eq.~\eqref{nlobkres} the opposite of the $\abar^2$ piece induced by Eq.~\eqref{shiftvirt}, that is
\begin{equation}
\label{extravirt}
-\frac{\abar^2 \Delta}{2\pi}
 \int 
 \frac{\dif^2 \bz \,(\bx\minus\by)^2}{(\bx \minus\bz)^2 (\bz \minus \by)^2}\,
 \left[\bar{S}_{\bx\bz}(\eta) \bar{S}_{\bz\by}(\eta) - \bar{S}_{\bx\by}(\eta) \right],
 \end{equation}
which nicely combines with the second term there. For our convenience, in the above we have defined the integral of the shift (weighted by the dipole kernel)
\begin{equation}
	\label{deltaint}
	\Delta = \frac{1}{\pi} \int 
	\frac{\dif^2 \bz \,(\bx\minus\by)^2}{(\bx \minus\bu)^2 (\bu \minus \by)^2}\,\Delta_{\bx\bu;r} = \frac{\pi^2}{6}.
\end{equation}
Therefore, it is clear that the single poles associated with Eq.~\eqref{extravirt} will precisely cancel the aforementioned single poles due to the last term in Eq.~\eqref{nlobkres}.

Staying at the level of matching to LO BK, the modification in Eq.~\eqref{extravirt} leads to the addition of the extra term
\begin{equation}
\label{lapmelextra}
 - \bar{T}(\omega,\gamma)\, \frac{\abar}{2\pi}
	\int \frac{\dif^2 \bz \,(\bx\minus\by)^2}{(\bx \minus\bz)^2 (\bz \minus \by)^2}
\left(\rme^{-2\omega \Delta_{\bx\bz;r}}-1\right)
\end{equation}
to the r.h.s.~of Eq.~\eqref{lapmel1}. The respective ``remainder'', like the one originating from the real terms and studied in Appendix \ref{sec:rem}, does not play any role in the asymptotics. One can perform the integration in Eq.~\eqref{lapmelextra} and the characteristic function analysis done in Sect.~\ref{sec:chi} remains unchanged, except that now we must replace Eq.~\eqref{chireal} by 
\begin{equation}
\label{chigammaomega}	
\chi(\gamma,\omega) = 
2 \psi(1) - \psi(\gamma + \omega) - \psi(\gamma) +
\frac{\psi(2\omega+1)-\psi(1)}{2}.
\end{equation}
It is straightforward to show that the positive solution to $\omega = \abar \chi(\gamma,\omega)$, with $\chi(\gamma,\omega)$ given in the above, does not contain any single poles at order $\abar^2$ as expected.

\section{Delay differential equations}
\label{app:delay}

In this Appendix we study a simple example of a delay differential equation, i.e.~an equation in which the derivative of the unknown function at a certain ``time'' depends on the values of the function at earlier times. To some extent, the problem that we consider here is realistic enough in the sense that it is the zero dimensional analog of the evolution equation discussed in the main text of the present work. The main features confirmed in this simple setup are: a) the delay (shift) slows down the evolution b) the coupling is not any more a good expansion parameter, but there is an effective parameter which depends also on the delay and c) the ``intercept'' determining the asymptotic exponential growth does not depend on the details of the initial condition. We shall study three variations of the same problem, which practically differ only on the way we start the evoltuion.

\smallskip

\noindent {\tt(i)} We start with the simplest case defined by
\begin{equation}
	\label{dfdy1}
	\frac{\dif f(Y)}{\dif Y} = \alpha f(Y - \Delta)
	\quad \text{with} \quad
	f(0)=f_0,
\end{equation} 
and where we shall assume that $\Delta>0$. This can be immediately solved and gives
\begin{equation}
	\label{f1sol}
	f(Y) = f_0 \exp(\omega Y),
\end{equation}
where the ``intercept'' $\omega$ is determined by the real solution to the transcendental equation
\begin{equation}
	\label{omega1}
	\omega = \alpha \exp(-\omega \Delta).
\end{equation}
Since $\Delta>0$, the above implies that $\omega < \alpha$. That is, the intercept in the delayed evolution is smaller than the intercept in the absence of a delay. It is also instructive to mention that an iterative solution to \eqn{omega1} can be constructed and reads  
\begin{equation}
	\label{omega1s}
	\omega = \alpha \left(1 - \Delta \alpha + 
	\frac{3}{2}\, \Delta^2 \alpha^2 + \cdots\right).
\end{equation}
This makes clear that the effective parameter is $\Delta \alpha$: even when the ``coupling'' $\alpha$ is small, the fixed order expansion in \eqn{omega1s} will not be valid when $\Delta \alpha \gtrsim 1$. (Notice that the effective parameter in the QCD problem is $\alpha \Delta^2$, where the additional factor of $\Delta$ is generated by the eventual transverse integration on the r.h.s.~of the evolution equation.)

\smallskip

\noindent {\tt(ii)} Second, we consider the problem
\begin{equation}
	\label{dfdy2}
	\frac{\dif f(Y)}{\dif Y} = \alpha \Theta(Y-\Delta) 
	f(Y - \Delta)
	\quad \text{with} \quad
	f(0)=f_0,
\end{equation}
and for which we would like to determine the solution when $Y>0$. This has to be solved interval by interval in $Y$ in steps of $\Delta$. When $0 \leq Y < \Delta$, the r.h.s.~vanishes due to the presence of the step function and thus $f(Y)$ remains constant, that is
\begin{equation}
	\label{f2sola}
	f(Y) = f_0 
	\quad \text{for} \quad
	0 \leq Y \leq \Delta.
\end{equation}
In the second interval $\Delta \leq Y < 2\Delta$, the r.h.s.~is determined by the solution to the previous interval and therefore it is constant. Requiring continuity of the solution when $Y=\Delta$, we readily obtain
\begin{equation}
	\label{f2solb}
	f(Y) = f_0 + \alpha f_0 (Y - \Delta) 
	\quad \text{for} \quad
	\Delta \leq Y \leq 2 \Delta.
\end{equation}
Let us explicitly do one last interval and find the solution when $2 \Delta \leq Y < 3\Delta$. The r.h.s.~is fixed by the branch in \eqn{f2solb} evaluated at $Y-\Delta$, and by integrating over $Y$ and requiring continuity at $Y=2\Delta$ we arrive at
\begin{equation}
	\label{f2solc}
	f(Y) = f_0 + \alpha f_0 (Y - \Delta)+ \frac{\alpha^2}{2} f_0 (Y - 2 \Delta)^2
	\quad \text{for} \quad
	2\Delta \leq Y \leq 3 \Delta.
\end{equation}
It becomes clear that the solution at any interval reads
\begin{equation}
	\label{f2sol}
	f(Y) = f_0 \sum_{\kappa=0}^{n} 
	\frac{\alpha^\kappa (Y - \kappa \Delta)^\kappa}{\kappa!}
	\quad \text{for} \quad
	n\Delta \leq Y \leq (n+1) \Delta.
\end{equation}
At first glance, it seems that \eqn{f2sol} does not have much in common with the solution of case {\tt (i)} given in \eqn{f1sol}. Nevertheless, when $Y$ is sufficiently large, the summation in \eqn{f2sol} is dominated by terms for which $\kappa$ is large, but smaller than $n$. Then the summation can be approximated by an integration and using Stirling's formula for the factorial, we find
\begin{equation}
	\label{f2solint}
	f(Y) \simeq f_0 \int \frac{\dif \kappa}{\sqrt{2\pi \kappa}}\,
	\exp\left\{\kappa \left[ \ln \left(\alpha Y - \kappa \alpha \Delta\right) - \ln \kappa +1 \right] \right\},
\end{equation}
which can be performed by employing the method of the steepest descend. Calling $\mcal{E}(\kappa)$ the exponent in the above, we can find the location of the saddle point by solving
\begin{equation}
	\label{saddle1}
	\mcal{E}'(\kappa_0) = 0 
	\,\,\Rightarrow\,\,
	\ln\frac{\alpha Y - \kappa_0 \alpha \Delta}{ \kappa_0} = 
	\frac{\kappa_0 \Delta}{Y - \kappa_0 \Delta}.
\end{equation}
It is not difficult to see in the above that $\kappa_0$ scales with $Y$ and, by using a notation convenient for our purposes, we find that
\begin{equation}
	\label{kappa0}
	\kappa_0 = \frac{\omega Y}{1 + \omega \Delta},
\end{equation}
where $\omega$ is precisely the one solving \eqn{omega1}. (Since $Y \simeq n \Delta$, notice that $\kappa_0$ is smaller than $n$ as it should.) The integration in \eqn{f2solint} becomes
\begin{equation}
	\label{f2solint1}
	f(Y) = \frac{f_0 \exp[\mcal{E}(\kappa_0)]}{\sqrt{2\pi \kappa_0}}
	\int \dif \kappa 
	\exp\left[-\frac{1}{2}\, \big|\mcal{E}''(\kappa_0)\big| (\kappa-\kappa_0)^2 \right]
	= \frac{f_0 \exp[\mcal{E}(\kappa_0)]}{\sqrt{\kappa_0  \big|\mcal{E}''(\kappa_0)\big|}}.
\end{equation}
We easily find
\begin{equation}
	\label{epsilons}
	\mcal{E}(\kappa_0) = \omega Y
	\quad \text{and} \quad 	\mcal{E}''(\kappa_0) = - \frac{(1 + \omega \Delta)^3}{\omega Y},
\end{equation}
so that finally the solution to \eqn{dfdy2} for large $Y$ is given by
\begin{equation}
	\label{f2solapp}
	f(Y) \simeq \frac{f_0 \exp(\omega Y)}{1 + \omega \Delta}.
\end{equation}

\smallskip

\noindent {\tt(iii)} Last, we take the case
\begin{equation}
	\label{dfdy3}
	\frac{\dif f(Y)}{\dif Y} = \alpha f(Y - \Delta)
	\quad \text{with} \quad
	f(Y \leq 0) = f_0.
\end{equation}
The construction of the exact solution is similar to the one in case {\tt (ii)} and one finds
\begin{equation}
	\label{f3sol}
	f(Y) = f_0 \sum_{\kappa=0}^{n+1} 
	\frac{\alpha^\kappa [Y - (\kappa-1) \Delta)]^\kappa}{\kappa!}
	\quad \text{for} \quad
	n\Delta \leq Y \leq (n+1) \Delta.
\end{equation}
It is not hard to be convinced that the asymptotic form of the above can be obtained by letting $Y \to Y + \Delta$ in the corresponding expression of case {\tt (ii)}, that is 
\begin{equation}
	\label{f3solapp}
	f(Y) \simeq \frac{f_0 \exp(\omega \Delta) \exp(\omega Y)}{1 + \omega \Delta}.
\end{equation}

\smallskip

Therefore we see that, although the exact solution to the three variations of the problem is different, the asymptotic solution is very similar. Eqs.~\eqref{f1sol}, \eqref{f2solapp} and \eqref{f3solapp} share the same ``intercept'' $\omega$, determined by \eqn{omega1} and they differ only in the overall prefactor.


\smallskip

\providecommand{\href}[2]{#2}\begingroup\raggedright\endgroup

\end{document}